\documentclass[aps,prd,reprint,onecolumn,nofootinbib,longbibliography,12pt,superscriptaddress]{revtex4-2}

\usepackage[utf8]{inputenc}
\usepackage{helvet}
\usepackage{ulem}

\usepackage{amsmath}
\usepackage{amsfonts}
\usepackage{amssymb}
\usepackage{mathtools}
\usepackage{latexsym}
\usepackage{tensor}
\usepackage{cancel}
\usepackage{stackengine}
\usepackage{siunitx}
\usepackage[colorlinks=true, linkcolor=blue, citecolor=blue, urlcolor=blue]{hyperref}
\usepackage{setspace}

\usepackage{array}
\usepackage{booktabs}
\usepackage{multirow}
\usepackage{framed}
\allowdisplaybreaks

\usepackage{graphicx}
\usepackage{caption}
\usepackage{subcaption}
\usepackage{wrapfig}

\usepackage{xcolor}

\usepackage{url}
\usepackage{hyperref}
\usepackage{cleveref}

\usepackage{eurosym}

\numberwithin{equation}{section}

\renewcommand{\theequation}{\arabic{section}.\arabic{equation}}

\crefname{section}{Sec.}{Secs.}
\Crefname{section}{Section}{Sections}
\crefname{figure}{Fig.}{Figs.}
\Crefname{figure}{Figure}{Figures}
\crefname{table}{Tab.}{Tabs.}
\Crefname{table}{Table}{Tables}
\crefname{equation}{Eq.}{Eqs.}
\Crefname{equation}{Equation}{Equations}

\newcommand{\be}{\begin{equation}}
\newcommand{\ee}{\end{equation}}
\newcommand{\bea}{\begin{eqnarray}}
\newcommand{\eea}{\end{eqnarray}}

\begin{document}

\title{Dimming of Photon Ring due to Photon-Axion Conversion around Kerr Black Holes}

\author{Rahul Dhyani}
\email{rahul.physics123@gmail.com}
\thanks{These authors contributed equally to this work.}
\affiliation{Department of Physics, School of Natural Sciences, Shiv Nadar Institution of Eminence (Deemed to be University), Tehsil Dadri, Gautam Buddha Nagar, Uttar Pradesh, 201314, India}

\author{Sauvik Sen}
\email{sauviksen.physics@gmail.com}
\thanks{These authors contributed equally to this work.}
\affiliation{Department of Physics, School of Natural Sciences, Shiv Nadar Institution of Eminence (Deemed to be University), Tehsil Dadri, Gautam Buddha Nagar, Uttar Pradesh, 201314, India}

\author{Indrani Banerjee}
\email{banerjeein@nitrkl.ac.in}
\affiliation{Department of Physics and Astronomy, National Institute of Technology, Rourkela, Odisha, 769008, India}

\author{Ashmita Chakraborty}
\email{ashmitachakraborty20@gmail.com}
\affiliation{Department of Physics, School of Natural Sciences, Shiv Nadar Institution of Eminence (Deemed to be University), Tehsil Dadri, Gautam Buddha Nagar, Uttar Pradesh, 201314, India}
\affiliation{Fachbereich Physik, Universität Hamburg, 22607 Hamburg, Germany}

\author{Arindam Chatterjee}
\email{arindam.chatterjee@snu.edu.in}
\affiliation{Department of Physics, School of Natural Sciences, Shiv Nadar Institution of Eminence (Deemed to be University), Tehsil Dadri, Gautam Buddha Nagar, Uttar Pradesh, 201314, India}

\date{\today}

\begin{abstract}
\begin{spacing}{0.95}
We investigate photon–axion conversion in the vicinity of rotating Kerr black holes where strong gravity traps photons on near-circular trajectories, effectively enhancing the path length. We explore the observable signatures of such a conversion near the photon region.
The process, driven by ambient magnetic fields, is significantly more efficient around supermassive black holes such as M87*, since the luminosity of photons increases with the mass of the BH.
By numerically evaluating photon path lengths (on which the conversion depends), we analyze how key parameters—photon frequency, axion mass, photon–axion coupling, magnetic field strength, plasma density, and black hole spin—affect the conversion probability and the resultant dimming of photon spectral luminosity.
We find that the conversion is most efficient at high frequencies (X-rays and gamma rays), while the frequency window associated with efficient conversion widens with an increase in the photon-axion coupling and a decrease in the electron density and the axion mass. The magnitude of dimming of the photon spectral luminosity depends primarily on the magnetic field, the photon-axion coupling and the BH spin. Our study reveals that rotating black holes generally exhibit enhanced dimming compared to static ones. Thus, if future telescopes achieving a resolution $\sim 10^{-5}$ arcsec in the X-ray/gamma-ray band detect a dimming of the photon spectral luminosity, then they can provide interesting constraints on the axion mass and its coupling with photons.
\end{spacing}
\end{abstract} %%%%%%%%%
 
\maketitle
\baselineskip=0.8\baselineskip
\tableofcontents
\section{Introduction}\label{Intro}
Recent advancements in astrophysical observations, particularly 
the first image of the supermassive black hole M87* captured by 
the Event Horizon Telescope (EHT)~\cite{EventHorizonTelescope:2019dse, EventHorizonTelescope:2019uob, EventHorizonTelescope:2019jan, EventHorizonTelescope:2019ths, EventHorizonTelescope:2019pgp, EventHorizonTelescope:2019ggy}, have opened up exciting new possibilities for testing fundamental physics in the strong gravity regime. This observation has provided insights into how photons behave near the event horizon of a compact object. In particular, it has established the presence of a photon region where light rays are temporarily trapped in unstable spherical photon orbits due to the strong gravity of the black hole (BH). Such photons are incident at nearly critical impact parameters, but since their orbits are unstable, a slight perturbation causes the photon to either plunge into the event horizon or escape to infinity. Since the photon region is very close to the event horizon, observations associated with the photon region can reveal crucial information regarding the nature of gravitational interaction in the strong field regime. This has only been possible recently with the advent of the Event Horizon Telescope. Apart from providing exciting opportunities to probe the nature of strong gravity, it also offers a unique observational window into new physics beyond the Standard Model, one such possibility being investigating the interaction between photons and axions in the vicinity of BHs.

Axion is a hypothetical particle which was originally proposed as 
a solution to the strong CP problem in quantum chromodynamics 
(QCD)~\cite{Peccei:1977hh,PhysRevD.16.1791,PhysRevLett.40.223,PhysRevLett.40.279}. In such models, the axion mass is directly related 
to its coupling constant. However, in various extensions of the 
Standard Model and in string theory, a broader class of pseudo-
scalar fields commonly known as axion-like particles (ALPs) are 
predicted~\cite{Svrcek:2006yi,Arvanitaki:2009fg}.
Unlike the QCD axion, these ALPs generally exhibit no fixed 
relationship between their mass and coupling constant, allowing 
them to assume a very light mass, potentially extending to 
extremely low values. Axions are of considerable cosmological 
interest ~\cite{Marsh:2015xka} since heavy axions can naturally 
support slow-roll inflation~\cite{Freese:1990rb,Kim:2004rp,Dimopoulos:2005ac}, light axions may constitute dark matter \cite{preskill:1983,Sikivie:1982i,Dine1983,Hui:2016ltb,Chadha-Day:2021szb}, and ultra-light axions with masses $\sim10^{-33}\,\mathrm{eV}$ can behave  effectively as Dark Energy \cite{Frieman:1995pm,Choi:1999xn,Copeland:2006wr,Marsh:2015xka}. 
This makes cosmology a powerful arena for probing axions and their 
mass. 

 In the presence of magnetic fields, axions (or, more generally ALPs) can convert into photons and vice versa through 
 their coupling, described by the interaction term $\mathcal{L}_{\text{int}} = -\frac{1}{4}g_{a\gamma} \, \phi \, F_{\mu\nu} 
 \tilde{F}^{\mu\nu}$, where $g_{a\gamma}$ is the axion–photon 
 coupling constant, $\phi$ is the axion field, $F_{\mu\nu}$ is the 
 electromagnetic field strength tensor, and $\tilde{F}^{\mu\nu}$ 
 is its dual~\cite{Maiani:1986,Raffelt:1988}. 
 Photon--axion conversion provides the theoretical foundation~\cite{Sikivie1983sk} for experimental searches for axions produced in the Sun~\cite{Cameron:1993bhr,Minowa:1998sj,CAST:2004gzq,CAST:2024eil}, in laboratory-based experiments~\cite{GammeVT-969:2007pci,Ehret:2010zz,Povey:2010hs,OSQAR:2013jqp,OSQAR:2015qdv,ALPSII:2025eri}, and for axions as dark matter candidates~\cite{Asztalos2010,Ouellet:2018beu,Crisosto:2019fcj,Gramolin:2020ict,Salemi:2021gck,Aybas:2021nvn,McAllister:2022ibe,QUAX:2023gop,MADMAX:2024jnp,QUAX:2024fut,Quiskamp:2024oet,ADMX:2025vom}. It has also been extensively explored in a broad range of cosmological and astrophysical contexts. It has been proposed 
 that high-energy photons from extragalactic sources can traverse 
 cosmological distances via conversion into axions and subsequent 
 reconversion into photons; without this mechanism, such photons 
 would be strongly attenuated through electron–positron pair 
 production on the extragalactic background light \cite{DeAngelis:2007dqd,Simet:2007sa,Conde2009,Mirizzi:2009aj,Meyer:2013pny}. This idea has been invoked to explain recent 
 observations of very-high-energy gamma-ray photons \cite{Galanti:2022chk,Troitsky:2022xso,Baktash:2022gnf,Lin:2022ocj,Gonzalez:2022opy,Nakagawa:2022wwm,Carenza:2022kjt,Galanti:2022xok}. Photon–axion conversion has also been suggested 
 to induce spectral distortions in the cosmic microwave background \cite{Yanagida1988,Tashiro:2013yea,Mirizzi:2009nq} and in the X-
 ray and gamma-ray spectra of high-energy astrophysical sources, 
 such as active galactic nuclei.~\cite{Hooper:2007bq,Hochmuth:2007hk,DeAngelis:2007wiw,HESS:2013udx,Fermi-LAT:2016nkz,Marsh:2017yvc,Zhang:2018wpc,Reynolds:2019uqt}. Such observations impose strong constraints on the photon-axion 
 coupling $g_{a\gamma}\lesssim 10^{-11}-10^{-10}~\rm {GeV}^{-1}$ 
 in the mass range $m_a\lesssim 10^{-5} $ eV ~\cite{Noordhuis:2022ljw,Dolan:2022kul,Dessert:2022yqq}.
 
 The photon-axion conversion has important implications in 
 astrophysics, particularly near compact objects with strong 
 magnetic fields, where such conversions can be efficient. 
 Supermassive black holes (SMBHs), such as M87*, harboring strong 
 magnetic field environments, can also act as natural laboratories 
 for this phenomenon~\cite{Nomura:2022zyy,Roy:2023rjk,Hazarika:2024nrj}. 
The Event Horizon Telescope has observed polarized synchrotron emission at 230 GHz from near the event horizon of M87*, indicating magnetic field strengths $\sim 1\text{--}30$ G, the average electron number density $n_e\sim 10^4-10^7 \rm {cm^{-3}}$ and the electron temperature $T_e \sim 10^{11}$K of the surrounding plasma  ~\cite{EventHorizonTelescope:2021srq}. Such fields are therefore expected to be common in the vicinity of black holes, motivating studies of photon–axion conversion in these environments. The photon propagation length required for conversion is typically comparable to or larger than the horizon radius of a supermassive black hole, which might suggest that a magnetic field must be sustained over similar radial scales. However, strong gravitational effects allow photons to remain near the black hole for extended periods. In particular, photons can orbit in the photon region on unstable circular trajectories before escaping to distant observers. During this phase, they propagate at an approximately constant radius, naturally maintaining the magnetic field along their paths. If such photons get converted to axions they are expected to cause a dimming of the photon ring of M87* observed by the EHT collaboration. 

In the present work we investigate photon-axion conversion around Kerr black holes. Previously such conversion has been studied in the static BH spacetimes \cite{Nomura:2022zyy,Roy:2023rjk}. The rotating Kerr BH has a photon region instead of a photon sphere for static BHs, where spherical photon orbits are possible. We numerically calculate the path length traversed by the photon near the photon region which are incident at impact parameters slightly deviated from the critical values. Thereafter, we study the probability of conversion of such photons to axions in the presence of magnetic fields as observed near M87*.  Such an investigation enables us to understand the role of the magnetic field, the electron number density, the photon frequency and the black hole spin on the photon-axion conversion. Furthermore, we also explore the role of the axion mass and the photon-axion coupling constant on the conversion probability. Such a conversion is most effective in the gamma-ray and X-ray band where measuring distortions in the photon ring spectrum opens a new avenue for constraining axion properties. If photon-axion conversion occurs in the photon region, it could leave observable imprints, such as frequency-dependent dimming or distortions in the brightness profile of the photon ring. Detecting such signatures would provide compelling evidence for new physics and deepen our understanding of both gravity and particle interactions in extreme environments.

The paper is organized as follows: In \cref{Sec2} we provide the theoretical framework for photon-axion conversion in the presence of magnetic fields. In \cref{S3} we derive the path length traversed by the photon in the photon region. The dimming of the photon ring due to conversion to axions is investigated in \cref{Sec4} and the resultant attenuation in the photon spectra is explored in \cref{S5}. We summarize the main findings of our work, discuss its implications and conclude with some scope for future work in \cref{S6}.

\section{Theoretical Framework}
\label{Sec2}
\subsection{Axion-photon conversion}

In this section, we briefly review the photon-axion conversion phenomenon in an external magnetic field. 

The Lagrangian for the photon-axion system is given by~\cite{Raffelt:1988} : 

\begin{equation}
    \mathcal{L} = -\frac{1}{4}F_{\mu\nu}{F}^{\mu\nu} - \frac{1}{2}\partial_\mu\phi\partial^\mu\phi -\frac{1}{2}m_a^2\phi^2 -\frac{1}{4}g_{a\gamma}\phi F_{\mu\nu}\tilde{F}^{\mu\nu} + \frac{\alpha^2}{90m_e^4}\left[(F_{\mu\nu}F^{\mu\nu})^2 + \frac{7}{4}(\tilde{F}_{\mu\nu}F^{\mu\nu})^2\right],
\end{equation} 
where $\alpha =\frac{e^2}{4\pi} = \frac{1}{137}$ is the fine-structure constant, $m_e = 511 \rm keV$ is the mass of the electron, $\phi$ is the axion field, $F_{\mu\nu}=\partial_\mu A_\nu - \partial_\nu A_\mu$ is the electromagnetic field strength tensor, $\tilde{F}_{\mu\nu} = \frac{1}{2}\epsilon_{\mu\nu\rho\sigma}F^{\rho\sigma}$ is its dual (with the totally antisymmetric tensor $\epsilon_{\mu\nu\rho\sigma}$ normalized as $\epsilon_{0123}=1$), $m_a$ is the axion mass and $g_{a\gamma}$ is the photon-axion coupling constant. We assume $\hbar=c=1$ in this section.
The last term, known as the Euler-Heisenberg effective Lagrangian, accounts for corrections in the presence of an external magnetic field. In what follows, we derive the photon--axion conversion probability from the above Lagrangian under the assumption that the produced axions are relativistic such that the photon frequency $\omega$ satisfies $\omega \gg m_a \,$.

The relativistic axion scenario is particularly relevant for high-energy photons such as X-rays and gamma rays, especially in the vicinity of compact objects like the supermassive black hole M87* \cite{Nomura:2022zyy,Roy:2023rjk}. In such environments, photon-axion conversion can be significantly enhanced by strong magnetic fields and low plasma densities.
The probability for a photon to convert into an axion after traversing a distance $ z $ through a magnetic field is derived in Appendix A and is given by \cite{Raffelt:1988}:
\begin{align}
P_{\gamma \to a}(z) &= \left( \frac{2\Delta_{\mathrm{M}}}{\Delta_{\mathrm{osc}}} \right)^2 \sin^2 \left( \frac{\Delta_{\mathrm{osc}} z}{2} \right), \label{eq:conversion_prob} \\
\Delta_{\mathrm{osc}}^2 &= \left( \Delta_{\mathrm{pl}} - \Delta_{\mathrm{vac}} - \Delta_\mathrm{a} \right)^2 + 4 \Delta_{\mathrm{M}}^2, \label{eq:oscillation}
\end{align}
where $\Delta_{\mathrm{osc}}$ is related to the total oscillation wave number, and the terms $\Delta_{\mathrm{M}}, \Delta_\mathrm{a}, \Delta_{\mathrm{pl}}, \Delta_{\mathrm{vac}}$ encapsulate the physical components of the system.

The mixing term $\Delta_{\mathrm{M}}$ arises from the axion-photon coupling in the presence of a magnetic field, and is expressed as
\begin{align}
\Delta_{\mathrm{M}} = \SI{9.8e-23}{\electronvolt} 
\left( \frac{g_{a\gamma}}{\SI{e-11}{\per\giga\electronvolt}} \right) 
\left( \frac{B}{\text{Gauss}} \right). \label{eq:delta_M}
\end{align}
This term is responsible for initiating the conversion and depends 
linearly on both the axion-photon coupling constant $ g_{a\gamma} $ 
and the strength of the magnetic field $ B $ perpendicular to the 
photon’s propagation direction.

The term $\mathrm{\Delta_a}$ accounts for the axion mass and 
suppresses the mixing term when the photon energy is low
\begin{align}
\mathrm{\Delta_a} = \SI{5.0e-22}{\electronvolt} 
\left( \frac{m_a}{\si{\nano\electronvolt}} \right)^2 
\left( \frac{\si{\kilo\electronvolt}}{\omega} \right). \label{eq:delta_a}
\end{align}
In \cref{eq:delta_a}, $m_a$ is the axion mass and $\omega$ is the photon frequency.
Plasma effects are encapsulated in the term $\Delta_{\mathrm{pl}}$, which originates from the effective mass acquired by the photon due to interactions with electrons in the plasma:
\begin{align}
\Delta_{\mathrm{pl}} = \SI{6.9e-25}{\electronvolt} 
\left( \frac{n_e}{\si{\per\cubic\centi\meter}} \right) 
\left( \frac{\si{\kilo\electronvolt}}{\omega} \right). \label{eq:delta_pl}
\end{align}
The electron number density \( n_e \) sets the plasma frequency via
\begin{align}
\omega_{\mathrm{pl}} &\equiv \sqrt{\frac{4 \pi e^2 n_e}{m_e}}, \notag \\
&= \SI{3.7e-11}{\electronvolt} \sqrt{\frac{n_e}{\si{\per\cubic\centi\meter}}}. \label{eq:plasma_freq}
\end{align}
 The requirement $\omega \gg \omega_{\mathrm{pl}}$ ensures that the photon can propagate through the medium.

Vacuum birefringence effects, arising from QED loop corrections in the presence of magnetic fields, contribute through the term
\begin{align}
\Delta_{\mathrm{vac}} = \SI{9.3e-29}{\electronvolt} 
\left( \frac{\omega}{\si{\kilo\electronvolt}} \right) 
\left( \frac{B}{\text{Gauss}} \right)^2. \label{eq:delta_vac}
\end{align}
Although typically small, this term can become relevant in extremely strong magnetic fields and extremely high photon frequencies.
To ensure the applicability of this formalism, three physical conditions must be satisfied. First, the nonlinear QED corrections encapsulated in the Euler–Heisenberg effective Lagrangian must remain small, which is guaranteed when
\[
\left( \frac{\alpha}{45 \pi} \right) \left( \frac{B}{B_{\mathrm{crit}}} \right)^2 \ll 1,
\]
where the critical magnetic field is given by
\[
B_{\mathrm{crit}} = \frac{m_e^2}{\sqrt{4 \pi \alpha}}\simeq 4\times 10^{13}\rm G.
\]
Second, as noted earlier, the photon energy must significantly exceed the plasma frequency, i.e., $\omega \gg \omega_{\mathrm{pl}}$, so that the photon is not screened by the plasma and finally the photon energy should exceed the axion mass, such that the relativistic axion assumption holds.
When these conditions are met, the probability expression in ~\cref{eq:conversion_prob} provides an estimate of photon-axion conversion in relativistic regimes relevant to astrophysical observations involving high-energy radiation in magnetized environments. For further details, one is referred to Appendix A (also see~\cite{Raffelt:1988,Masaki:2017aea,Nomura:2022zyy}).

\begin{figure}[!htbp]
\centering

% -------- Row 1 --------
\begin{subfigure}[b]{0.32\textwidth}
    \centering
    \includegraphics[width=\linewidth]{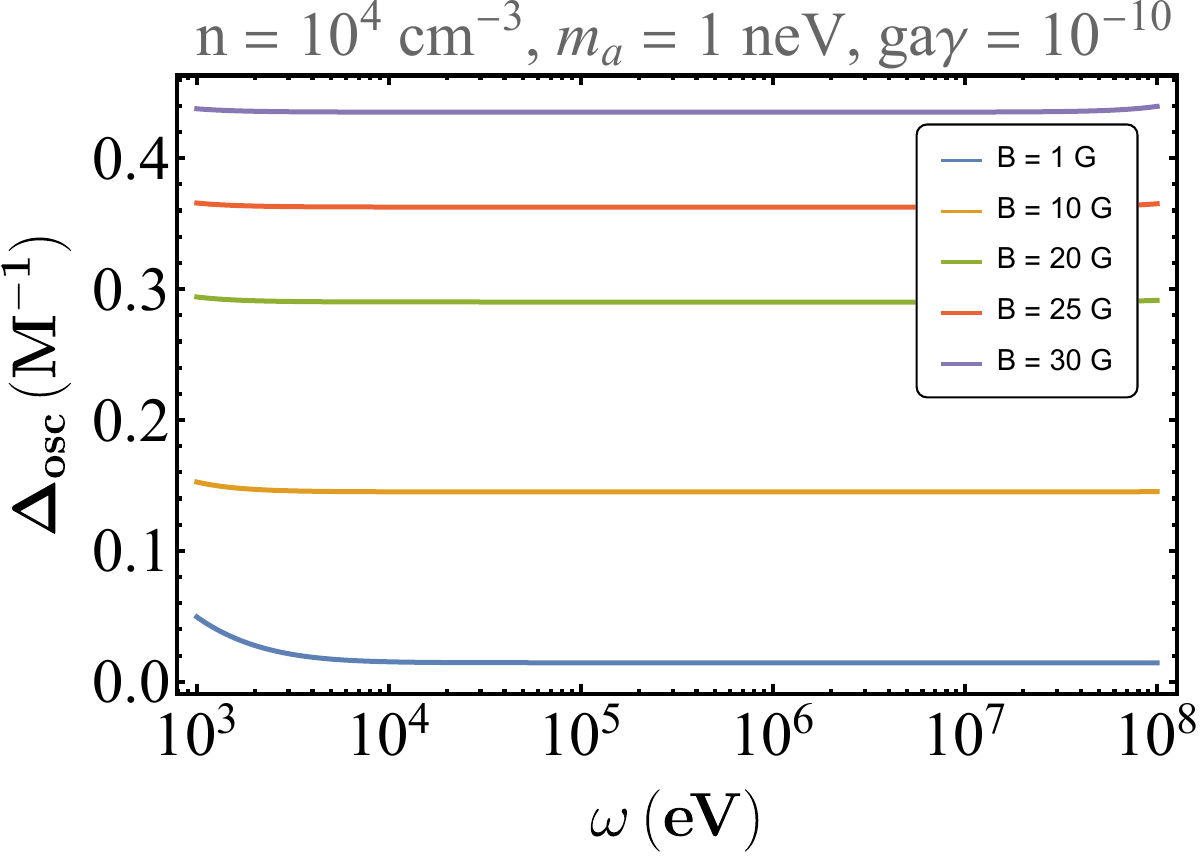}
        \caption{}
        \label{1a}
\end{subfigure}\hfill
\begin{subfigure}[b]{0.32\textwidth}
    \centering
    \includegraphics[width=\linewidth]{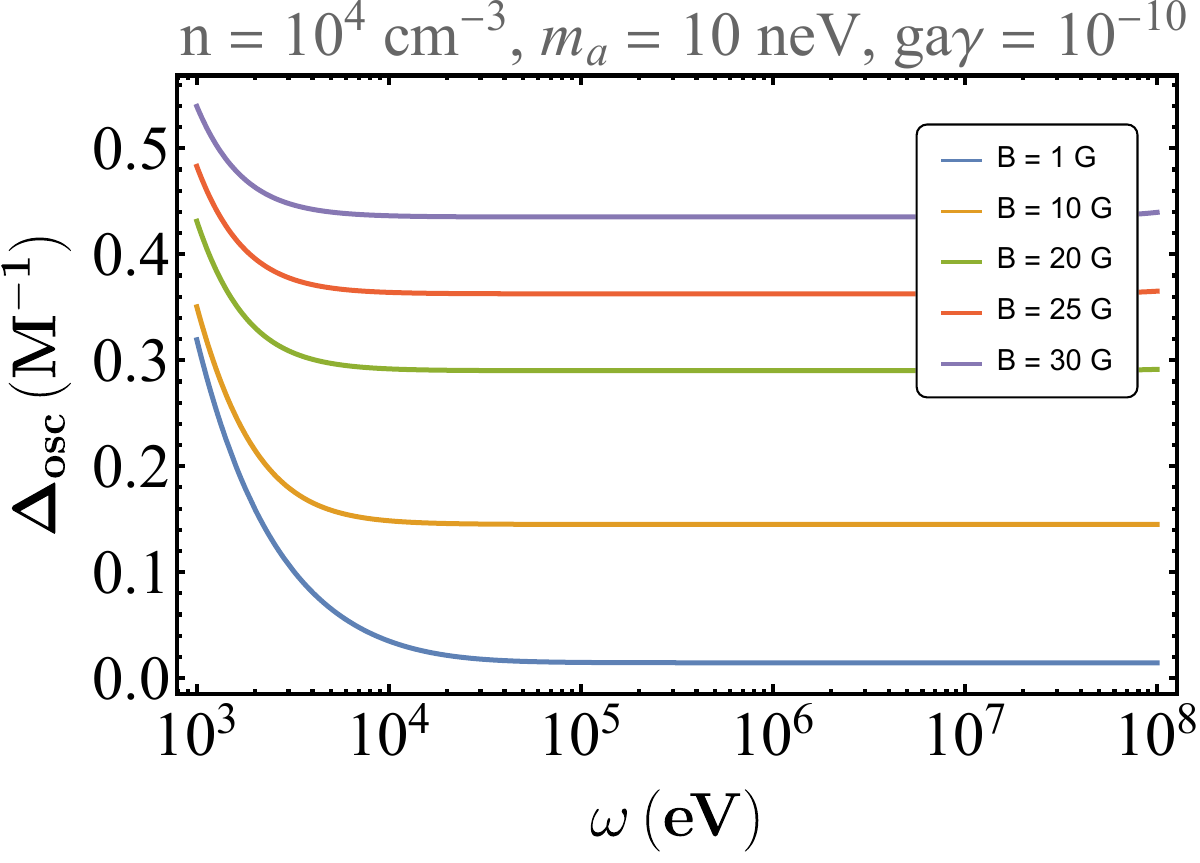}
    \caption{}
    \label{1b}
\end{subfigure}\hfill
\begin{subfigure}[b]{0.32\textwidth}
    \centering
    \includegraphics[width=\linewidth]{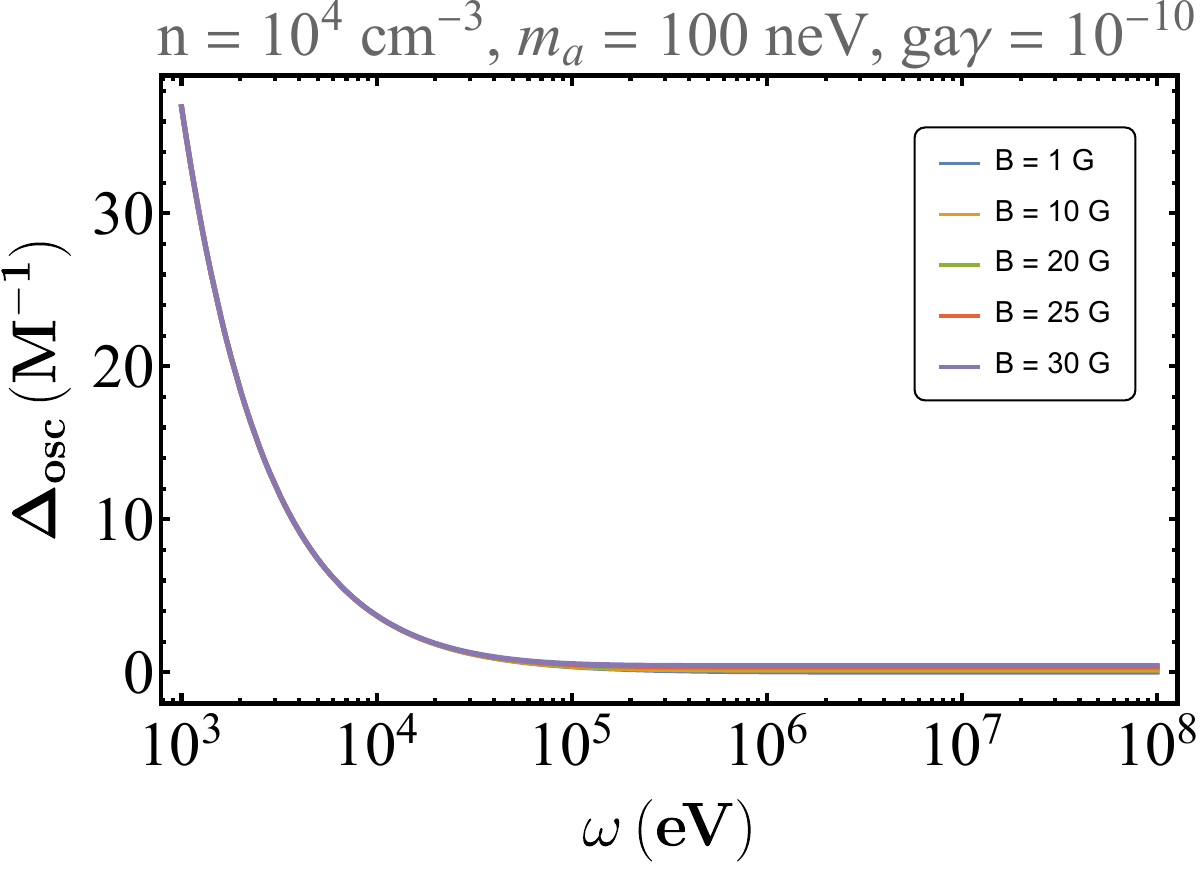}
    \caption{}
    \label{1c}
\end{subfigure}

\vspace{0.3cm}

% -------- Row 2 --------
\begin{subfigure}[b]{0.32\textwidth}
    \centering
    \includegraphics[width=\linewidth]{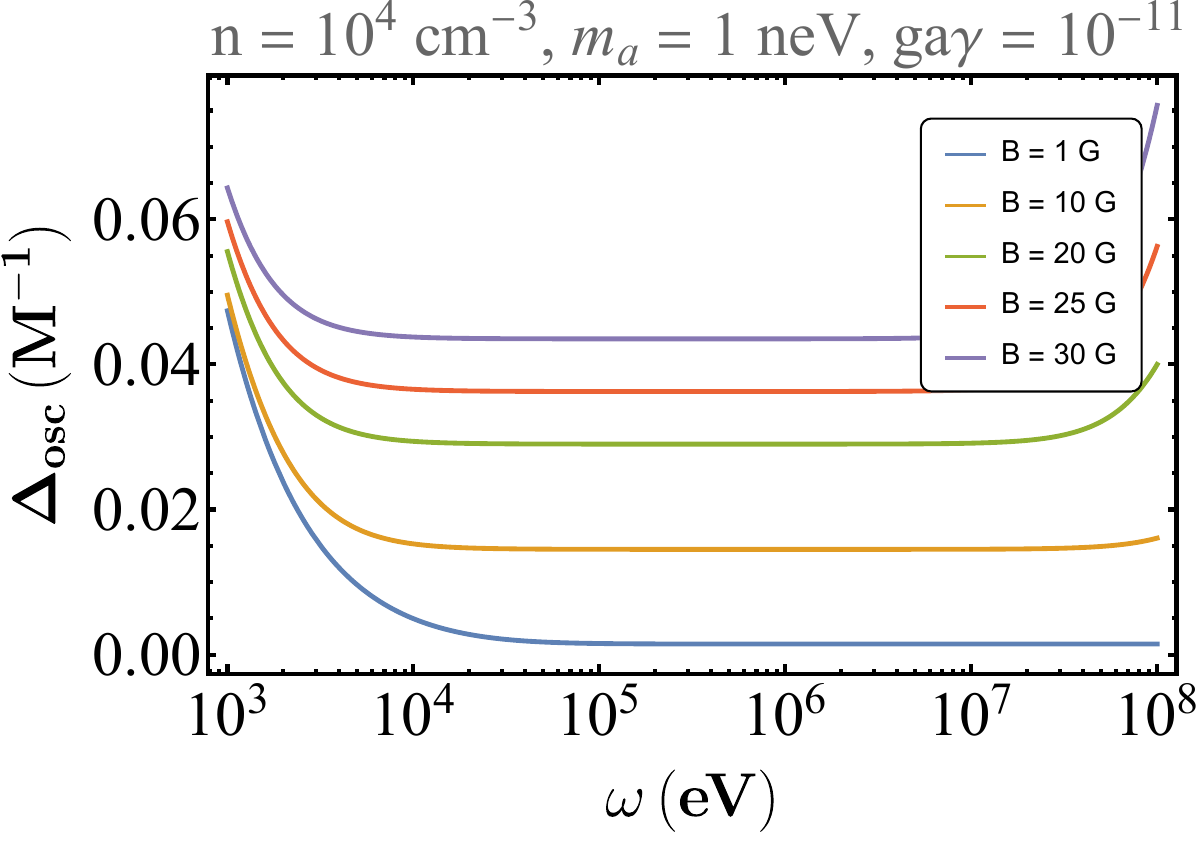}
    \caption{}
    \label{1d}
\end{subfigure}\hfill
\begin{subfigure}[b]{0.32\textwidth}
    \centering
    \includegraphics[width=\linewidth]{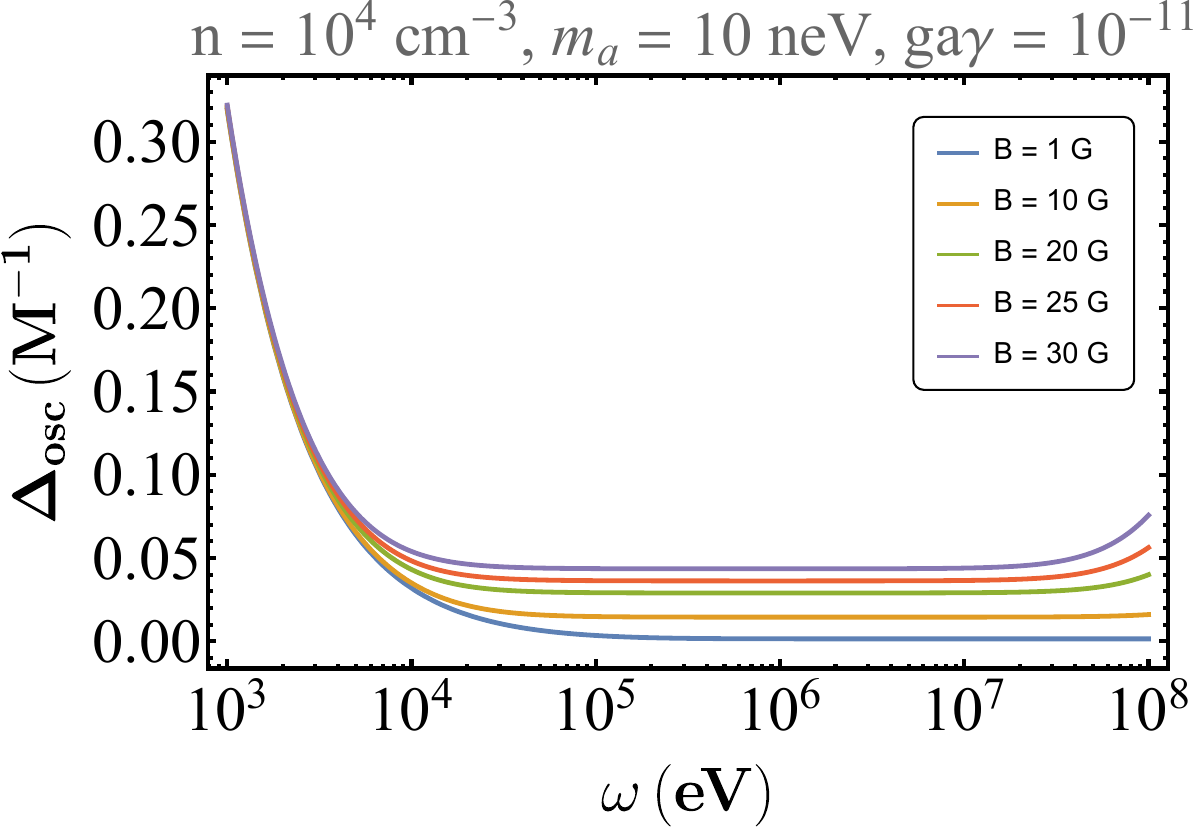}
    \caption{}
    \label{1e}
\end{subfigure}\hfill
\begin{subfigure}[b]{0.32\textwidth}
    \centering
    \includegraphics[width=\linewidth]{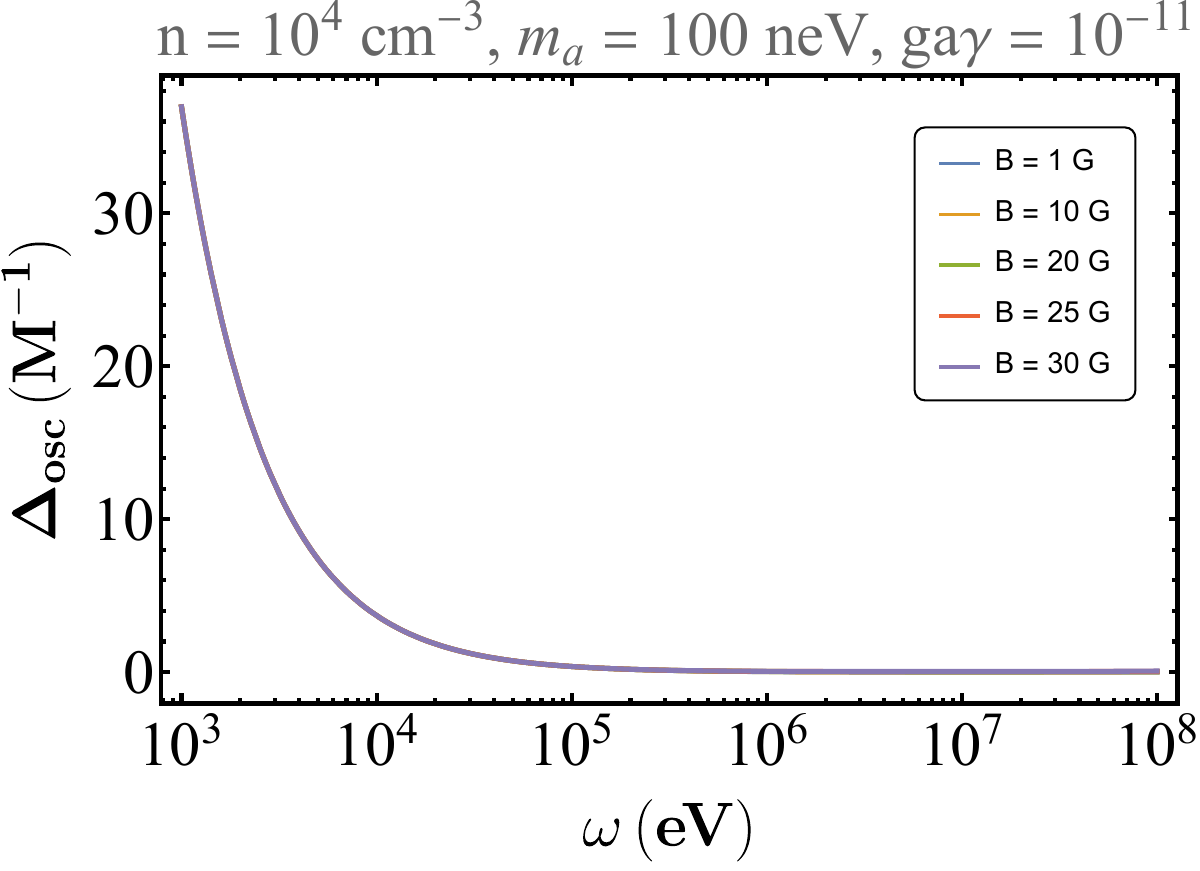}
    \caption{}
    \label{1f}
\end{subfigure}

\vspace{0.3cm}
% ---------------------

\vspace{0.3cm}

\caption{The above figure illustrates the variation of $\Delta_{\mathrm{osc}}$ with photon frequency $\omega$ for different magnetic fields, axion mass and photon-axion coupling. The electron number density is fixed to $n_e=10^4~cm^{-3}$}.
\label{fig:delta_osc_mag_field}
\end{figure}

\begin{figure}[!htbp]
\centering

% -------- Row 1 --------
\begin{subfigure}[b]{0.32\textwidth}
    \centering
    \includegraphics[width=\linewidth]{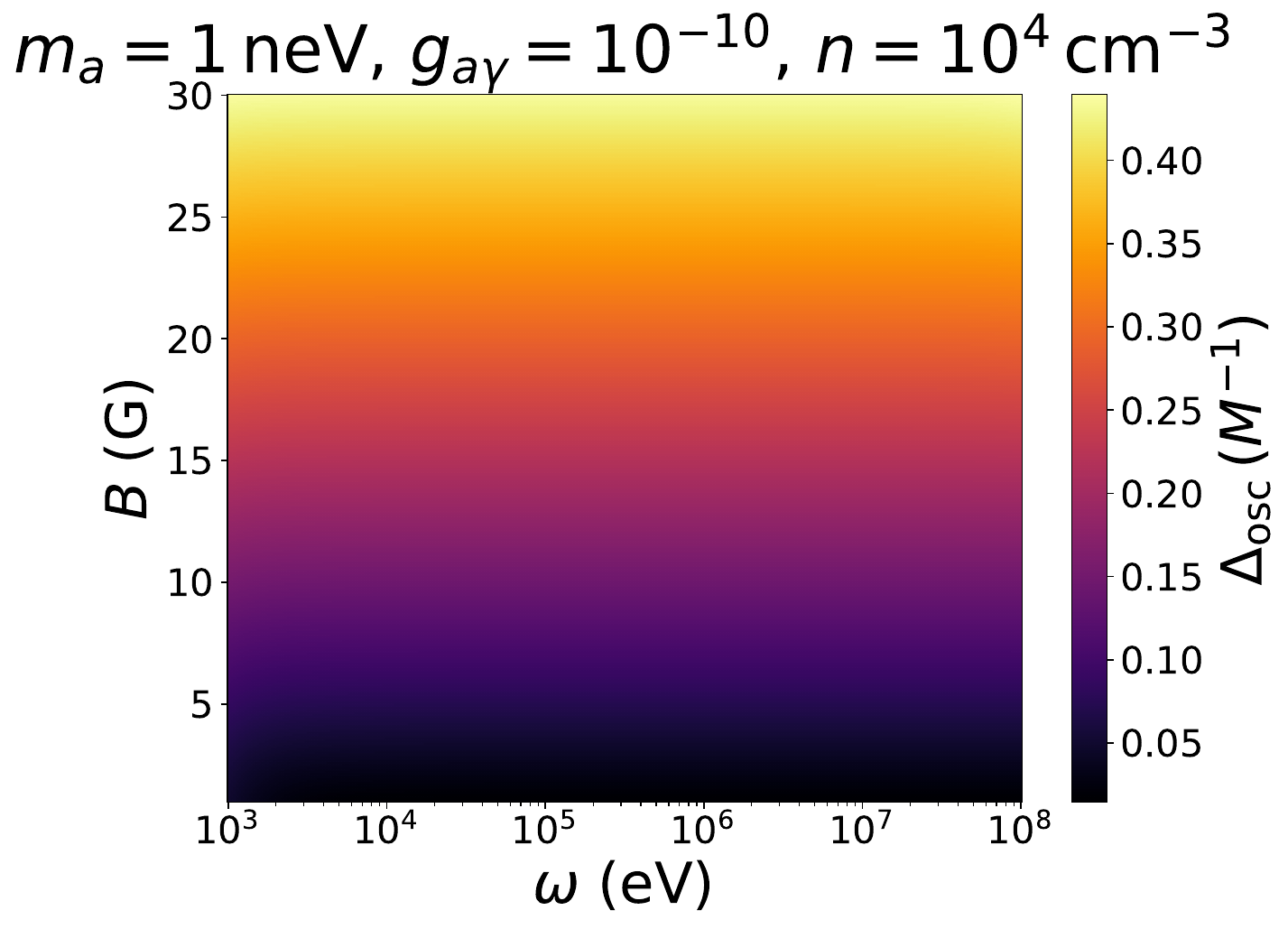}
    \caption{}
\end{subfigure}\hfill
\begin{subfigure}[b]{0.32\textwidth}
    \centering
    \includegraphics[width=\linewidth]{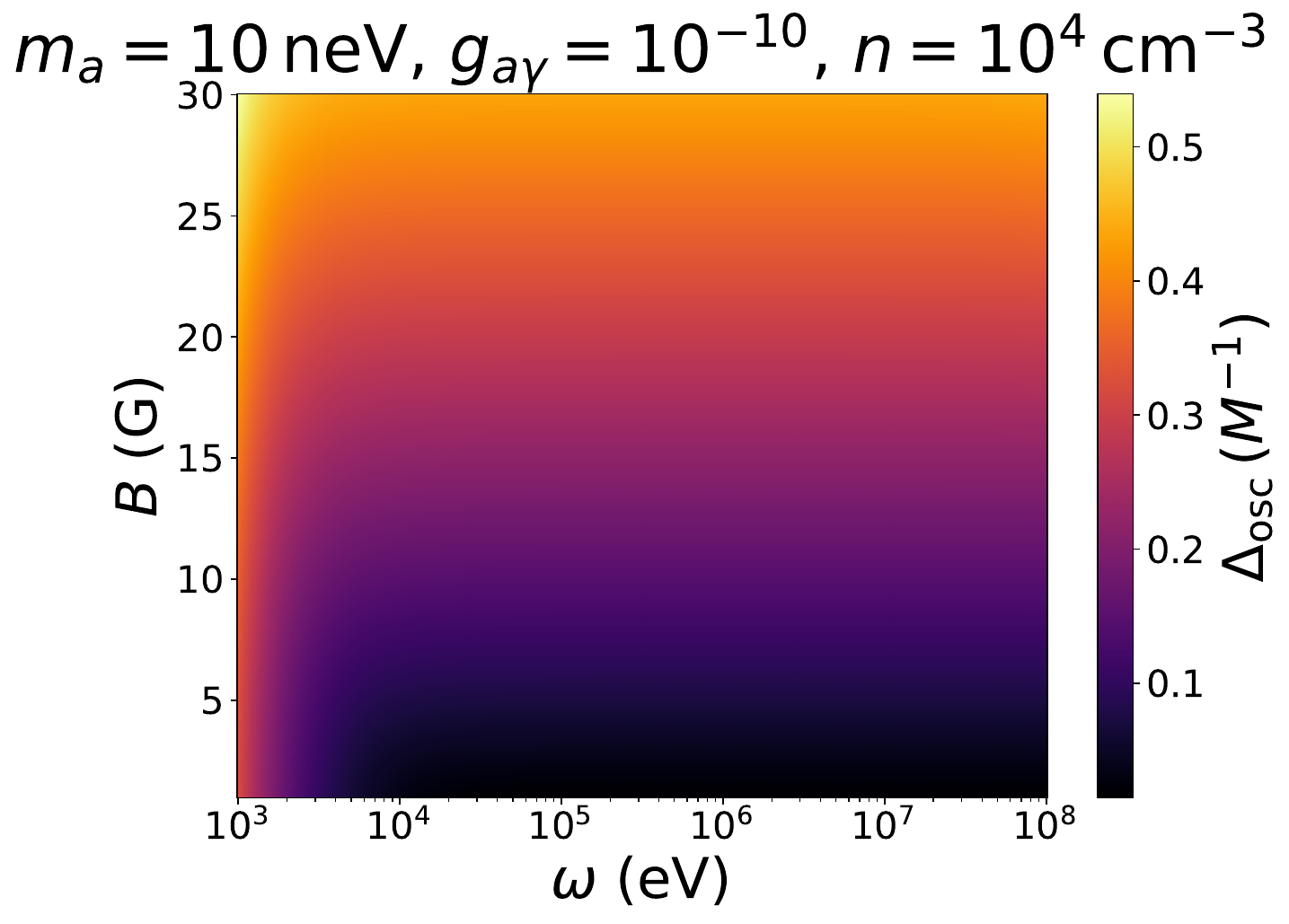}
    \caption{}
\end{subfigure}\hfill
\begin{subfigure}[b]{0.32\textwidth}
    \centering
    \includegraphics[width=\linewidth]{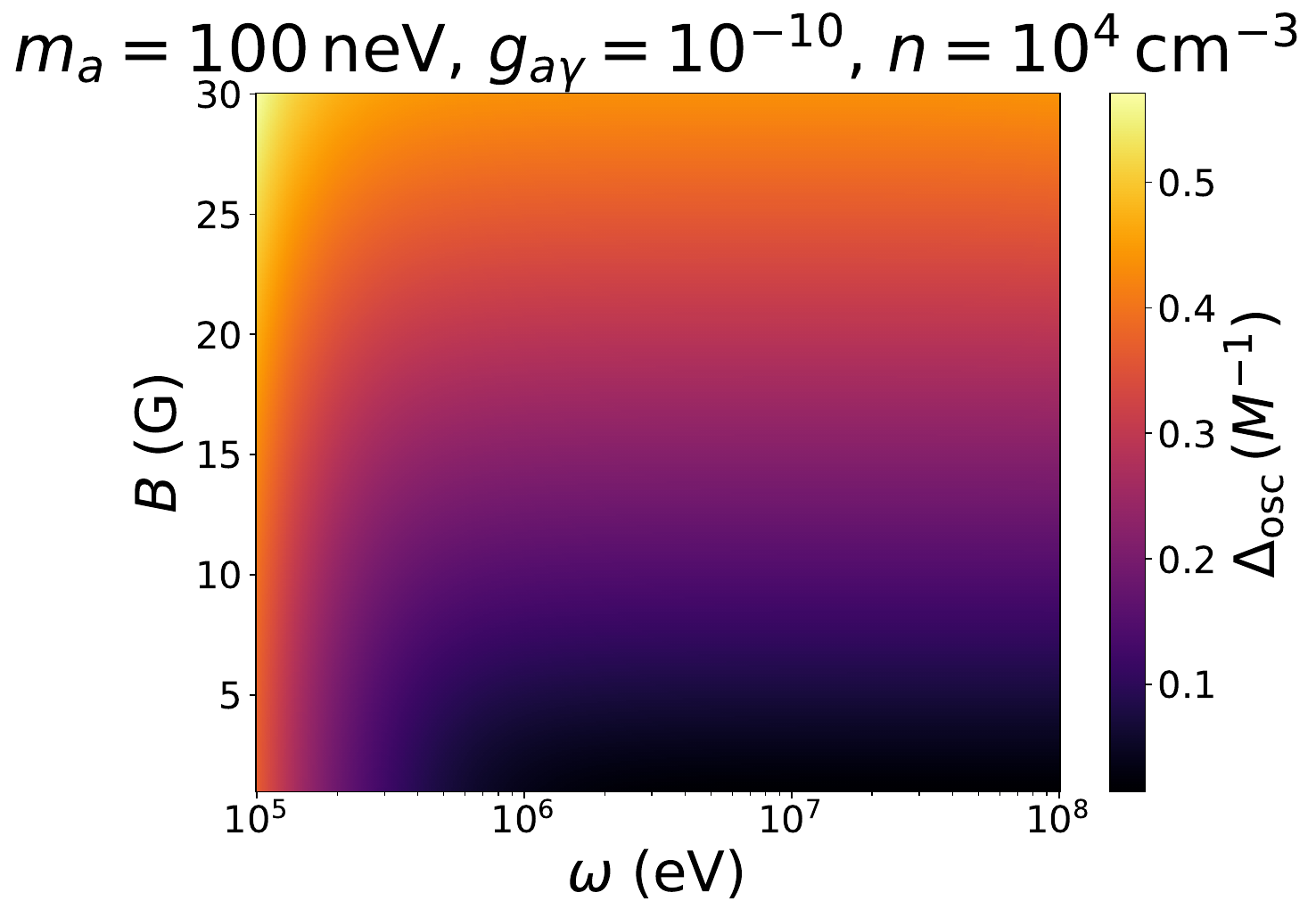}
    \caption{}
\end{subfigure}

\vspace{0.3cm}

% -------- Row 2 --------
\begin{subfigure}[b]{0.32\textwidth}
    \centering
    \includegraphics[width=\linewidth]{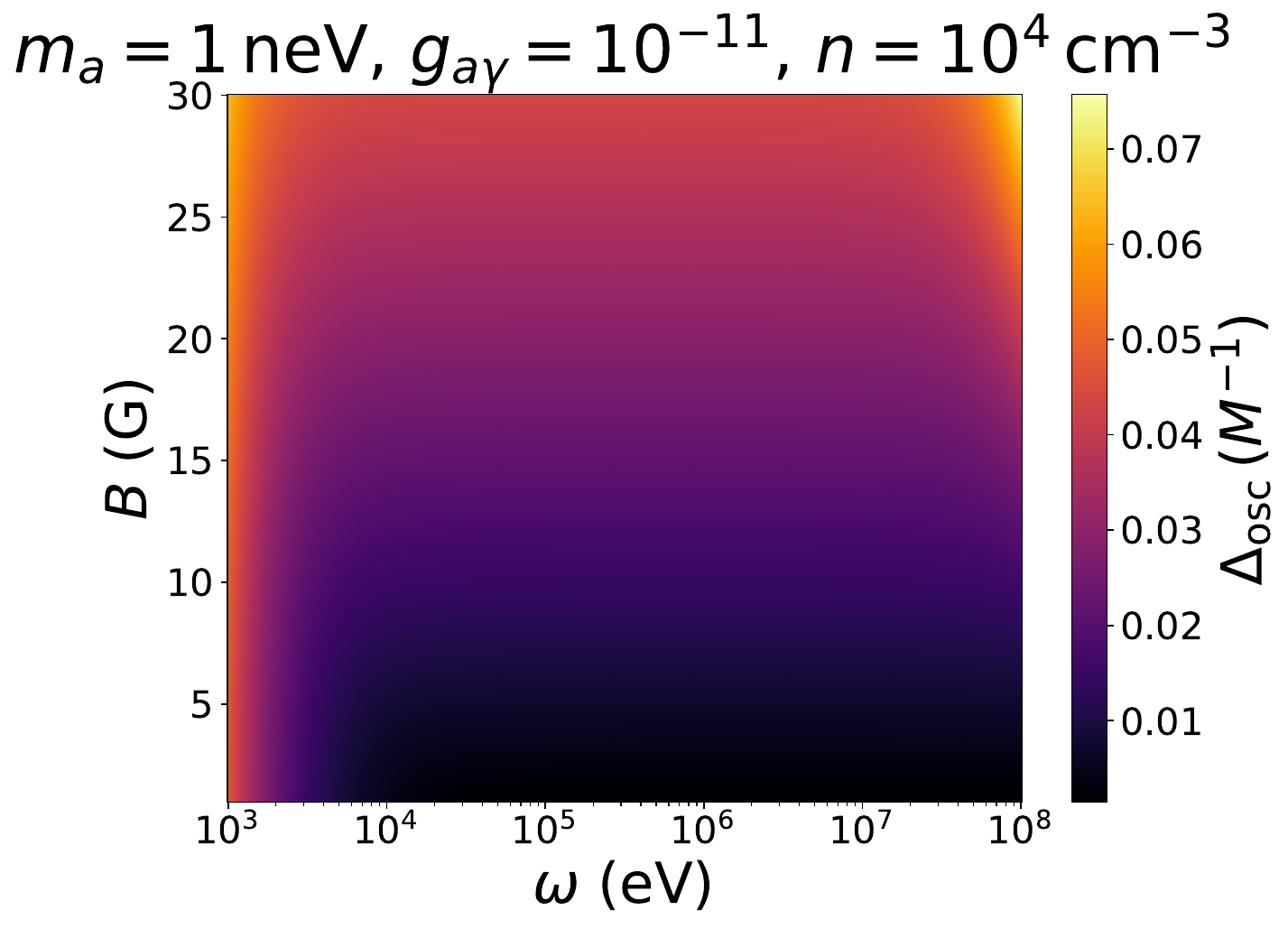}
    \caption{}
\end{subfigure}\hfill
\begin{subfigure}[b]{0.32\textwidth}
    \centering
    \includegraphics[width=\linewidth]{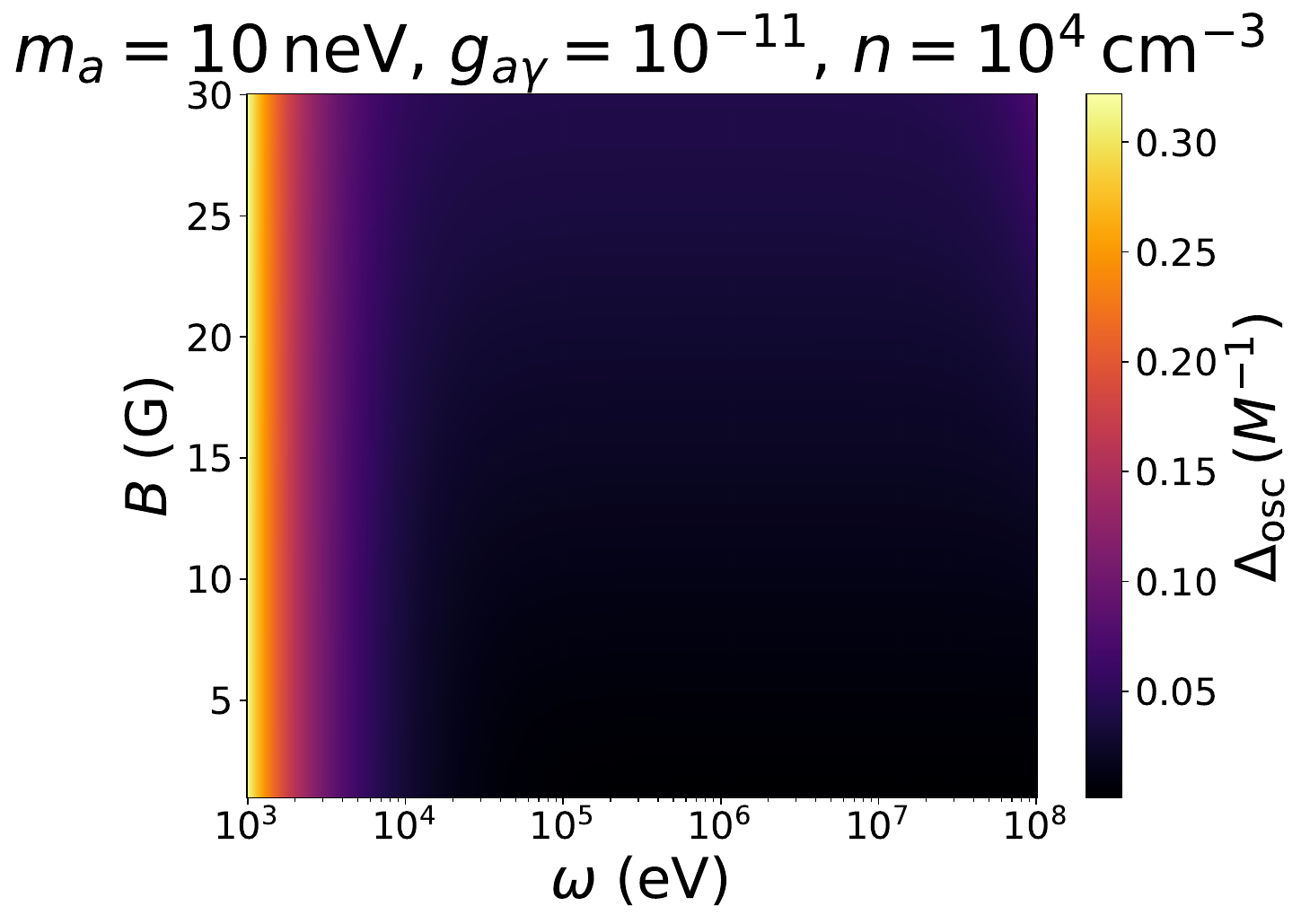}
    \caption{}
\end{subfigure}\hfill
\begin{subfigure}[b]{0.32\textwidth}
    \centering
    \includegraphics[width=\linewidth]{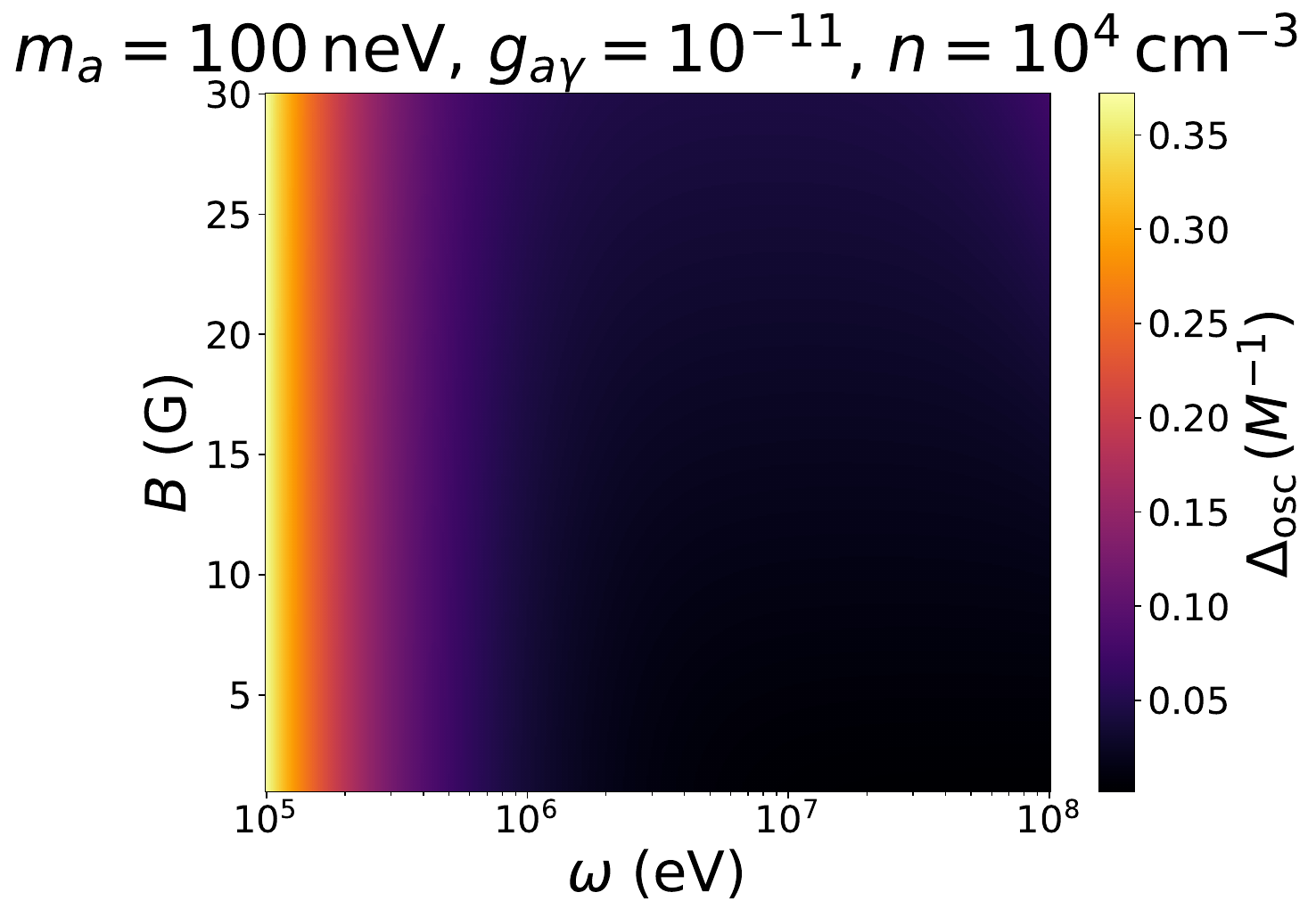}
    \caption{}
\end{subfigure}

\vspace{0.3cm}
% ---------------------

\vspace{0.3cm}

\caption{Density plots showing the continuous variation of $\Delta_{\mathrm{osc}}$ with magnetic field and photon frequency. }
\label{fig:delta_osc}
\end{figure}

\cref{fig:delta_osc_mag_field} and \cref{fig:delta_osc} depict the variation of $\mathrm{\Delta_{osc}}$ with the photon frequency and the magnetic field. The three different sub-figures in the upper panel are associated with three different axion masses with photon-axion coupling $g_{a\gamma}\simeq 10^{-10}\rm~{GeV}^{-1}$ while the subfigures in the lower panel are associated with three different axion masses but now plotted with a lower photon-axion coupling ($g_{a\gamma}\simeq 10^{-11}\rm~{GeV}^{-1}$). From \cref{fig:delta_osc_mag_field}
we note that $\mathrm{\Delta_{osc}}$ is insensitive to the photon frequency over a certain range of $\omega$ which depends on the axion mass and the photon-axion coupling constant. This can happen only when $\mathrm{\Delta_{osc}}$ is dominated by the mixing term $\mathrm{\Delta_M}$ which is independent of frequency. With low axion-mass ($m_a\sim \rm 1 ~neV$) and high photon-axion coupling ($g_{a\gamma}\sim 10^{-10}$) (Fig 1a), $\mathrm{\Delta_{osc}}\sim 2\mathrm{\Delta_M}$ in the entire frequency range considered in \cref{fig:delta_osc_mag_field}. Since $\mathrm{\Delta_M}$ increases linearly with the magnetic field, the constant value of $\mathrm{\Delta_{osc}}$ increases with an increase in $B$ in the frequency regime where $\mathrm{\Delta_{osc}}$ is a constant.

As we increase the axion-mass from $\mathcal{O}(1)\mathrm{neV}$ to higher values keeping $g_{a\gamma}$ fixed to $10^{-10}~\rm {GeV}^{-1}$, $\mathrm{\Delta_{osc}}$ is dominated by $\left| \Delta_{\rm pl} - \Delta_a \right|$ in the low frequency regime ($\sim 10^3 -10^4 ~\rm eV$) and hence there is a decrease in $\mathrm{\Delta_{osc}}$ with increasing $\omega$ in the aforementioned frequency range beyond which it again becomes constant due to the dominance of $\mathrm{\Delta_M}$ in $\mathrm{\Delta_{osc}}$ with an increase in frequency. This increase in $\mathrm{\Delta_{osc}}$ with decreasing frequency becomes substantially pronounced when axion mass is increased to $m_a\sim 100~\rm neV$ (Fig 1c and 1f) when changes in $\mathrm{\Delta_{osc}}$ due to a changing magnetic field become practically indistinguishable due to the dominance of $\mathrm{\Delta_a}$ (which is independent of $B$) in $\mathrm{\Delta_{osc}}$. Since $\mathrm{\Delta_a}$ is independent of $g_{a\gamma}$, therefore the low frequency behavior of Fig 1(c) and 1(f) are nearly the same.
From intermediate frequencies $\Delta_{M}$ does take over, but the magnitude of change it causes in the constant value of $\mathrm{\Delta_{osc}}$ is negligible relative of the magnitude of $\mathrm{\Delta_{osc}}$ at low frequencies. Hence, in Fig. 1(c) and 1(f), the change in $\mathrm{\Delta_{osc}}$ with $B$ is practically invisible.
When $g_{a\gamma}$ is decreased to $10^{-11}\rm ~GeV^{-1}$, at very high frequencies, $\mathrm{\Delta_{osc}}$ is dominated by $\mathrm{\Delta_{vac}}$ at higher magnetic fields, which is evident from the slight increase in $\mathrm{\Delta_{osc}}$ at high frequencies in Fig. 1(d).

In \cref{fig:delta_osc} we have shown the density plots showing continuous variation of $\mathrm{\Delta_{osc}}$ with magnetic field and photon frequency for different axion masses and photon-axion coupling.
We have varied the magnetic field in the range $1\lesssim B\lesssim 30$G (as reported for M87*) and the photon frequency in suitable range such that
$\mathrm{\Delta_{osc}}\approx 2\mathrm{\Delta_M}$ and the
the magnitude of $\mathrm{\Delta_{osc}}$ is between $0.01-0.5$. As a consequence for $m_a\sim 100~$neV the frequency range in taken to be $10^5-10^8~$eV instead of $10^3-10^8~$eV as considered for other masses (see Fig. 2(c) and Fig. 2(f)). The reason for focussing on such magnitude of $\mathrm{\Delta_{osc}}$ in \cref{fig:delta_osc} stems from the fact that for such values of $\mathrm{\Delta_{osc}}$ the oscillatory behavior of $\sin^2(\frac{z\mathrm{\Delta_{osc}}}{2})$ term becomes conspicuous in the conversion probability $P_{\gamma\to a}$ for $g_{a\gamma}\sim 10^{-10}~\rm GeV^{-1}$ which will be relevant for later discussion. Since the path length $z$ traversed by the photon plays a significant role in the conversion probability, in the next section we provide a framework to calculate the photon path length in the photon region of the Kerr spacetime.

\subsection{Path length traversed by photons in the Kerr photon region}
\label{S3}

In the last section we have mentioned that the conversion of photons to axions depends crucially on the path length $z$ traversed by the photons in the presence of the external magnetic field (\cref{eq:conversion_prob}). We further note from \cref{fig:delta_osc} that $\mathrm{\Delta_{osc}}^{-1}$, which appears in the conversion probability, is typically of the order of the gravitational radius of the Kerr BH. The conversion is most efficient when $\mathrm{\Delta_{osc}}\approx 2\mathrm{\Delta_M}$ and $z \mathrm{\Delta_{osc}}\sim (2n+1)\pi$ (where $n$ is an integer) which in turn implies that the photons need to travel distances $\sim \mathrm{\Delta_{osc}}^{-1}$ and a strong magnetic field needs to be maintained over this distance. However, when photons are travelling in the photon region of the Kerr BH, they undergo strong gravitational lensing due to which they go round the BH several times before escaping to infinity. Further, if they are incident at impact parameters very near the critical values they orbit around the BH several times which increases their path length. For such photons, conversion to axion can easily happen as it does not require magnetic field to be maintained over a large radial distance. If magnetic field in the photon region is high enough then the conversion to axions can efficiently take place. Since it has been reported \cite{EventHorizonTelescope:2021srq} that the BH M87* harbors strong magnetic field in its vicinity, therefore if dimming of the photon ring of M87* is observed, then it may be attributed to conversion to axions. In what follows we will calculate the path length of photons incident at near critical impact parameters orbiting the Kerr BH in the photon region.

The Kerr metric is given by
\begin{multline}
    ds^2 = -\left(1-\frac{2Mr}{\Sigma}\right)dt^2 + \frac{\Sigma}{\Delta} dr^2 + \Sigma d\theta^2 \\
    +\left(r^2 + a^2 +\frac{2Mra^2\sin^2\theta}{\Sigma}\right)\sin^2\theta d\phi^2 - 2\left(\frac{2Mr}{\Sigma}\right)a\sin^2\theta dtd\phi,
\end{multline}
where $\Delta = r^2+a^2-2Mr$ and $\Sigma = r^2 + a^2 \cos^2 \theta$. Here $a$ is the spin parameter, $M$ is the mass of the black hole, and the metric has been defined in the Boyer-Lindquist coordinates ($t,r,\theta,\phi$). In this section we have used geometrized units $G=c=1$.
There are four conserved quantities that can be defined for the Kerr metric
\begin{eqnarray}
p_{\mu}p^{\mu}&=&0 ~(\rm ~for ~photons),\\
    E&=&-p_t \label{E},\\
    L&=&p_\phi,\\
    Q&=& p_{\theta}^2-\cos^2 \theta \;(a^2 p_t^2-p_\phi^2\;\mathrm{cosec}^2 \theta).   
\end{eqnarray}
$E$ is the specific energy of observer at infinity, wherein the Kerr metric approaches flatness for $r\rightarrow \infty$,  $L$ is the specific angular of the photon and $Q$ is the Carter constant. 
The geodesic equations in the Kerr background are
\begin{align}\label{geodesic}
\frac{\Sigma}{E}\; p^t&= -a \,(a\sin^2\theta-\lambda)+\frac{r^2+a^2}{\Delta}(r^2+a^2-a\lambda),\\
\frac{\Sigma}{E}\; p^r&= \pm\sqrt{\mathcal{R}(r)}, \label{eq:Radial_geodesic} \\ 
\frac{\Sigma}{E}\; p^\theta&= \pm\sqrt{\Theta(\theta)}, \label{eq:Theta_geodesic} \\
\frac{\Sigma}{E}\; p^\phi&= -\left(a-\frac{\lambda}{\sin^2\theta}\right)+\frac{a}{\Delta}(r^2+a^2-a\lambda),
\label{eq:Phi_geodesic}
\end{align}
where $p^\mu = \frac{dx^\mu}{d \sigma}$ are the four momentum with $x^\mu = t, r, \theta, \phi$ and $\sigma$ is the affine parameter. 
The right-hand side of (\cref{eq:Radial_geodesic}) and (\cref{eq:Theta_geodesic}) act as the radial and the polar potentials, respectively which are given by,
\begin{align}
    \mathcal{R}(r)&=(r^2+a^2-a\lambda)^2-\Delta[\eta+(a-\lambda)^2], \label{eq:Radial_potential}\\
\Theta(\theta)&=\eta+a^2\cos^2\theta-\lambda^2\cot^2\theta. \label{eq:Theta_potential}    
\end{align}
In the above equations $\lambda=L/E$ and $\eta=Q/E^2$ refer to the two impact parameters of the photon in the Kerr background.
The plus and minus signs accompanying the right-hand side of (\cref{eq:Radial_geodesic}) and (\cref{eq:Theta_geodesic}) indicate the signature of $p^r$ and $p^\theta$, respectively. 
The initial conditions of the photon when it starts coming from the far-off source define its sign. As a turning point is reached in each case, the sign of the potential flips. By tracking the change in this sign accordingly, while solving the geodesic equation, the correct path length can be estimated.
There are two sets of turning points. One set arises for the radial case and the other for the polar one. As a turning point is reached the relevant momentum (radial or polar) of the photon goes to zero and as the photon continues its trajectory its momentum changes sign signifying the reversal in direction.  

We numerically solve the coupled null geodesic equations to obtain photon trajectories in the exterior spacetime of a Kerr black hole, employing the \textit{backward ray tracing method}. In this approach, instead of propagating the photons forward in time from the emission region, we integrate the geodesic equations \textit{backward} from the location of a distant observer situated at an asymptotically flat region ($r \gg M$). Each geodesic is initialized with a given direction on the observer’s image plane, corresponding to specific conserved quantities of motion—namely, the impact parameters $\lambda = L_z/E$ (related to the azimuthal angular momentum) and $\eta = Q/E^2$ (associated with the Carter constant). 
By integrating along the affine parameter $\sigma$, we effectively trace the photon backward through the curved spacetime, allowing us to map each observed light ray to its potential point of origin near the black hole—whether it escapes from the accretion flow, grazes the photon sphere, or plunges into the event horizon. This time-reversed formulation is computationally efficient, since all rays reaching the observer are, by construction, those that contribute to the observable image, eliminating the need to evolve a large number of photons that would otherwise miss the detector.

\begin{figure}[!htbp]
    \centering
        \includegraphics[width=.99 \textwidth]{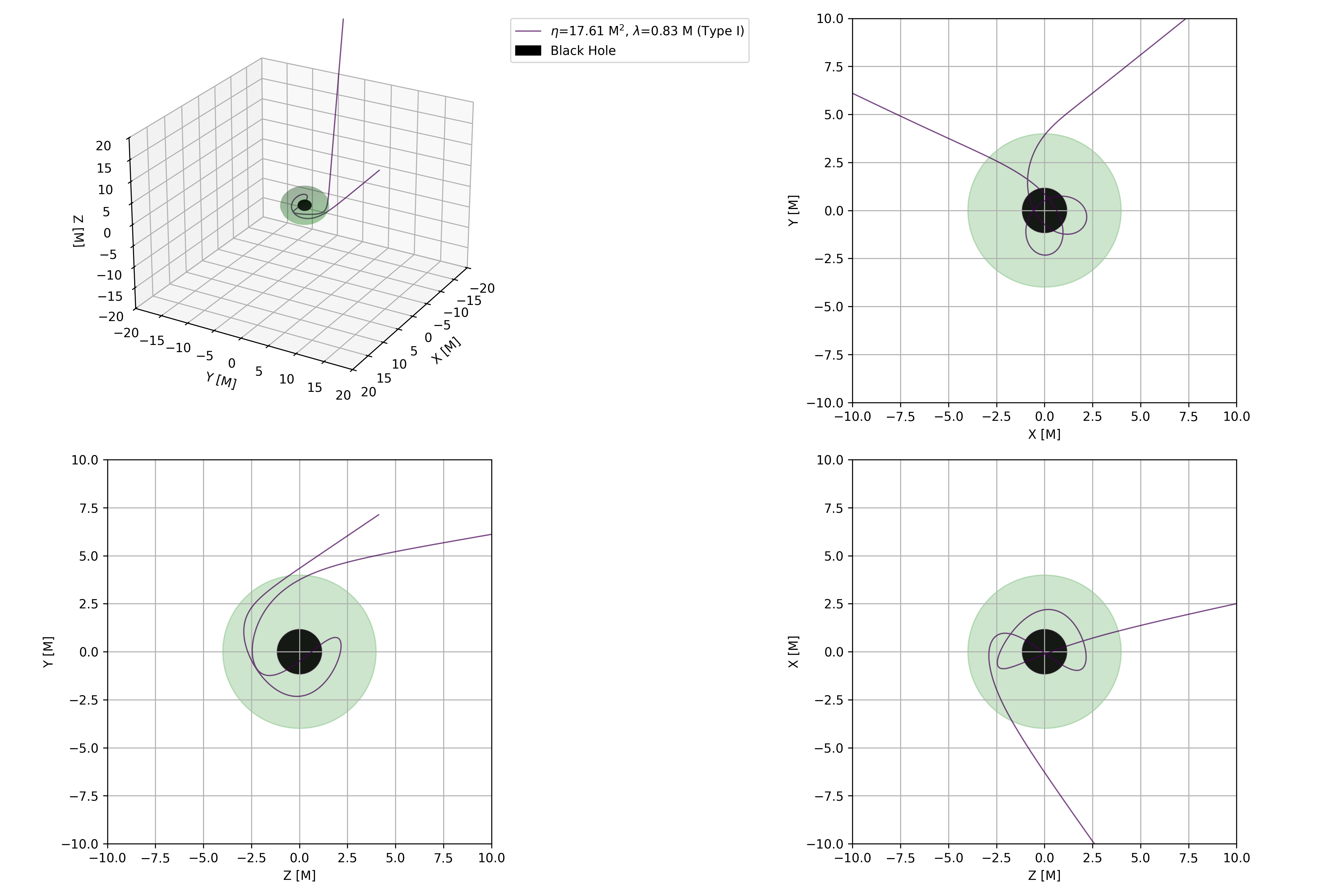} 
\caption{The above figure illustrates a characteristic photon trajectory with impact parameters $\eta=17.61 M^2$ and $\lambda=0.83M$ around the Kerr BH with spin $a=0.99$. The Top Left panel represents the three dimensional photon trajectory, while the Top Right and Bottom panels respectively represent the projection of the above-mentioned photon trajectory along the  $X-Y$, $X-Z$, and $Y-Z$ planes. The green and the black shaded region represent the photon region and the outer horizon respectively.}
    \label{fig:trajectory}
\end{figure}

The coupled ordinary differential equations governing the evolution of $r(\sigma)$, $\theta(\sigma)$, $\phi(\sigma)$, and $t(\sigma)$ are integrated using  Runge–Kutta method, which ensures numerical stability and precision, particularly in the strong-field regime where the metric gradients and curvature terms vary rapidly. Photon trajectories, computed for a particular set of impact parameters $(\lambda, \eta)$ and spin $a$, are shown in ~\cref{fig:trajectory}. These demonstrate the behavior of light in the Kerr geometry, including multiple winding orbits around the black hole, gravitational lensing, and scattering by the rotating spacetime. Small variations in the initial conditions lead to markedly different outcomes—such as capture, deflection, or escape—reflecting the sensitive dependence on initial data that characterizes unstable photon orbits. The spin parameter $a$ plays a central role: higher spins shift the locations of critical orbits, alter the extent of the photon region, and introduce frame-dragging effects that produce characteristic asymmetries in the observed photon ring.

The coordinates in which we observe the black hole image  
are a pair of orthogonal coordinates ($\alpha,\beta$) also known as the screen coordinates. The $\alpha-\beta$ plane is perpendicular to the line of sight and in terms of the critical impact parameters, they are expressed as 
\begin{align}
(\alpha,\beta) = \left(-\displaystyle\frac{\tilde{\lambda}}{\sin \theta_0}, \pm\sqrt{\tilde{\eta}+a^2 \cos^2 \theta_0 - \tilde{\lambda}\cot^2 \theta_0}\right),
\label{screen}
\end{align}
where $\theta_0$ is the inclination angle. The critical impact parameters are given by \cite{Teo:2003ltt},
\begin{align}
    \tilde{\lambda}&=-\frac{\tilde{r}^3-3M\tilde{r}^2+a^2 \tilde{r}+a^2 M}{a(\tilde{r}-M)}\nonumber, \\
    \tilde{\eta}&=-\frac{\tilde{r}^3(\tilde{r}^3-6M\tilde{r}^2+9M^2\tilde{r}-4a^2M)}{a^2(\tilde{r}-M)^2}.
    \label{critIP}
\end{align}
When the photons are incident at the critical impact parameters they can orbit around the BH in spherical photon orbits in the photon region ($\tilde{r_1}\leq \tilde{r}\leq \tilde{r_2}$) which extends from $\tilde{r}_1$ to  $\tilde{r}_2$ which are given by,
\begin{align}
    \tilde{r_1}&=2M\Big[1+\cos\Big(\frac{2}{3}\cos^{-1}\Big(-\frac{|a|}{M}\Big)\Big)\Big]\nonumber, \\
    \tilde{r_2}&=2M\Big[1+\cos\Big(\frac{2}{3}\cos^{-1}\Big(\frac{|a|}{M}\Big)\Big)\Big],
\end{align}
note that the extent of the photon region is determined by the BH's spin. Thus, for an extremal Kerr BH the photon region extends from $M\leq \tilde{r}\leq 4M$ while for a Schwarzschild BH the photon region reduces to a photon sphere located at $\tilde{r}=3M$.

\begin{figure}[!htbp]
    \centering
    \begin{subfigure}[b]{0.45\textwidth}
        \centering
        \includegraphics[width=\textwidth]{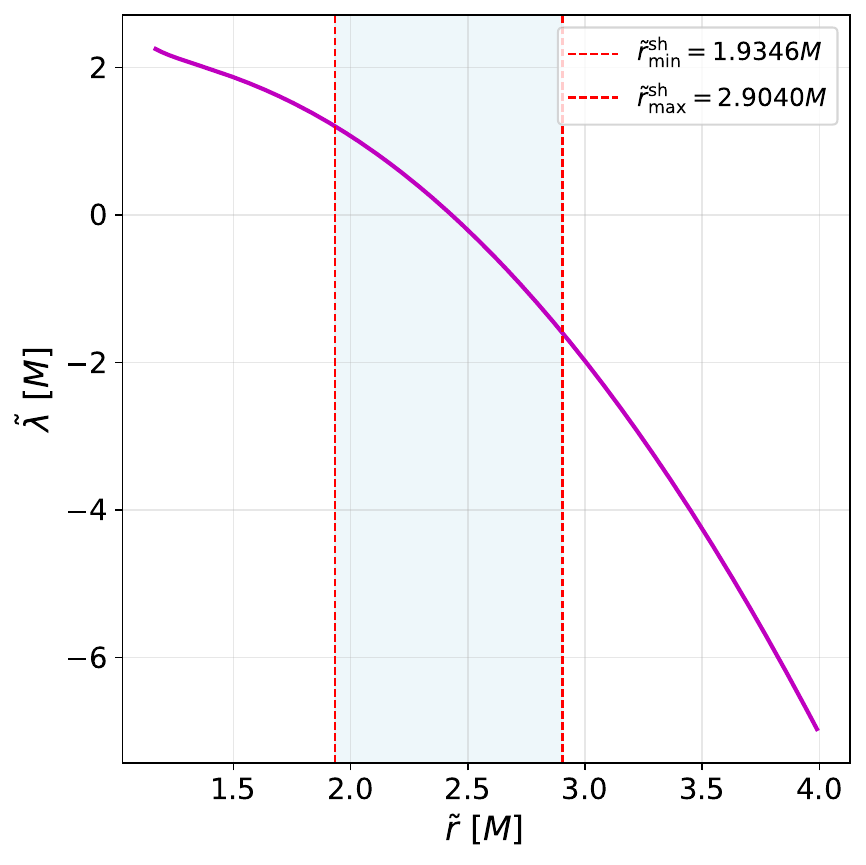}
        \caption{$\tilde{\lambda}$ vs $\tilde{r}$}
        \label{fig:combined:a}
    \end{subfigure}
    \hfill
    \begin{subfigure}[b]{0.45\textwidth}
        \centering
        \includegraphics[width=\textwidth]{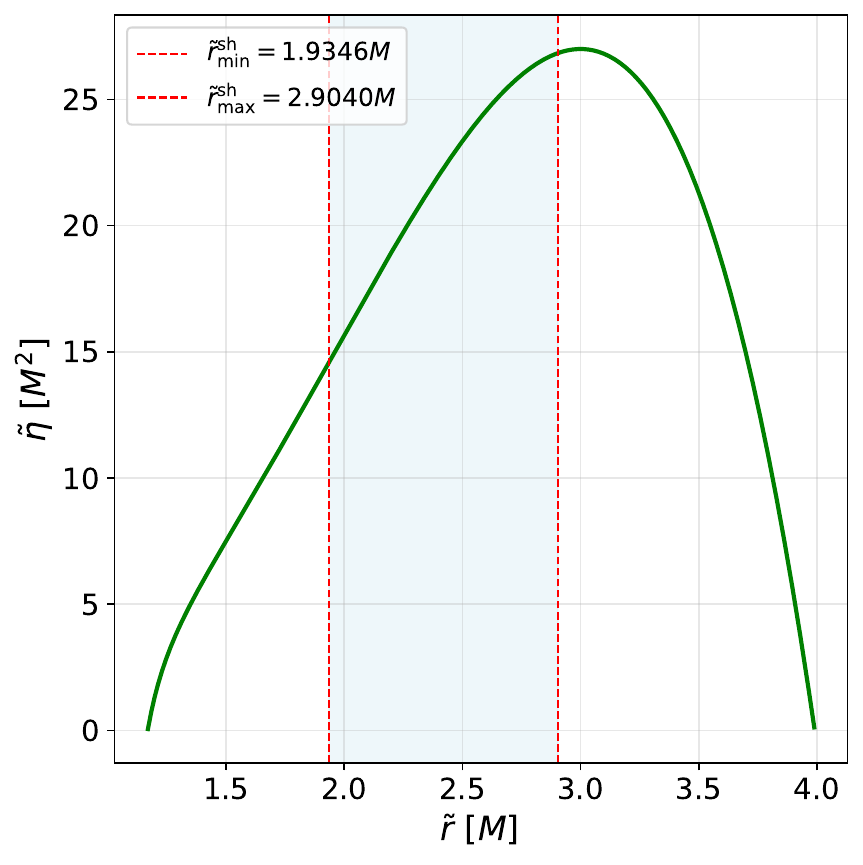}
        \caption{$\tilde{\eta}$ vs $\tilde{r}$}
        \label{fig:combined:b}
    \end{subfigure}
   
    \begin{subfigure}[b]{0.45\textwidth}
        \centering
        \includegraphics[width=\textwidth]{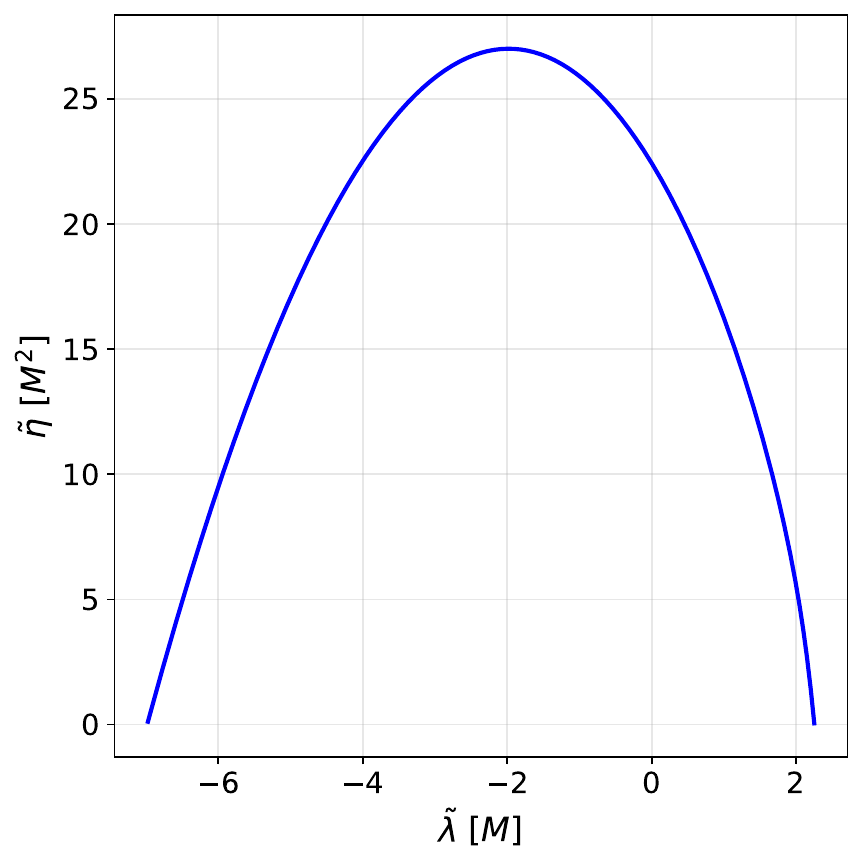}
        \caption{$\tilde{\eta}$ vs $\tilde{\lambda}$}
        \label{fig:combined:c}
    \end{subfigure}
    \hfill
    \begin{subfigure}[b]{0.45\textwidth}
        \centering
        \includegraphics[width=\textwidth]{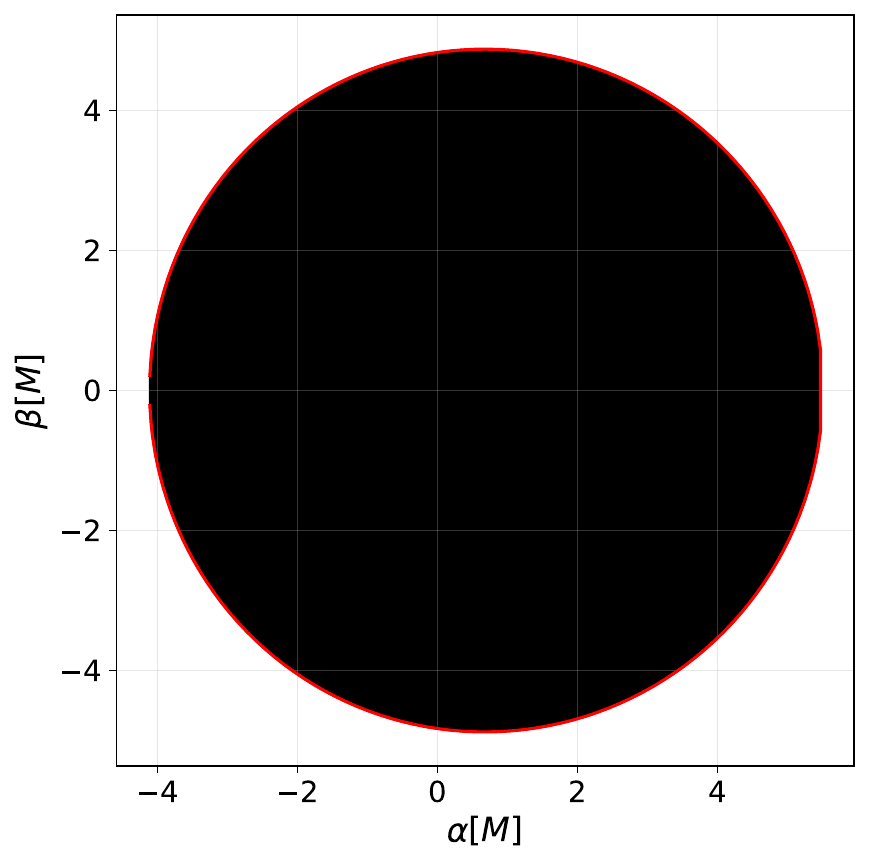}
        \caption{$\beta$ vs $\alpha$}
        \label{fig:combined:d}
    \end{subfigure}
   
    \caption{The above figure shows the variation of 
    (a) $\tilde{\lambda}$ with $\tilde{r}$, 
    (b) $\tilde{\eta}$ with $\tilde{r}$, and 
    (c) $\tilde{\eta}$ with $\tilde{\lambda}$ 
    in the photon region of a Kerr black hole with spin $a=0.99$. 
    In \cref{fig:combined:d} we show the outline of the critical curve for the aforementioned Kerr black hole, observed at an inclination angle of $\theta_0=17^\circ$. 
    The blue shaded region in \cref{fig:combined:a} and \cref{fig:combined:b} refers to the photon region accessible to the observer at inclination angle $\theta_0\simeq 17^\circ$.
    }
   
    \label{fig:combined}
\end{figure}
In \cref{fig:combined:a} and \cref{fig:combined:b} we show the variation of the magnitude of the critical impact parameters $\tilde{\lambda}$ and $\tilde{\eta}$ as a function of the radius of the spherical photon orbits in the photon region for a Kerr BH of spin $a\simeq 0.99$.
Note that the entire photon region is only accessible to the observer if the inclination angle is $\theta_0=\pi/2$. For lower inclination angles only a part of the photon region is visible to the observer since one needs to ensure that $\beta$ is real. In these figures we take $\theta_0\simeq 17^\circ$ following the inclination angle of M87* which has been imaged by the EHT collaboration. The blue shaded region refers to the photon region which is ``visible" to the observer at $\theta_0\simeq 17^\circ$. In \cref{fig:combined:c} we show the variation of $\tilde{\lambda}$ with $\tilde{\eta}$ in the photon region of the Kerr BH with $a\simeq0.99$ while in \cref{fig:combined:d} we present the contour of the shadow cast by a Kerr BH of spin $a\simeq 0.99$ observed at $\theta_0\simeq 17^\circ.$

In dimming calculations, the spatial path traveled by the photon in and around the photon region is important, as longer paths enhance the probability of photon-to-axion conversion in the presence of magnetic fields. To calculate the path length of the photon, we decompose the Kerr metric into the $(3+1)$ form using the ADM formalism \cite{Arnowitt:1959ah}. The lapse and shift functions are computed to be
\begin{eqnarray}\label{ADM}
N^2 &=& \frac{\Sigma \Delta}{\Sigma(r^2+a^2)+2Mra^2\sin^2\theta}, \\
N^\phi &=& \frac{-2Mra}{\Sigma(r^2+a^2)+2Mra^2\sin^2\theta}.
\end{eqnarray}
Using these lapse and shift functions, the metric decomposes as
\begin{equation}
  g_{\mu\nu} =
  \begin{bmatrix}
  -(N^2 - N^{\phi}N_{\phi}) & 0 & 0 & N_{\phi} \\
  0 & \gamma_{rr} & 0 & 0\\
  0 & 0 & \gamma_{\theta\theta} & 0\\
  N_{\phi} & 0 & 0 & \gamma_{\phi\phi}
    \end{bmatrix},
    \label{induced-metric}
\end{equation}
where \(N_{\phi} = \gamma_{\phi\phi} N^\phi\). As a result, the induced 3-metric on spatial slices gives the spatial line element
\begin{equation}\label{spatial_ds2}
    ds^2_3 = \gamma_{ij}\,dx^idx^j,
\end{equation}
with \(\gamma_{ij} \equiv g_{ij} = \operatorname{diag}\,(g_{rr},g_{\theta\theta},g_{\phi\phi})\). Here \(\mu,\nu = t,r,\theta,\phi\) and \(i,j = r,\theta,\phi\).

The spatial path length \(z\) traveled by the photon is then estimated by integrating along the null geodesic
\begin{eqnarray}
    z &=& \int ds_3
      = \int_{in}^{out} \sqrt{\gamma_{rr}\left(\frac{dr}{d\sigma}\right)^2+\gamma_{\theta\theta}\left(\frac{d\theta}{d\sigma}\right)^2+\gamma_{\phi\phi}\left(\frac{d\phi}{d\sigma}\right)^2}\;d\sigma,
      \label{z}
\end{eqnarray}
where \(\sigma\) is the affine parameter. We use the geodesic equations mentioned \cref{S3} to evaluate the path length. Note that the path length depends crucially on the impact parameters which appear in the geodesic equations. The integration limits, $in$ and $out$, correspond to the points where the photon trajectory enters and exits the photon region---where unstable spherical photon orbits are possible. This region is of primary interest for axion-photon conversion, since photons entering the photon region at near critical impact parameters can orbit around the BH several times before escaping to infinity, thereby leading to an enhancement in the path length.

\subsection{Dimming of the photon ring}
\label{Sec4}
In the context of photon propagation near a black hole's photon region, we consider the number of photons approaching the photon sphere with impact parameters $\lambda$ and $\eta$ close to their critical values, $\tilde{\lambda}$ and $\tilde{\eta}$, per unit time $t$, unit frequency $\omega$, and unit impact parameters. This is denoted by:

\begin{equation}
\frac{d^4 N}{dt \, d\omega \, d\lambda \, d\eta},
\label{eq:photon_rate}
\end{equation}
which represents the number of photons per unit time, frequency, and impact parameter space. The rate of photon-to-axion conversion in the photon sphere region, per unit time and unit frequency, is given by integrating over a small range of impact parameters around the critical values:

\begin{equation}
\frac{d^2 N_{\gamma \to a}}{dt \, d\omega} = \frac{1}{2} \int_{\tilde{\lambda}}^{\tilde{\lambda}(1+\delta\lambda)} \int_{\tilde{\eta}}^{\tilde{\eta}(1+\delta\eta)} \frac{d^4 N}{dt \, d\omega \, d\lambda \, d\eta} \, P_{\gamma \to a}(z(\lambda, \eta)) \, d\lambda \, d\eta,
\label{eq:conversion_rate}
\end{equation}
where $\tilde{\lambda}$ and $\tilde{\eta}$ correspond to the critical impact parameters discussed in \cref{S3}, $P_{\gamma \to a}(z(\lambda, \eta))$ is the probability of a photon converting to an axion (\cref{eq:conversion_prob}) as a function of the path length $z$, which depends on the impact parameters $\lambda$ and $\eta$ through the geodesic equations as discussed in \cref{S3}. The factor of $1/2$ stems from the fact that conversion to axions can only happen to those photons whose polarization is parallel to the external magnetic field. The conversion probability is given as:

\begin{equation*}
P_{\gamma \to a}(z) = \left( \frac{\mathrm{\Delta_M}}{\Delta_{\text{osc}}/2} \right)^2 \sin^2 \left( \frac{\Delta_{\text{osc}}}{2} z \right),
% \label{eq:conversion_probability}
\end{equation*}
where $\mathrm{\Delta_M}$ represents the mixing amplitude, and $\Delta_{\text{osc}}$ is the oscillation wavenumber and $z$ is the effective propagation distance in the photon region, evaluated using \cref{z}.
Assuming that the differential photon rate $\frac{d^4 N}{dt \, d\omega \, d\lambda \, d\eta}$ is approximately constant over the small intervals $\Delta\lambda$ and $\Delta \eta$ around the critical values, we can evaluate it at the critical impact parameters, yielding:

\begin{equation}
\frac{d^2 N_{\gamma \to a}}{dt \, d\omega} = \frac{1}{2} \left. \frac{d^4 N}{dt \, d\omega \, d\lambda \, d\eta} \right|_{\tilde{\lambda}, \tilde{\eta}} \int_{\tilde{\lambda
}}^{\tilde{\lambda}(1+\delta\lambda)} \int_{\tilde{\eta}}^{\tilde{\eta}(1+\delta\eta)} P_{\gamma \to a}(z(\lambda, \eta)) \, d\lambda \, d\eta,
\label{eq:simplified_conversion_rate}
\end{equation}
where the vertical bar with subscript $\tilde{\lambda}, \tilde{\eta}$ denotes evaluation at the critical impact parameters. Therefore, the fraction of photons entering into the photon region that gets converted to axions is given by,
\begin{equation}
\text{Dimming fraction} = \frac{d^2 N_{\gamma \to a}}{dt \, d\omega}\big/{\frac{d^2 N}{dt \, d\omega}}=\frac{1}{2} \frac{1}{d\lambda d\eta}\int_{\tilde{\lambda}}^{\tilde{\lambda}(1+\delta\lambda)} \int_{\tilde{\eta}}^{\tilde{\eta}(1+\delta\eta)}P_{\gamma \to a}(z(\lambda, \eta)) \, d\lambda \, d\eta,
\label{DF}
\end{equation}
where,
\begin{align}
 \frac{d^2 N}{dt \, d\omega}=  \frac{d^4 N}{dt \, d\omega \, d\lambda \, d\eta} \Bigg|_{\tilde{\lambda}, \tilde{\eta}}d\lambda d\eta.
\end{align}
To determine the dimming fraction with high precision, we analyze the trajectories of photons whose impact parameters $(\lambda,\eta)$ are perturbed slightly away from their critical values $(\tilde{\lambda},\tilde{\eta})$. Following the formalism for Kerr null geodesics established in Ref.~\cite{Gralla:2019xty}, we linearize the radial motion in the vicinity of the unstable spherical photon orbit $\tilde{r}$.
The radial position and impact parameters are expanded around their critical values as
\begin{equation}
r = \tilde{r}(1 + \delta r), \quad
\lambda = \tilde{\lambda}(1 + \delta \lambda), \quad
\eta = \tilde{\eta}(1 + \delta \eta),
\label{eq:perturbations}
\end{equation}
where we consider the near-critical regime $\delta r,\,\delta\lambda,\,\delta\eta \ll 1$. Consistent with the second-order nature of the radial potential $\mathcal{R}(r)$ near its maximum, we adopt the scaling ~\cite{Gralla:2019xty}
\begin{equation}
\delta r^2 \sim \delta\lambda \sim \delta\eta.
\end{equation}
In this limit, the radial potential admits the quadratic approximation ~\cite{Gralla:2019xty}
\begin{equation}
\mathcal{R} \approx C_r \delta r^2 - \delta B,
\label{eq:potential_approx}
\end{equation}
where
\begin{equation}
\delta B = C_\eta \delta\eta + C_\lambda \delta\lambda,
\end{equation}
quantifies the nature of the radial potential experienced by the photon incident at near critical impact parameters. The coefficients $\{C_r, C_\eta, C_\lambda\}$, which encode the nature of the spacetime curvature near the spherical photon orbit $\tilde{r}$, are given by,
\begin{subequations}
\begin{align}
C_r &= \frac{4\tilde{r}^3}{(\tilde{r} - M)^2}
\left[ \tilde{r}^3 - 3M\tilde{r}(\tilde{r} - M) - a^2M \right], \\[1ex]
C_\eta &= -\frac{2\tilde{r}^3}{a^2(\tilde{r} - M)^2}
\Bigl[ \tilde{r}^5 - 8M\tilde{r}^4 + \tilde{r}^3(a^2 + 21M^2)
- M\tilde{r}^2(10a^2 + 18M^2) \nonumber \\
&\hspace{3.5cm}
+ 17a^2M^2\tilde{r} - 4a^4M \Bigr], \\[1ex]
\begin{split}
C_\lambda &= \frac{2\tilde{r}^2}{a^2(\tilde{r} - M)^2}
\Bigl[ \tilde{r}^6 - 8M\tilde{r}^5 + \tilde{r}^4(2a^2 + 21M^2)
- M\tilde{r}^3(10a^2 + 18M^2) \\
&\quad + a^2\tilde{r}^2(a^2 + 10M^2)
+ a^2M\tilde{r}(6M^2 - 2a^2)
- 3a^4M^2 \Bigr].
\end{split}
\end{align}
\end{subequations}
Since we attribute the dimming of the observed photon ring to photon-axion conversion, we need to consider photons incident at suitable impact parameters which contribute to the critical curve. For this purpose we consider critical curves of Kerr BHs observed at $\theta_0\simeq17^\circ$ such that it can be related to M87* in case a photon ring dimming is observed for this source in future EHT observations. In \cref{fig:combined:d} we have shown the critical curve of a Kerr BH with $a\simeq 0.99$ observed at $\theta_0\simeq 17^\circ$.  In \cref{tab:dimming_results_17_sorted} we report the spherical photon orbit radius and the critical impact parameters corresponding to different screen coordinates of the critical curve. This is accomplished by inverting \cref{screen} given $(\alpha,\beta)$ and by inverting \cref{critIP} from the critical impact parameters thus obtained from $(\alpha,\beta)$.

\begin{table}[!htbp]
\centering
\caption{Dimming calculation results for: $a/M = 0.99$, $\theta_0 = 17^\circ$, $B = 30~\mathrm{G}$, $n_e=10^{4}~\mathrm{cm^{-3}}$, $g_{a\gamma} = 1.0\times10^{-11}~\mathrm{GeV^{-1}}$, $m_a = 10.0~\mathrm{neV}$, and $\omega = 1.0\times10^{6}~\mathrm{eV}$. Columns show certain representative $(\alpha,\beta)$ on the critical curve and the corresponding $(\tilde{r},\tilde{\lambda}, \tilde{\eta})$, about which we perturb and hence obtain the average path length $z_{\rm avg}$ ($1\sigma$), and the corresponding dimming due to photon-axion conversion.}
\label{tab:dimming_results_17_sorted}
\resizebox{\textwidth}{!}{%
\begin{tabular}{cccccccc}
\hline
\textbf{Critical Point} & ${\alpha}$ & ${\beta}$ & $\tilde{r}~(M)$ & $\tilde{\lambda}~(M)$ & $\tilde{\eta}~(M^2)$ & \textbf{Path Length} $(z_{\rm avg})(M)$ & \textbf{Dimming (\%)} \\
\hline
1 & $-2.068$ & $4.004$ & $2.205$ & $0.605$  & $19.05$   & $31 \pm 2$ & $29$ \\
2 & $-3.101$ & $-3.008$ & $2.076$ & $-0.479$  & $24.40$   & $33 \pm 2$ & $26$ \\
3 & $-2.081$ & $ 3.995$ & $2.206$ & $ 0.433$  & $20.11$   & $36 \pm 2$ & $27$ \\
4 & $-1.532$ & $ 4.334$ & $2.268$ & $ 1.201$  & $14.58$   & $45 \pm 3$ & $26$ \\
5 & $ 1.641$ & $ 4.779$ & $2.587$ & $-0.480$  & $24.41$   & $33 \pm 2$ & $27$ \\
6 & $ 3.233$ & $-4.137$ & $2.726$ & $-0.945$  & $25.78$   & $21 \pm 3$ & $32$ \\
7 & $ 5.045$ & $-2.080$ & $2.871$ & $-0.553$  & $26.66$   & $32 \pm 3$ & $26$ \\
8 & $ 5.416$ & $ 0.945$ & $2.899$ & $-1.578$  & $26.82$   & $29 \pm 4$ & $29$ \\
\hline
\end{tabular}%
}
\end{table}

For the numerical evaluation of the dimming fraction, we integrate the photon--axion conversion probability $P_{\gamma \to a}$ as given in \cref{DF} over the perturbed impact parameter space. We consider small deviations from the critical impact parameters reported in \cref{tab:dimming_results_17_sorted}. The integration is performed over fractional perturbations
\begin{equation}
\delta\lambda,\ \delta\eta \in [10^{-2},\,10^{-3}],
\end{equation}
which correspond to photon trajectories sufficiently close to the critical curve to contribute to the photon ring while remaining within the domain of validity of the near-photon-orbit expansion. In order to ensure an accurate estimation of the dimming percentage we employ a high-density sampling grid with approximately $10^2 \times 10^2$ points. For each combination of $\delta\lambda, \delta\eta$ we need to ensure that $\delta{r}\approx\sqrt{\delta B/C_r}$ such that the radial potential (given in \cref{eq:potential_approx}) $\mathcal{R}\gtrsim0$, which in turn follows from the radial geodesic equation (\cref{eq:Radial_geodesic}). For each $\tilde{\lambda}, \tilde{\eta}$ in \cref{tab:dimming_results_17_sorted} we consider $10^2 \times 10^2$ $\delta\lambda, \delta\eta$ sets and obtain the path length traversed by the photon corresponding to each set of the impact parameters using \cref{z} which in turn enables us to obtain the dimming fraction using \cref{DF}. In \cref{tab:dimming_results_17_sorted} we report the dimming percentage for a photon of frequency $\omega\simeq 10^6 ~\rm eV$ traveling through a plasma with electron number density $n_e=10^{4}\rm cm^{-3}$, converting to an axion of mass $m_a\simeq 10\rm neV$ and photon-axion coupling $g_{a\gamma}\sim 10^{-11}\rm GeV^{-1}$ in presence of a magnetic field of strength $B\simeq \rm 30 G$  near a Kerr BH of spin $a/M\simeq 0.99$ observed at $\theta_0\simeq 17^\circ$. \cref{tab:dimming_results_2} and \cref{tab:dimming_results_3} report the dimming percentage for the same $\omega, n_e, B, g_{a\gamma},m_a$ but near a Kerr BH of $a/M\simeq 0.6$ and $a/M\simeq 0.3$ respectively. 
Note that the path length calculation is independent of $B$, $n_e$, $m_a$, $\omega$ and $g_{a\gamma}$ since the path length is sensitive only to the background spacetime. In what follows we investigate the role of the spin, axion-mass, the photon-axion coupling, the magnetic field and the electron density on the photon-axion conversion probability and hence the dimming percentage.

\iffalse
To ensure convergence and to accurately capture the rapid oscillations of the conversion probability,
\begin{equation}
P_{\gamma \to a}(z) \propto \sin^2\!\left( \frac{\Delta_{\mathrm{osc}}}{2} z \right),
\end{equation}
we employ a high-density sampling grid with approximately $10^3 \times 10^3$ points. This resolution is essential because the effective propagation distance $z$ diverges logarithmically as the impact parameters approach their critical values so some of the trajectories will be rejected if those fell inside. Physically, this divergence corresponds to an increasing number of orbital windings around the black hole, which leads to a rapid growth in the oscillation frequency of the integrand. For each entry in the ~\cref{tab:dimming_results_17_sorted,tab:dimming_results_2,tab:dimming_results_3} same process is repeated.
\fi

\begin{table}[!htbp]
\centering
\caption{Dimming calculation results for parameters: $a/M = 0.6$, $n_e=10^{4}~\mathrm{cm^{-3}}$, $\theta_0 = 17^\circ$, $B = 30~\mathrm{G}$, $g_{a\gamma} = 1.0\times10^{-11}~\mathrm{GeV^{-1}}$, $m_a = 10.0~\mathrm{neV}$, and $\omega = 1.0\times10^{6}~\mathrm{eV}$. Columns show certain representative $(\alpha,\beta)$ on the critical curve, and the corresponding $(\tilde{r}, \tilde{\lambda}, \tilde{\eta})$, about which we perturb and hence obtain the average path length $z_{\rm avg}$ ($1\sigma$), and the corresponding dimming from photon-axion conversion.}
\label{tab:dimming_results_2}
\resizebox{\textwidth}{!}{%
\begin{tabular}{cccccccc}
\hline
\textbf{Critical Point} & ${\alpha}$ & ${\beta}$ & $r_c~(M)$ & $\tilde{\lambda}$ & $\tilde{\eta}$ & \textbf{Avg. Path Length} $(z_{\rm avg})~(M)$ & \textbf{Dimming (\%)} \\
\hline
1 & $-4.075$ & $ 2.478$ & $2.644$ & $ 1.191$ & $20.998$ & $21 \pm 2$ & $26$ \\
2 & $-2.444$ & $ 4.246$ & $2.719$ & $ 0.714$ & $23.161$ & $19 \pm 3$ & $24$ \\
3 & $ 2.677$ & $ 4.538$ & $2.941$ & $-0.783$ & $26.818$ & $20 \pm 3$ & $27$ \\
4 & $ 5.156$ & $-1.708$ & $3.042$ & $-1.507$ & $26.901$ & $18 \pm 2$ & $25$ \\
5 & $ 3.693$ & $-3.853$ & $2.983$ & $-1.080$ & $26.985$ & $20 \pm 3$ & $26$ \\
6 & $ 1.850$ & $ 4.873$ & $2.907$ & $-0.540$ & $26.547$ & $21 \pm 4$ & $26$ \\
7 & $-1.835$ & $-4.593$ & $2.747$ & $ 0.537$ & $23.842$ & $21 \pm 5$ & $25$ \\
8 & $-4.690$ & $-0.517$ & $2.615$ & $ 1.371$ & $20.056$ & $24 \pm 3$ & $25$ \\
\hline
\end{tabular}%
}
\end{table}

\begin{table}[!htbp]
\centering
\caption{Dimming calculation results for parameters: $a/M = 0.3$, $\theta_0 = 17^\circ$, $B = 30~\mathrm{G}$, $n_e=10^{4}~\mathrm{cm^{-3}}$, $g_{a\gamma} = 1.0\times10^{-11}~\mathrm{GeV^{-1}}$, $m_a = 10.0~\mathrm{neV}$, and $\omega = 1.0\times10^{6}~\mathrm{eV}$. Columns show certain representative $(\alpha,\beta)$ on the critical curve, and the corresponding $(\tilde{r}, \tilde{\lambda}, \tilde{\eta})$, about which we perturb and hence obtain the average path length $z_{\rm avg}$ ($1\sigma$), and the corresponding dimming from photon-axion conversion.}
\label{tab:dimming_results_3}
\resizebox{\textwidth}{!}{%
\begin{tabular}{cccccccc}
\hline
\textbf{Critical Point} & ${\alpha}$ & ${\beta}$ & $\tilde{r}~(M)$ & ${\tilde{\lambda}}$ & ${\tilde{\eta}}$ & \textbf{Avg. Path Length} $(z_{\rm avg})~(M)$ & \textbf{Dimming (\%)} \\
\hline
1 & $-4.799$ & $-1.400$ & $2.861$ & $ 0.784$ & $25.063$ & $14 \pm 2$ & $23$ \\
2 & $-2.854$ & $ 4.189$ & $2.901$ & $ 1.440$ & $22.780$ & $15 \pm 2$ & $22$ \\
3 & $ 2.071$ & $ 4.812$ & $3.000$ & $ 1.155$ & $23.888$ & $14 \pm 1$ & $20$ \\
4 & $ 1.610$ & $-4.969$ & $2.991$ & $ 0.714$ & $25.255$ & $16 \pm 2$ & $24$ \\
5 & $-4.969$ & $ 0.496$ & $2.858$ & $ 1.453$ & $22.740$ & $16 \pm 2$ & $24$ \\
6 & $ 0.060$ & $ 5.171$ & $2.961$ & $-0.606$ & $27.000$ & $07 \pm 6$ & $19$ \\
7 & $-2.028$ & $-4.678$ & $2.919$ & $ 1.294$ & $22.944$ & $11 \pm 2$ & $23$ \\
8 & $-0.814$ & $ 5.076$ & $2.943$ & $-0.471$ & $26.983$ & $18 \pm 4$ & $22$ \\
\hline
\end{tabular}%
}
\end{table}

\begin{figure}[!htbp]
\centering

\begin{subfigure}[b]{0.32\textwidth}
  \centering
  \includegraphics[width=\textwidth]{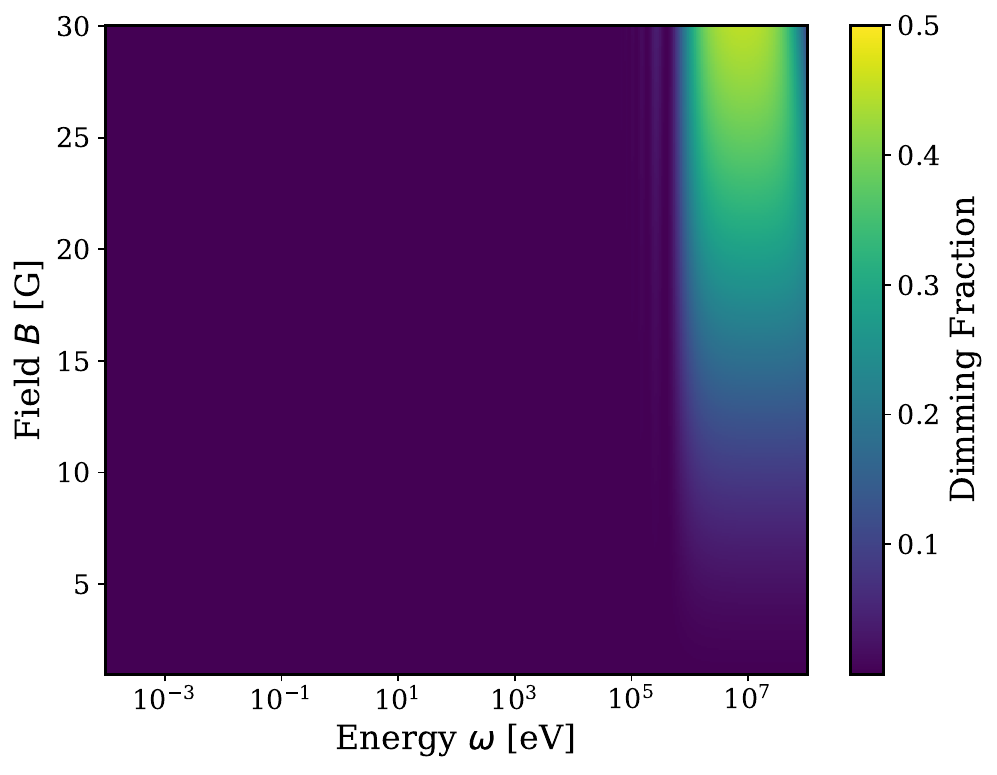}
  \caption{$m_a = 10^{-7}$ eV, $a = 0.3$}
  \label{fig:dimming1_a}
\end{subfigure}
\hfill
\begin{subfigure}[b]{0.32\textwidth}
  \centering
  \includegraphics[width=\textwidth]{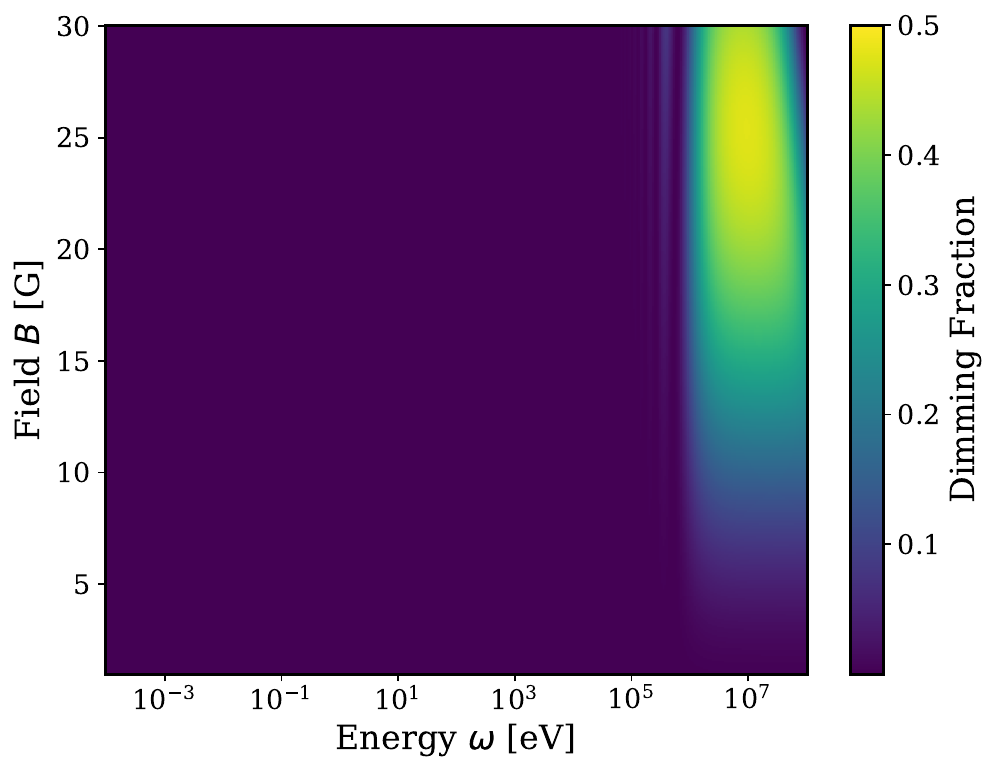}
  \caption{$m_a = 10^{-7}$ eV, $a = 0.6$}
  \label{fig:dimming1_b}
\end{subfigure}
\hfill
\begin{subfigure}[b]{0.32\textwidth}
  \centering
  \includegraphics[width=\textwidth]{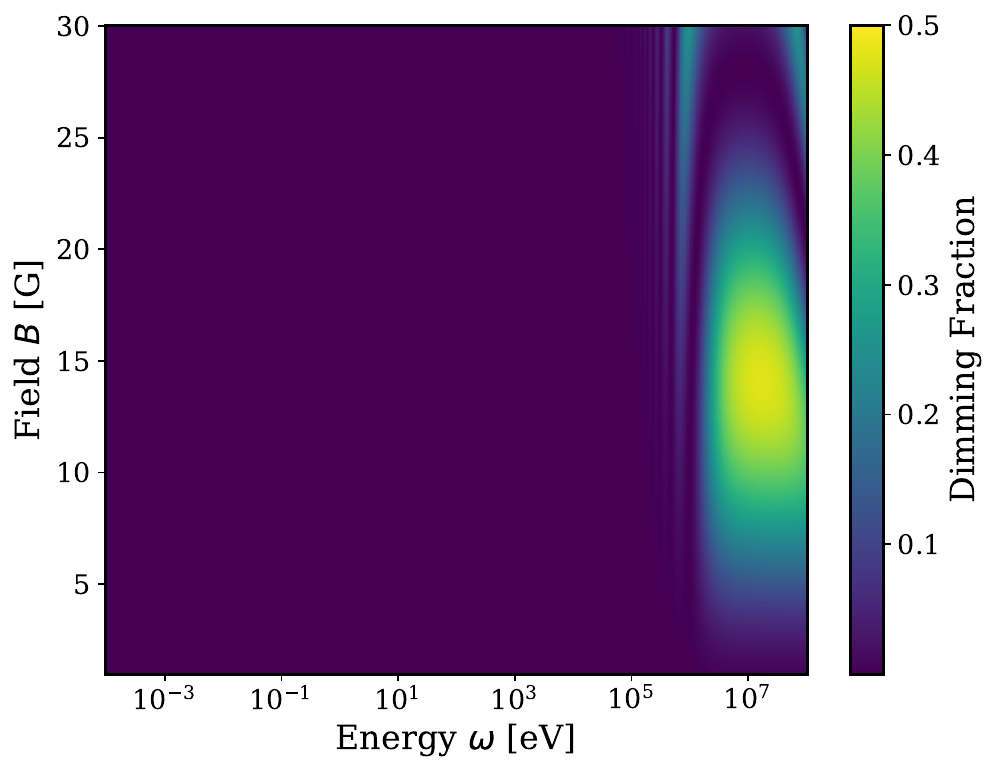}
  \caption{$m_a = 10^{-7}$ eV, $a = 0.99$}
  \label{fig:dimming1_c}
\end{subfigure}

\bigskip

\begin{subfigure}[b]{0.32\textwidth}
  \centering
  \includegraphics[width=\textwidth]{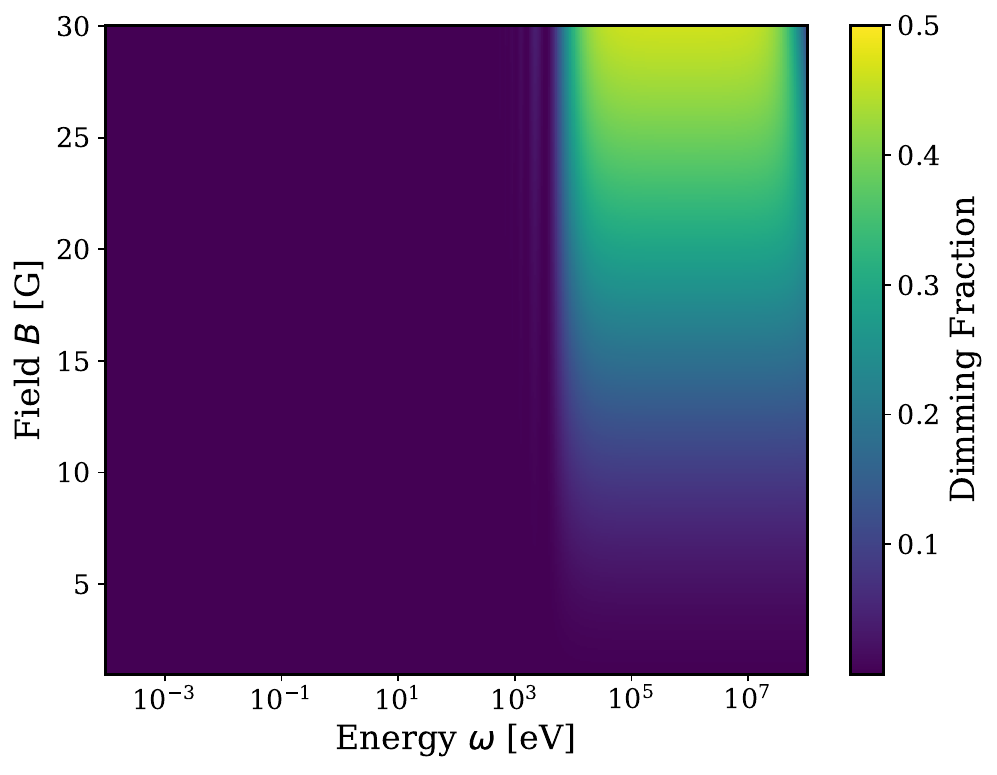}
  \caption{$m_a = 10^{-8}$ eV, $a = 0.3$}
  \label{fig:dimming1_d}
\end{subfigure}
\hfill
\begin{subfigure}[b]{0.32\textwidth}
  \centering
  \includegraphics[width=\textwidth]{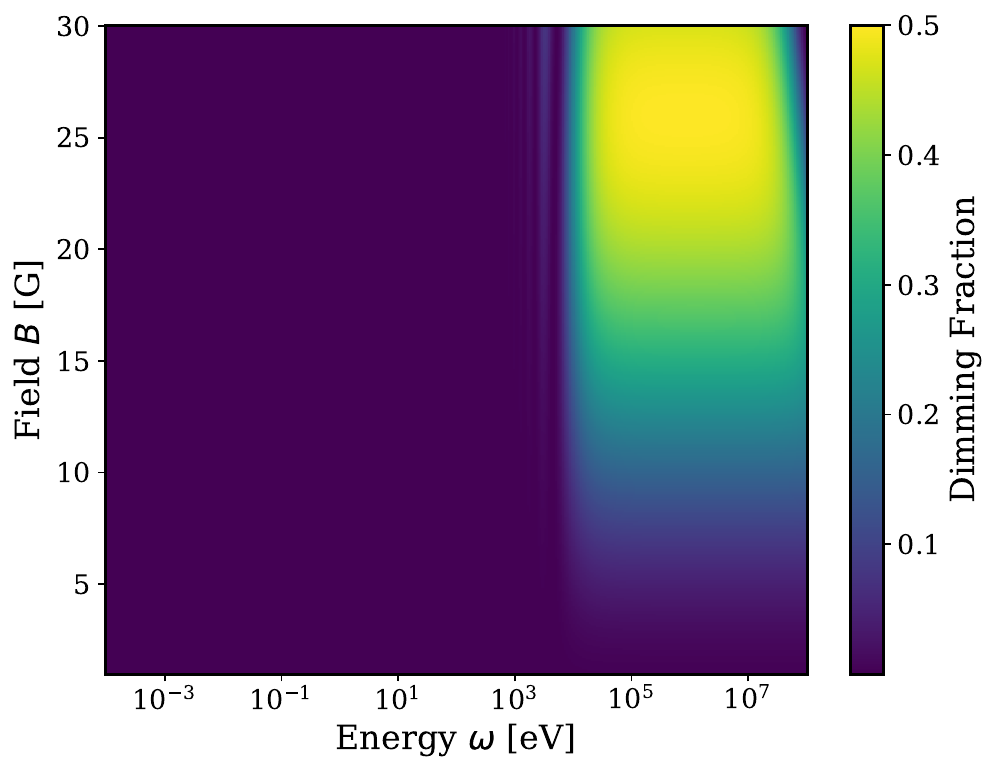}
  \caption{$m_a = 10^{-8}$ eV, $a = 0.6$}
  \label{fig:dimming1_e}
\end{subfigure}
\hfill
\begin{subfigure}[b]{0.32\textwidth}
  \centering
  \includegraphics[width=\textwidth]{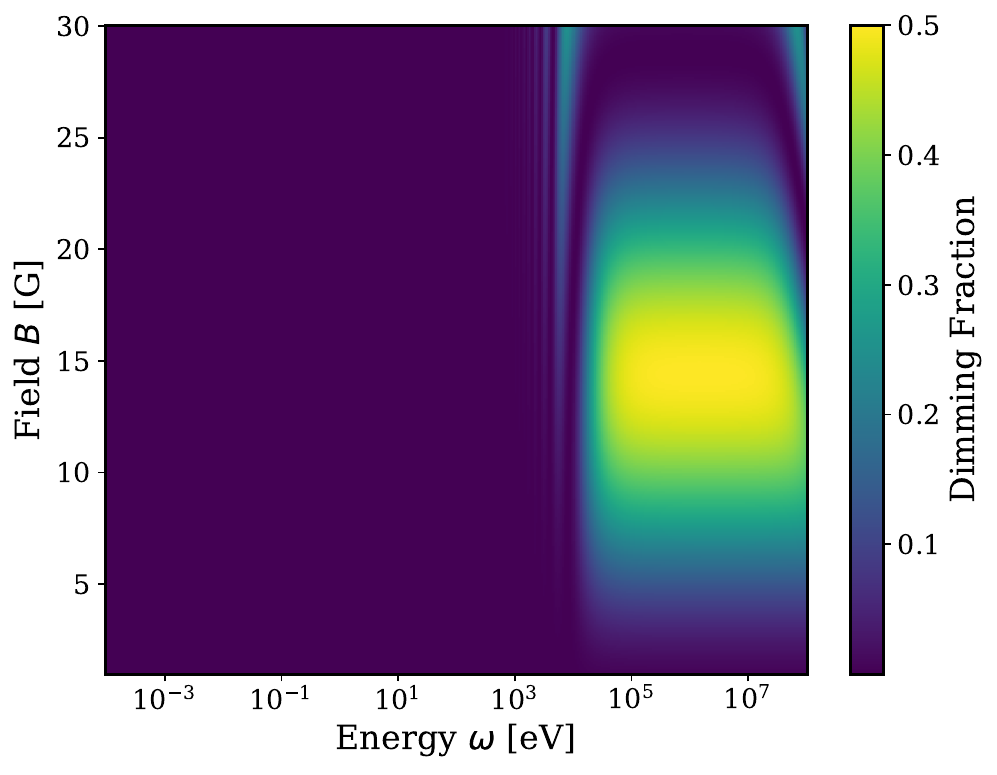}
  \caption{$m_a = 10^{-8}$ eV, $a = 0.99$}
  \label{fig:dimming1_f}
\end{subfigure}

\begin{subfigure}[b]{0.32\textwidth}
  \centering
  \includegraphics[width=\textwidth]{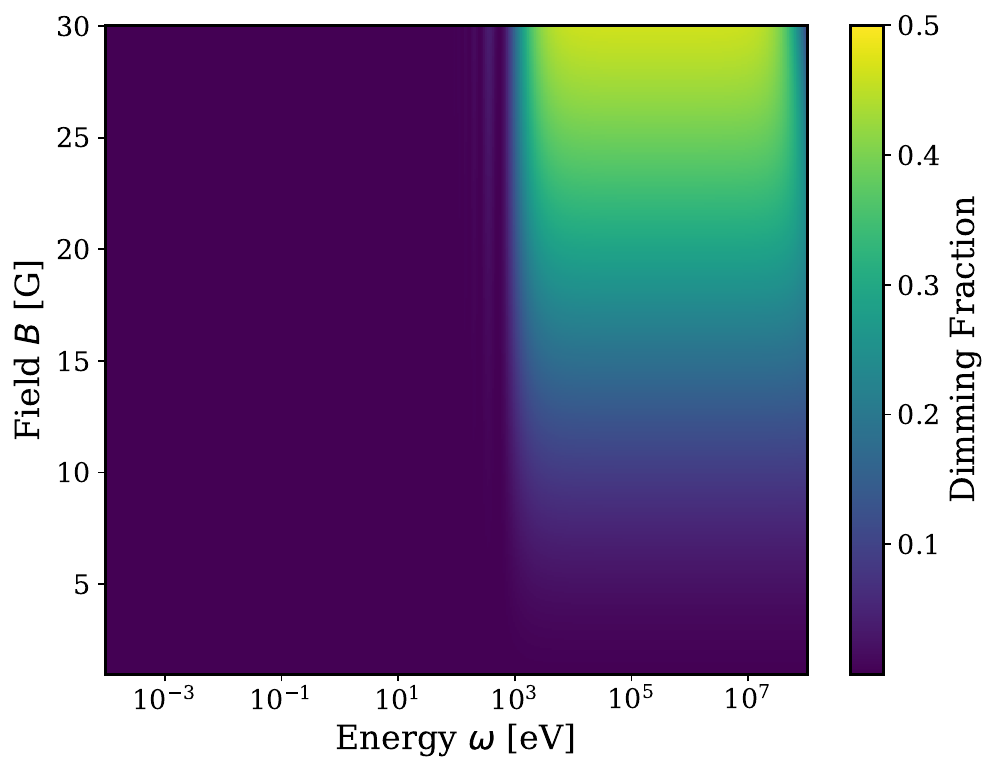}
  \caption{$m_a = 10^{-10}$ eV, $a = 0.3$}
  \label{fig:dimming1_g}
\end{subfigure}
\hfill
\begin{subfigure}[b]{0.32\textwidth}
  \centering
  \includegraphics[width=\textwidth]{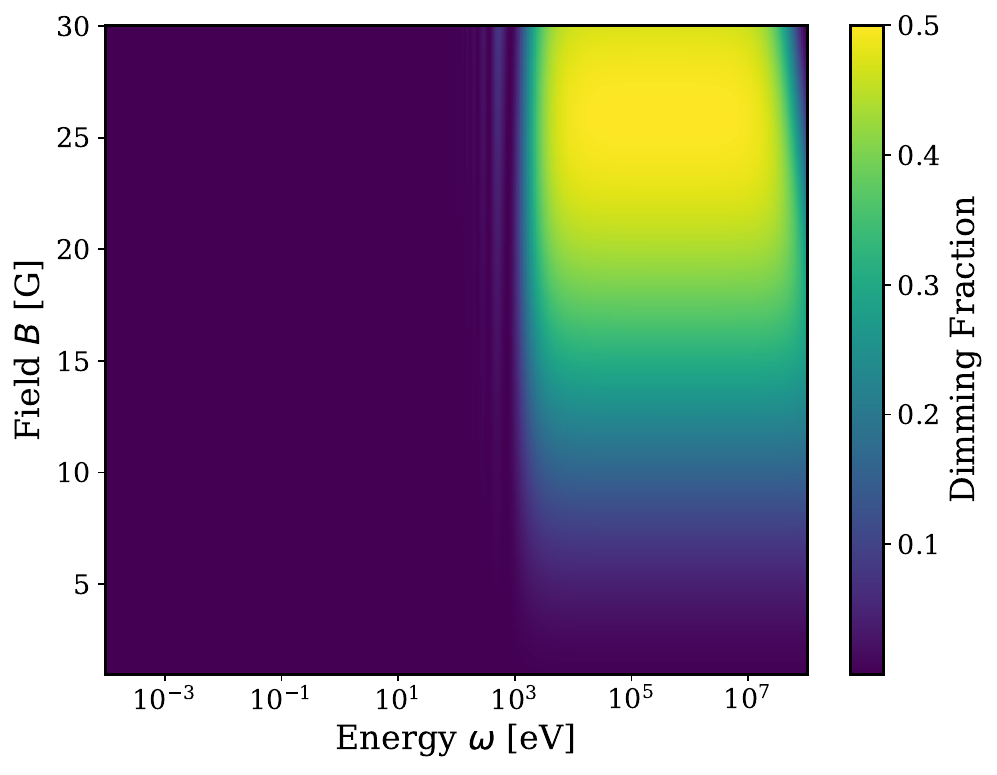}
  \caption{$m_a = 10^{-10}$ eV, $a = 0.6$}
  \label{fig:dimming1_h}
\end{subfigure}
\hfill
\begin{subfigure}[b]{0.32\textwidth}
  \centering
  \includegraphics[width=\textwidth]{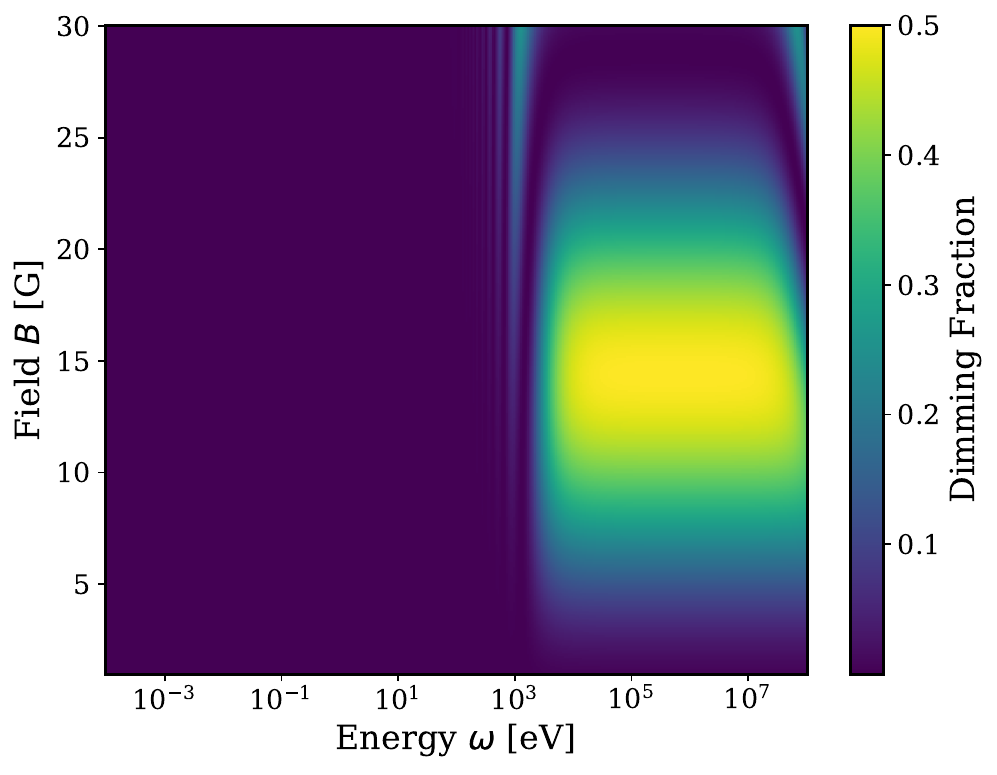}
  \caption{$m_a = 10^{-10}$ eV, $a = 0.99$}
  \label{fig:dimming1_i}
\end{subfigure}

\caption{Variation of dimming fraction with magnetic field strength and photon frequency corresponding to $g_{a\gamma} = 10^{-11}$ GeV$^{-1}$, and number density of electron $n_e = 10^{4}$ $\text{cm}^{-3}$ . In each horizontal panel we consider three different BH spins for a fixed axion mass while in each vertical panel we consider variation in dimming percentage with axion mass for a fixed BH spin.}
\label{fig:dimming_g1}

\end{figure}

\begin{figure}[!htbp]
\centering

\begin{subfigure}[b]{0.32\textwidth}
  \centering
  \includegraphics[width=\textwidth]{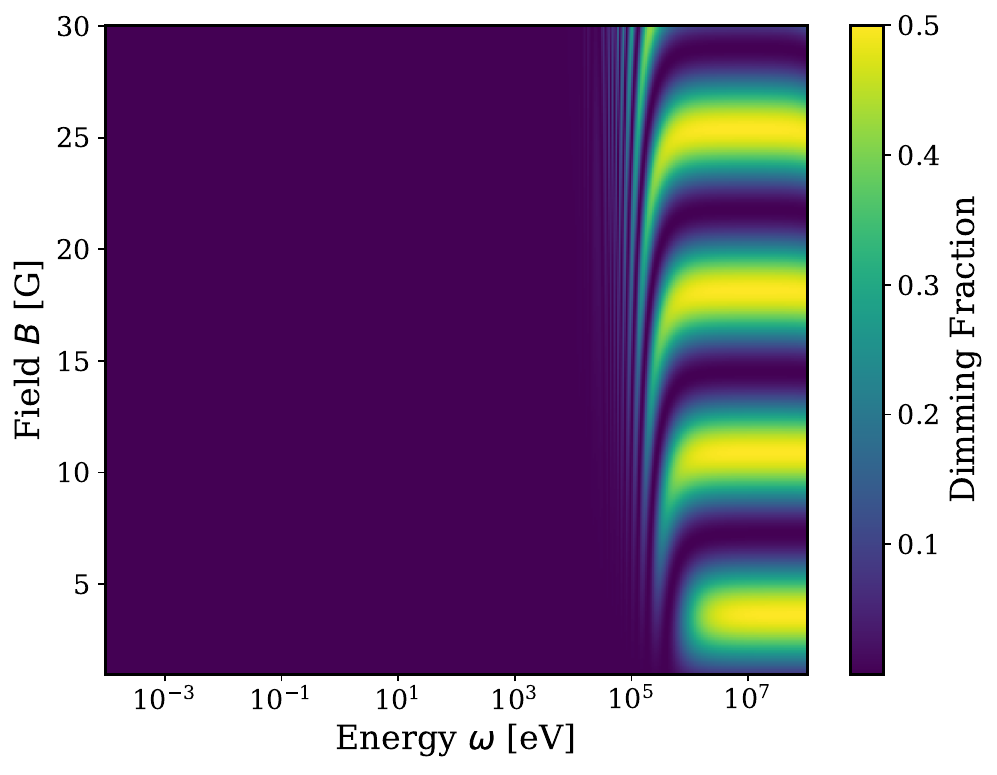}
  \caption{$m_a = 10^{-7}$ eV, $a = 0.3$}
  \label{fig:dimming2_a}
\end{subfigure}
\hfill
\begin{subfigure}[b]{0.32\textwidth}
  \centering
  \includegraphics[width=\textwidth]{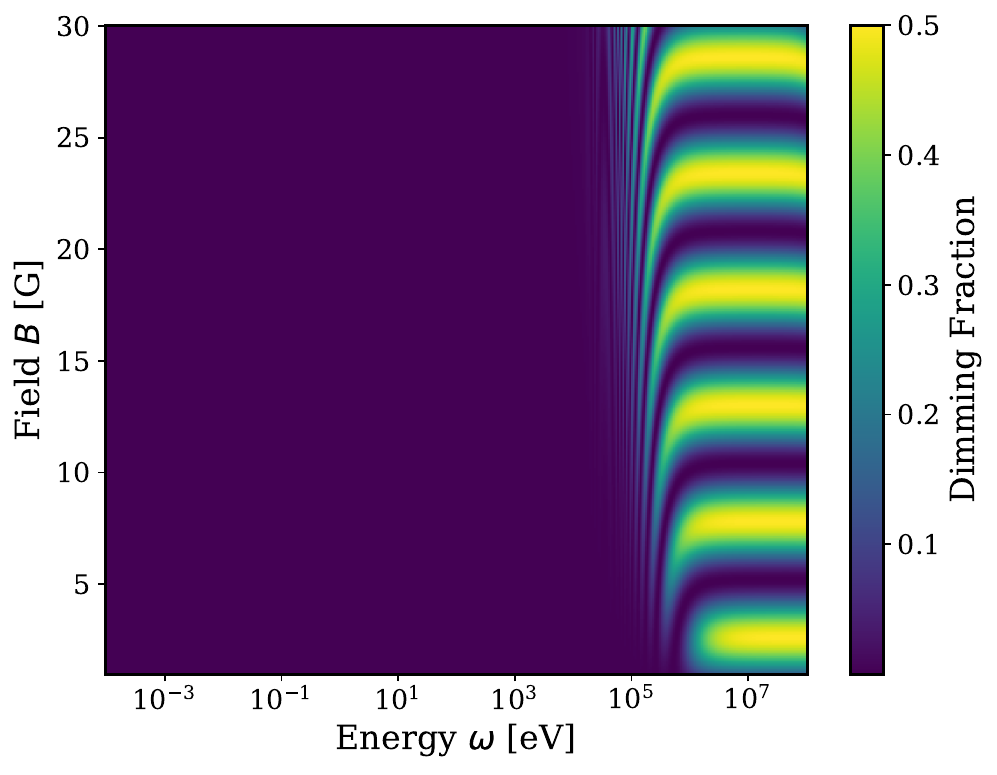}
  \caption{$m_a = 10^{-7}$ eV, $a = 0.6$}
  \label{fig:dimming2_b}
\end{subfigure}
\hfill
\begin{subfigure}[b]{0.32\textwidth}
  \centering
  \includegraphics[width=\textwidth]{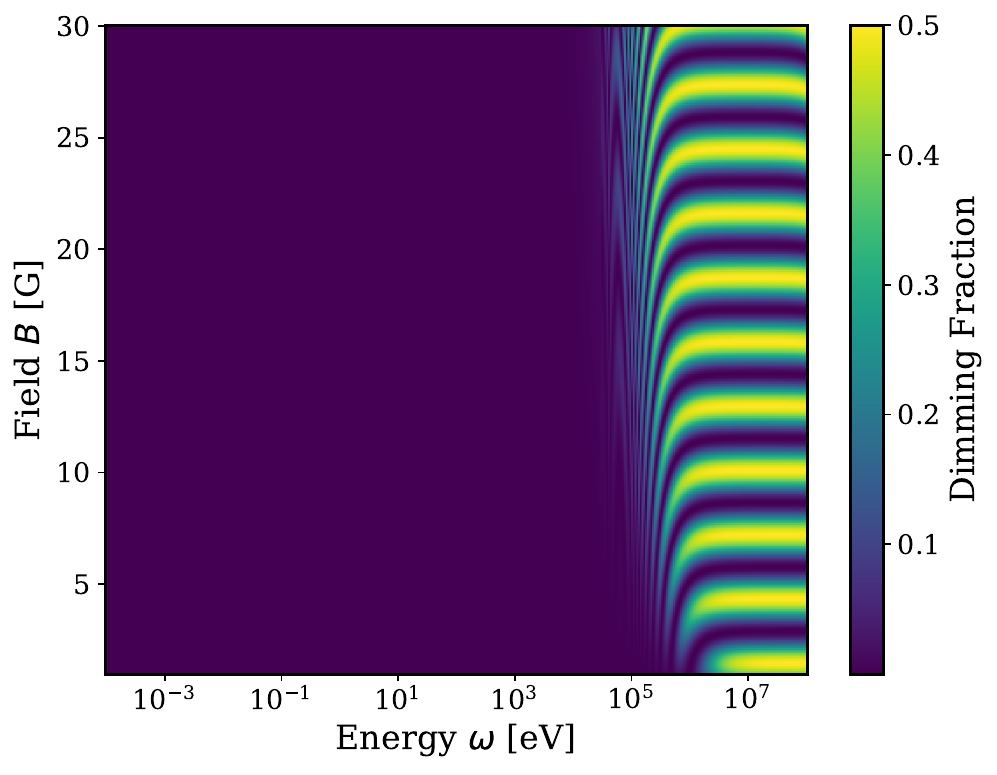}
  \caption{$m_a = 10^{-7}$ eV, $a = 0.99$}
  \label{fig:dimming2_c}
\end{subfigure}

\bigskip

\begin{subfigure}[b]{0.32\textwidth}
  \centering
  \includegraphics[width=\textwidth]{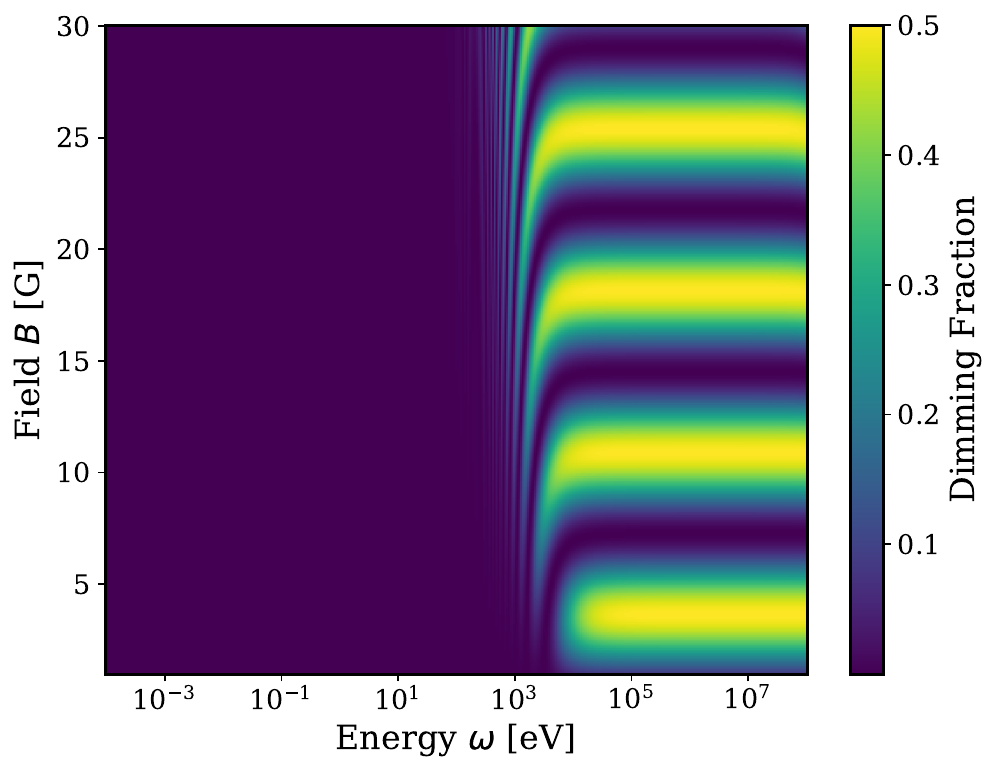}
  \caption{$m_a = 10^{-8}$ eV, $a = 0.3$}
  \label{fig:dimming2_d}
\end{subfigure}
\hfill
\begin{subfigure}[b]{0.32\textwidth}
  \centering
  \includegraphics[width=\textwidth]{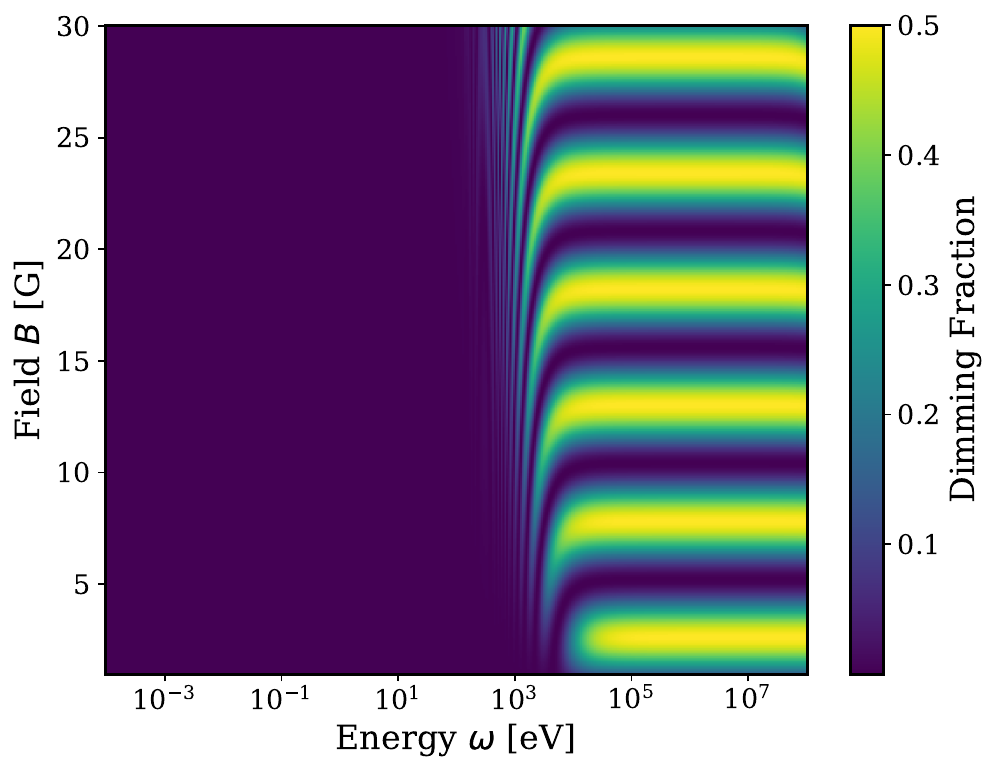}
  \caption{$m_a = 10^{-8}$ eV, $a = 0.6$}
  \label{fig:dimming2_e}
\end{subfigure}
\hfill
\begin{subfigure}[b]{0.32\textwidth}
  \centering
  \includegraphics[width=\textwidth]{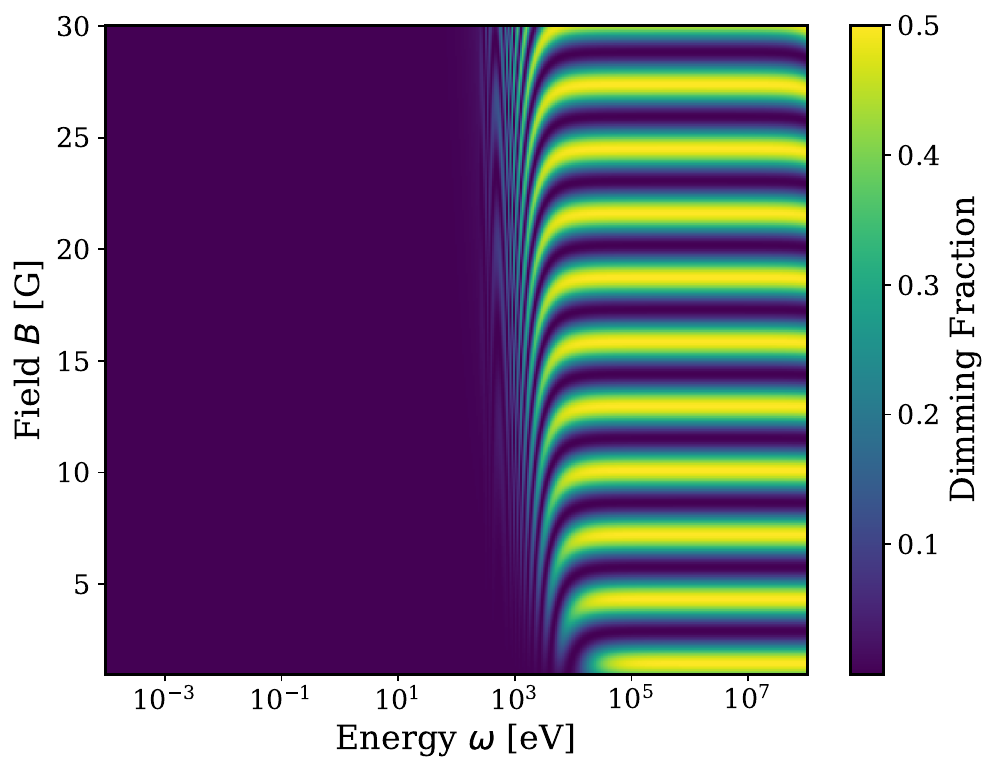}
  \caption{$m_a = 10^{-8}$ eV, $a = 0.99$}
  \label{fig:dimming2_f}
\end{subfigure}

\begin{subfigure}[b]{0.32\textwidth}
  \centering
  \includegraphics[width=\textwidth]{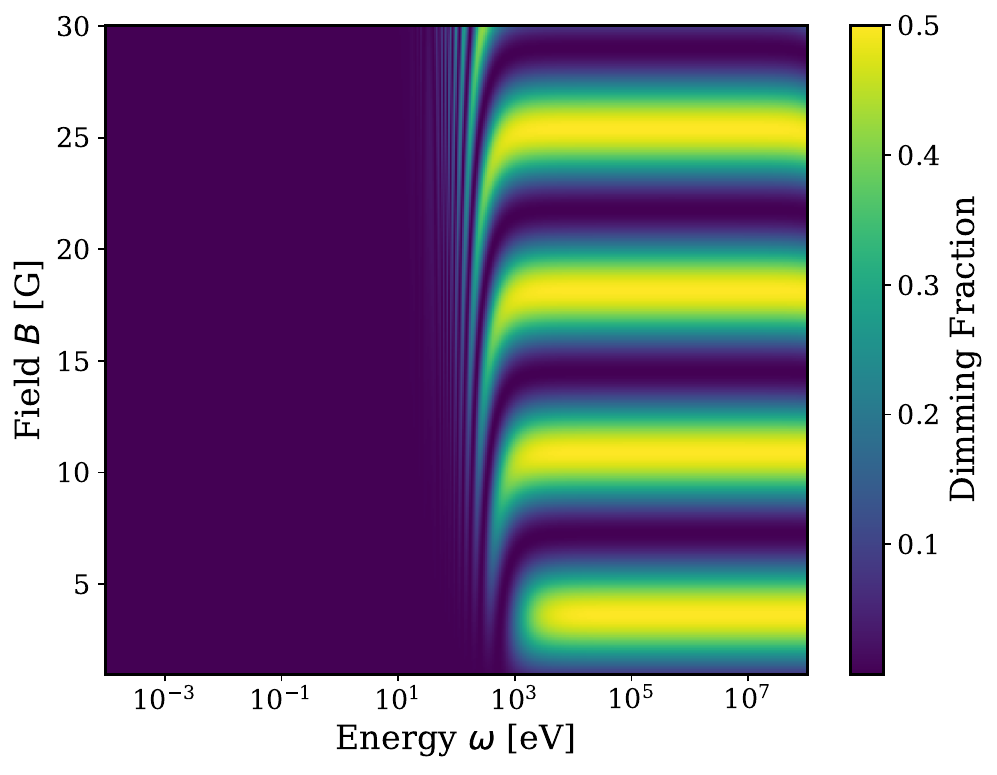}
  \caption{$m_a = 10^{-10}$ eV, $a = 0.3$}
  \label{fig:dimming2_g}
\end{subfigure}
\hfill
\begin{subfigure}[b]{0.32\textwidth}
  \centering
  \includegraphics[width=\textwidth]{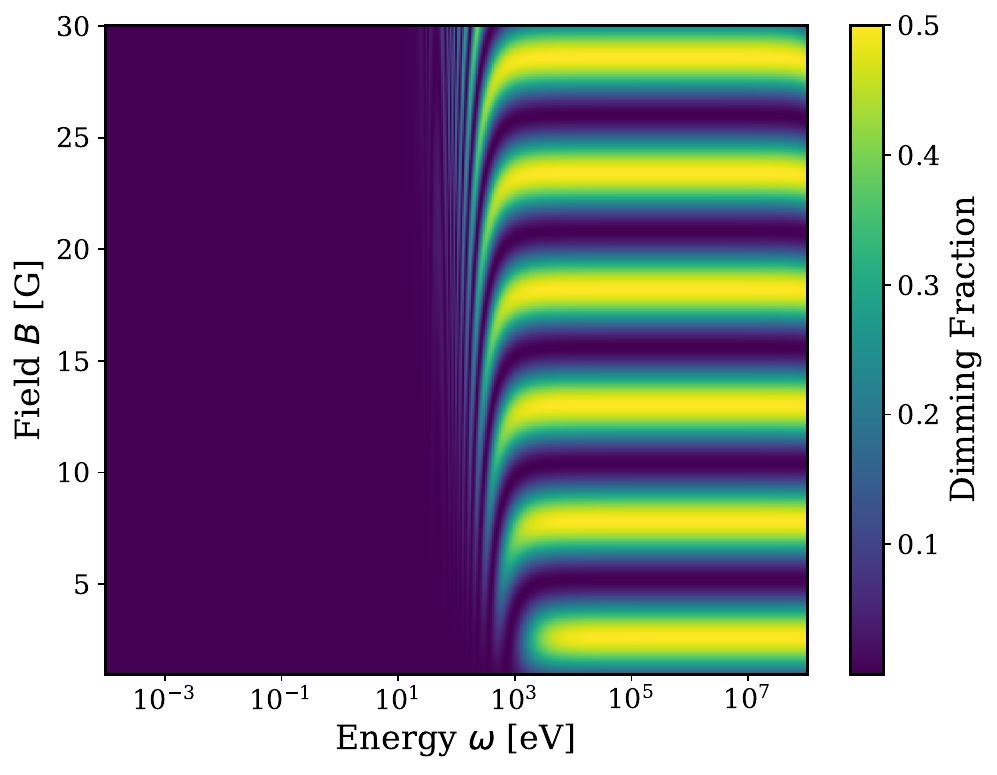}
  \caption{$m_a = 10^{-10}$ eV, $a = 0.6$}
  \label{fig:dimming2_h}
\end{subfigure}
\hfill
\begin{subfigure}[b]{0.32\textwidth}
  \centering
  \includegraphics[width=\textwidth]{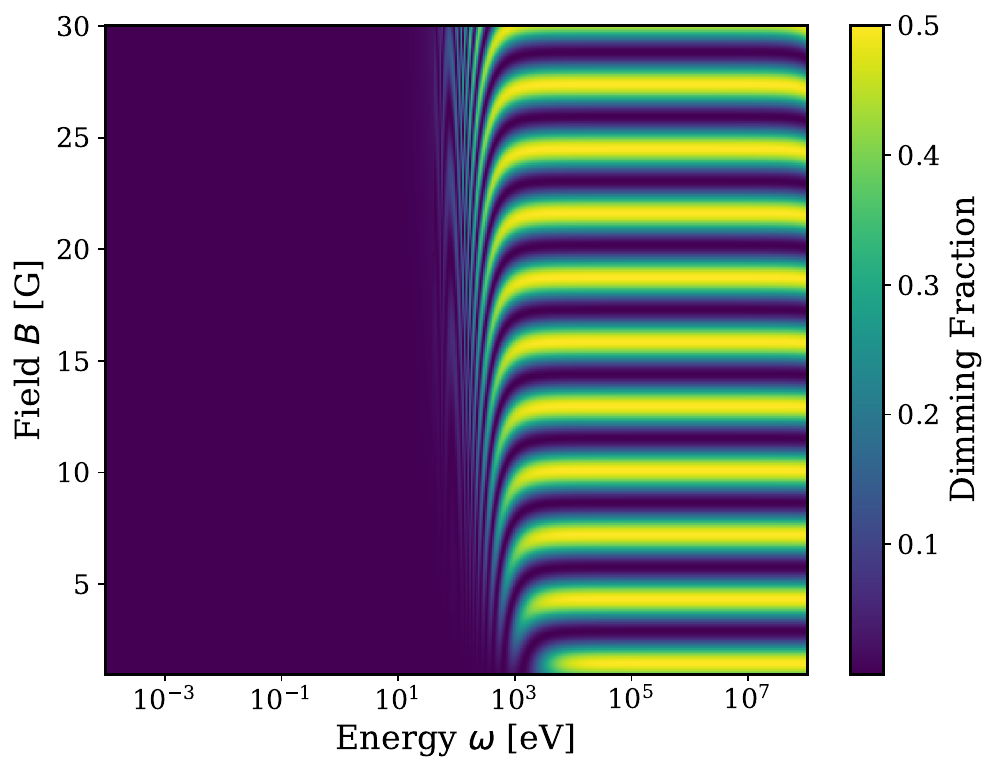}
  \caption{$m_a = 10^{-10}$ eV, $a = 0.99$}
  \label{fig:dimming2_i}
\end{subfigure}

\caption{Variation of dimming fraction with magnetic field strength and photon frequency corresponding to $g_{a\gamma} = 10^{-10}$ GeV$^{-1}$, and number density of electron $n_e = 10^{4}$ $\text{cm}^{-3}$ . In each horizontal panel we consider three different BH spins for a fixed axion mass while in each vertical panel we consider variation in dimming percentage with axion mass for a fixed BH spin.}
\label{fig:dimming_g2}

\end{figure}

In \cref{fig:dimming_g1} and \cref{fig:dimming_g2} we illustrate the dependence of the dimming percentage and the resulting dimming of astrophysical sources (in particular M87*), on the magnetic field strength and the photon energy corresponding to photon-axion coupling $g_{a\gamma}=10^{-11}\rm GeV^{-1}$ and $g_{a\gamma}=10^{-10}\rm GeV^{-1}$, respectively. We explore the dimming percentage in the two aforesaid figures that leads to formation of axions of mass 100 neV, 10 neV and 0.1 neV near Kerr BHs of spin $a=0.3$, $a=0.6$ and $a=0.99$ surrounded by plasma of electron  number density $n_e=10^{4}\rm cm^{-3}$. We note from \cref{fig:dimming_g1} and \cref{fig:dimming_g2}
that the dimming percentage suddenly increases when the photon frequency exceeds a certain cut-off. This is because,
above the cut-off frequency (approximately $ \omega\gtrsim 10^4\rm eV$, for \cref{1a}), $\mathrm{\Delta_{osc}}\simeq 2\mathrm{\Delta_M}$ (see \cref{eq:oscillation}). As a consequence, the amplitude term in the photon-axion conversion probability (\cref{eq:conversion_prob}) is nearly unity in this frequency range while below this frequency $\mathrm{\Delta_{osc}}\simeq \mathrm{\Delta_a}$ or $\mathrm{\Delta_{osc}}\simeq \mathrm{\Delta_{pl}}$ such that the amplitude term in the conversion probability is significantly lesser than unity. 

We further note from \cref{fig:dimming_g1} and \cref{fig:dimming_g2} that the cut-off frequency above which there is efficient conversion depends mainly on the axion mass and decreases with a decrease in the axion mass. This is because, at lower photon frequencies, for a fixed electron density, $\mathrm{\Delta_{osc}}\simeq\mathrm{\Delta_a}$. This relation holds over a larger frequency range when the axion mass is large (see \cref{eq:delta_a}). Thus, the transition from $\mathrm{\Delta_{osc}}\simeq\mathrm{\Delta_a}$ to $\mathrm{\Delta_{osc}}\simeq 2\mathrm{\Delta_M}$ happens at a higher frequency if the axion mass is increased, as a consequence the cut-off frequency above which there is efficient conversion increases with an increase in the axion mass.

We further note from \cref{fig:dimming_g1} that near low to intermediate spin Kerr BHs ($a=0.3$ to $a=0.6$), the photon-axion conversion can happen efficiently above the cut-off frequency if the magnetic field is sufficiently strong. The figure also reveals that near a slowly rotating BH stronger magnetic fields are required for efficient conversion which in turn leads to an enhanced dimming fraction. Conversely, near a rapidly rotating Kerr BH efficient conversion can happen even at intermediate magnetic fields(\cref{fig:dimming1_c}, \cref{fig:dimming1_f} and \cref{fig:dimming1_i}).
This is because in the conversion probability (\cref{eq:conversion_prob}) there is a sinusoidal term which contributes maximally when $\mathrm{\Delta_{osc}}z\sim (2n+1)\pi$. Above the cut-off frequency $\mathrm{\Delta_{osc}}\simeq 2\mathrm{\Delta_M}$, which increases with an increase in the magnetic field and the photon-axion coupling. In the vicinity of a rapidly rotating Kerr BH, the photon spends more time in the photon region which leads to an enhanced path length $z$ (see \cref{tab:dimming_results_17_sorted}, \cref{tab:dimming_results_2} and \cref{tab:dimming_results_3}). Thus, for a given photon-axion coupling, $\mathrm{\Delta_{osc}}z\sim (2n+1)\pi$ can be achieved at lower magnetic fields if the path length proportionally increases (\cref{eq:conversion_prob}), which is what happens for a Kerr BH with near maximal spin.

Conversely, efficient conversion can also be achieved above the cut-off frequency at magnetic fields $\sim1$G, if the desired magnitude of $\mathrm{\Delta_{osc}}\simeq 2\mathrm{\Delta_M}$ can be achieved by an enhancement of the photon-axion coupling (\cref{fig:dimming_g2}). Under such circumstances, one can get large dimming fraction above the cut-off frequency even for low magnetic fields and also near slowly rotating BHs since $\mathrm{\Delta_{osc}}$ increases with an increase in $g_{a\gamma}$.
The oscillatory behavior of the sinusoidal term in the conversion probability (\cref{eq:conversion_prob}) becomes more apparent when the photon-axion coupling is increased by an order of magnitude (\cref{fig:dimming_g2}). 
From \cref{fig:delta_osc_mag_field} it is clear that when the photon frequency is approximately in the range $10^4 \rm eV \lesssim \omega\lesssim 10^8\rm eV$,  with an increase in $g_{a\gamma}$, $\mathrm{\Delta_{osc}}$ increases almost by an order and hence as one varies magnetic field in the range 1-30 G (as inferred for M87* \cite{EventHorizonTelescope:2021srq}), the path lengths are such that $\mathrm{\Delta_{osc}}z\sim (2n+1)\pi$ are achieved more frequently, for different choices of the integer $n$. Thus, the conversion probability (and hence the dimming fraction) more rapidly increases and decreases with an increase in the magnetic field, as $g_{a\gamma}$ is increased to $g_{a\gamma}\simeq 10^{-10}\rm GeV ^{-1}$ (\cref{fig:dimming_g2}) from $g_{a\gamma}\simeq 10^{-11}\rm GeV ^{-1}$ (\cref{fig:dimming_g1}). As path length increases near a rapidly rotating Kerr BH, the rate of change of the dimming fraction increases with an increase in spin as is evident from \cref{fig:dimming2_a}- \cref{fig:dimming2_c}, \cref{fig:dimming2_d}- \cref{fig:dimming2_f}, and \cref{fig:dimming2_g}- \cref{fig:dimming2_i}. By comparing \cref{fig:dimming_g1} and \cref{fig:dimming_g2} we note that the cut-off frequency from where the conversion becomes efficient marginally decreases with an increase in the photon-axion coupling, for a given axion mass. This is because $\Delta_{M}$ increases with an increase in $g_{a\gamma}$ and hence the transition from $\mathrm{\Delta_{osc}}\simeq \left| \Delta_{\rm pl} - \Delta_a \right|$ to $\mathrm{\Delta_{osc}}\simeq 2\mathrm{\Delta_M}$ occurs at a lower frequency 
(see \cref{eq:delta_a} and \cref{eq:delta_M}).

From previous discussion it is clear that the conversion is most efficient when $\mathrm{\Delta_{osc}}\simeq 2\mathrm{\Delta_M}$ which in turn implies $(\mathrm{\Delta_{pl}}-\mathrm{\Delta_{vac}}-\mathrm{\Delta_a})^2\ll 4\mathrm{\Delta_M}^2$ in \cref{eq:oscillation}. This can be achieved in the following ways \cite{Nomura:2022zyy}:
\begin{itemize}
    \item $\mathrm{\Delta_a}\ll\mathrm{\Delta_M}$, $\mathrm{\Delta_{pl}}\ll\mathrm{\Delta_M}$ and $\mathrm{\Delta_{vac}}\ll\mathrm{\Delta_M}$ which implies the following inequalities:
    \begin{align}
        5.1\Big(\frac{m_a}{\rm neV}\Big)^2\ll \Big(\frac{g_{a\gamma}}{10^{-11}\rm GeV^{-1}}\Big)\Big(\frac{\omega}{\rm keV}\Big)\Big(\frac{B}{G}\Big),
        \label{ineq1}
    \end{align}
    \begin{align}
        7.0\times 10^{-3}\Big(\frac{n_e}{\rm cm^{-3}}\Big)\ll \Big(\frac{g_{a\gamma}}{10^{-11}\rm GeV^{-1}}\Big)\Big(\frac{\omega}{\rm keV}\Big)\Big(\frac{B}{G}\Big),
        \label{ineq2}
    \end{align}
    \begin{align}
  \Big(\frac{\omega}{\rm keV}\Big)\Big(\frac{B}{G}\Big)\ll 1.1\times10^6\Big(\frac{g_{a\gamma}}{10^{-11}\rm GeV^{-1}}\Big).
  \label{ineq3}
    \end{align}

\cref{ineq1} and \cref{ineq2} gives a lower bound  while \cref{ineq3} provides an upper bound on the photon frequency $\omega$. These inequalities also indicate that for magnetic fields and electron density as observed for M87*, at X-ray/gamma-ray frequencies the above inequalities are satisfied and hence efficient conversion can be generally realized for high frequency photons. 
  \item $(\mathrm{\Delta_{pl}}-\mathrm{\Delta_{vac}}-\mathrm{\Delta_a})^2\ll 4\mathrm{\Delta_M}^2$ can also be achieved when $\mathrm{\Delta_{pl}}\simeq \mathrm{\Delta_{vac}}\gg \mathrm{\Delta_a}$ which implies
  \begin{align}
      \Big(\frac{\omega}{\rm keV}\Big)^2\Big(\frac{B}{G}\Big)^2\simeq 7.4\times 10^3\Big(\frac{n_e}{\rm cm^{-3}}\Big),
    \end{align}
\begin{align}
  \Big(\frac{\omega}{\rm keV}\Big)^2  \Big(\frac{B}{G}\Big)^2\gg5.4\times10^6\Big(\frac{m_a}{\rm neV}\Big)^2
\end{align}
The above conditions also provide a lower bound on $\omega$ and are generally satisfied for high frequency photons.

\begin{figure}[!htbp]
\centering

% ---------- Row 1 ----------
\begin{subfigure}[b]{0.32\textwidth}
  \centering
  \includegraphics[width=\textwidth]{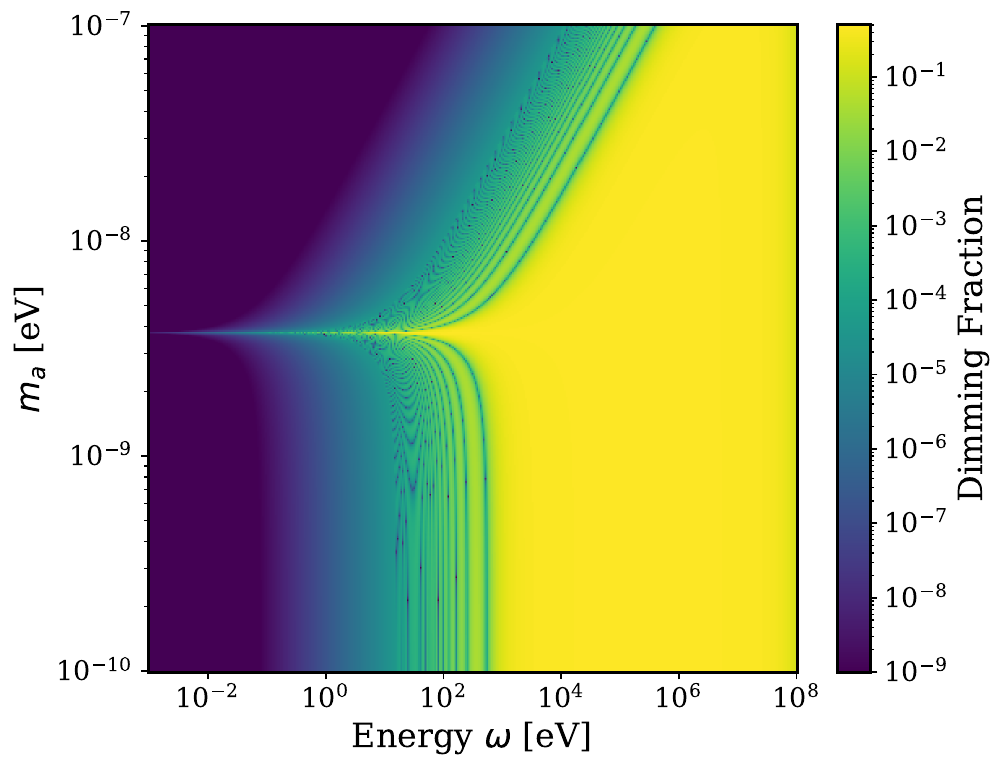}
  \caption{$a = 0.3$}
\end{subfigure}
\hfill
\begin{subfigure}[b]{0.32\textwidth}
  \centering
  \includegraphics[width=\textwidth]{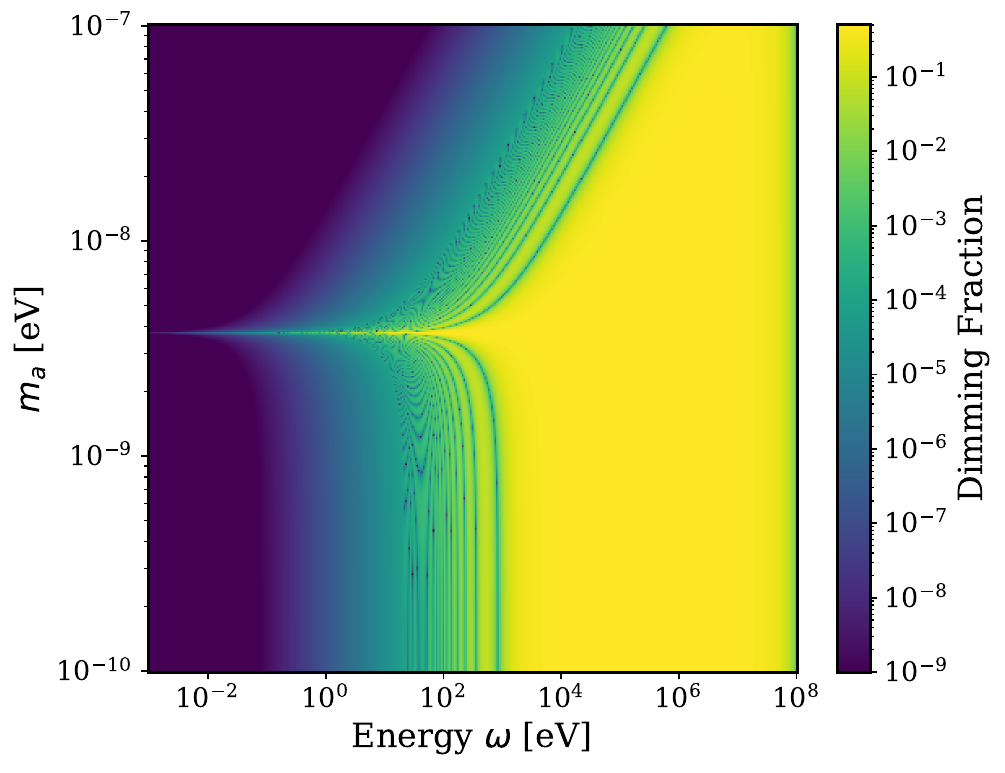}
  \caption{$a = 0.6$}
\end{subfigure}
\hfill
\begin{subfigure}[b]{0.32\textwidth}
  \centering
  \includegraphics[width=\textwidth]{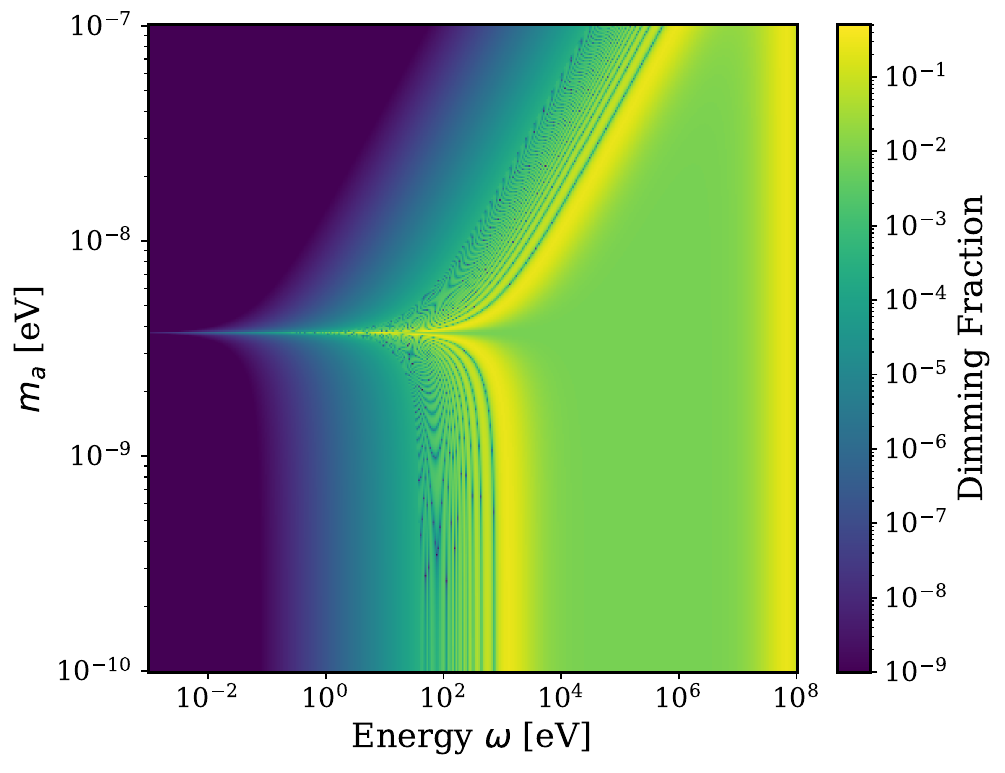}
  \caption{$a = 0.99$}
\end{subfigure}

\vspace{0.4cm}

% ---------- Row 2 ----------
\begin{subfigure}[b]{0.32\textwidth}
  \centering
  \includegraphics[width=\textwidth]{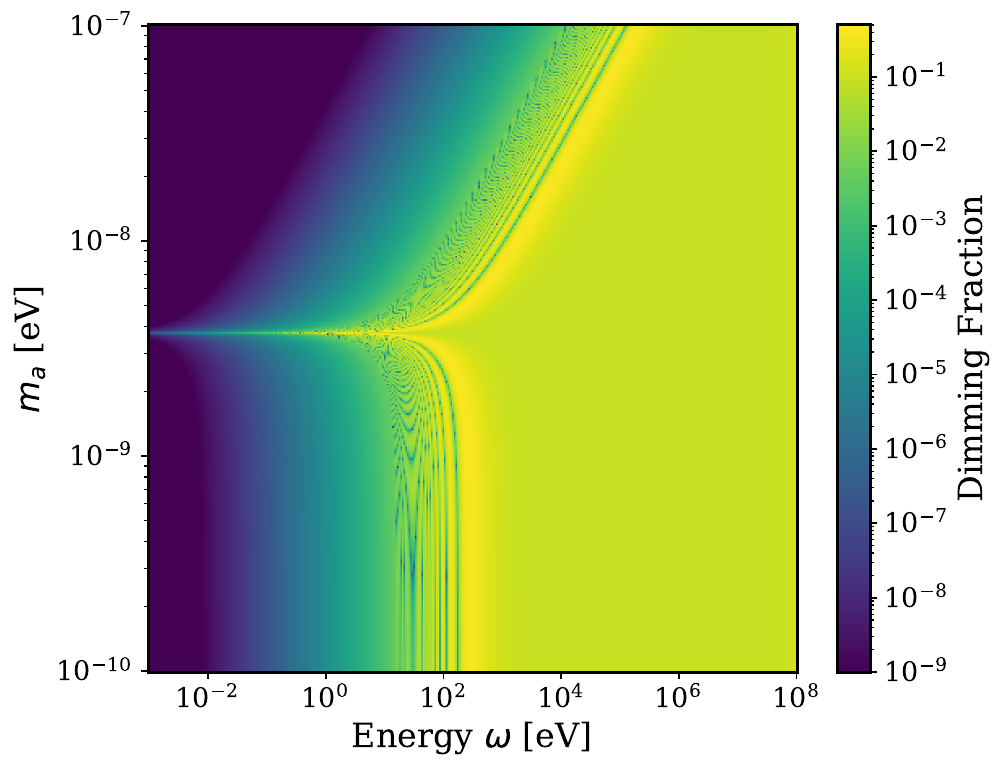}
  \caption{$a = 0.3$}
\end{subfigure}
\hfill
\begin{subfigure}[b]{0.32\textwidth}
  \centering
  \includegraphics[width=\textwidth]{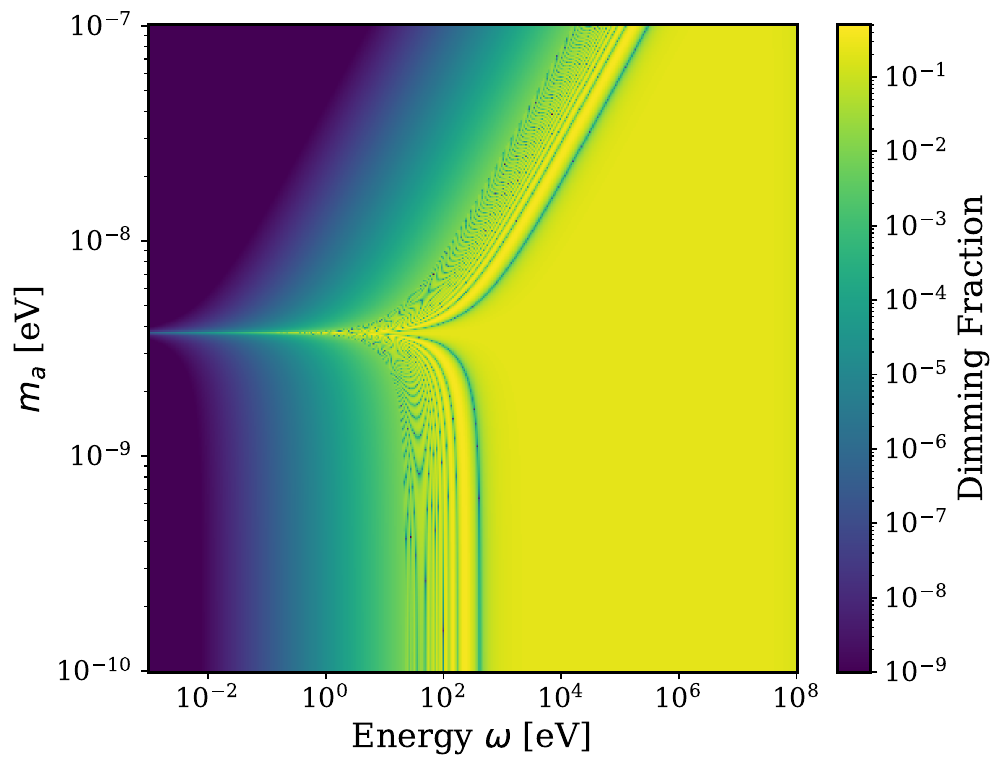}
  \caption{$a = 0.6$}
\end{subfigure}
\hfill
\begin{subfigure}[b]{0.32\textwidth}
  \centering
  \includegraphics[width=\textwidth]{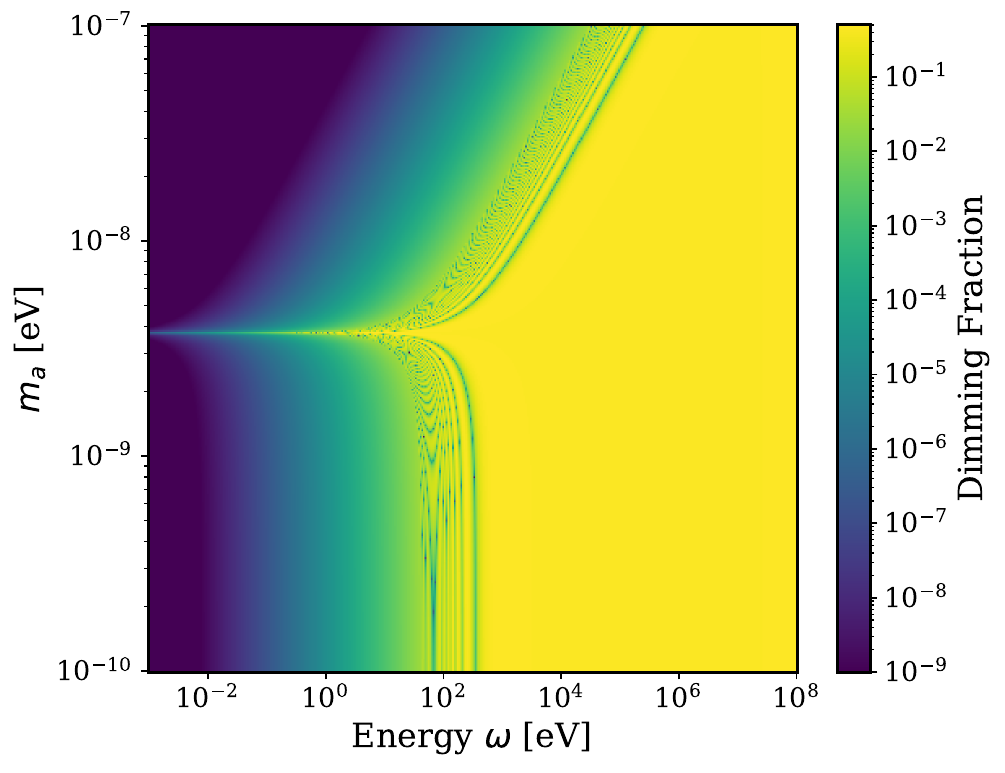}
  \caption{$a = 0.99$}
\end{subfigure}

\caption{
The above figure depicts the variation of the dimming fraction with axion mass and photon energy in the vicinity of Kerr BHs with spin $a=0.3,0.6, 0.99$. 
Top row: $g_{a\gamma} = 10^{-11}\,\mathrm{GeV}^{-1}$.
Bottom row: $g_{a\gamma} = 10^{-10}\,\mathrm{GeV}^{-1}$.
The magnetic field is fixed to $B = 30\,\mathrm{G}$ and the electron number density to
$n_e = 10^{4}\,\mathrm{cm}^{-3}$.
}
\label{Fig7}
\end{figure}

\item $(\mathrm{\Delta_{pl}}-\mathrm{\Delta_{vac}}-\mathrm{\Delta_a})^2\ll 4\mathrm{\Delta_M}^2$ can also be achieved when $\mathrm{\Delta_{pl}}\simeq \Delta_{a}\gg \mathrm{\Delta_{vac}}$. These conditions imply:
\begin{align}
 \Big(\frac{m_a}{\rm neV}\Big)^2\simeq 1.4\times 10^{-3}\Big(\frac{n_e}{\rm cm^{-3}}\Big),
 \label{mreso}
\end{align}
\begin{align}
   \Big(\frac{\omega}{\rm keV}\Big)^2 \Big(\frac{B}{G}\Big)^2\ll5.4\times10^6\Big(\frac{m_a}{\rm neV}\Big)^2,
\end{align}
which in turn provides an upper bound on the photon frequency $\omega$ depending on the magnetic field and the axion mass. Thus, if this condition is satisfied, efficient conversion can be achieved  at lower frequencies (e.g. radio frequencies) only if the axion assumes the resonant mass given in \cref{mreso} which in turn depends on the electron number density.
\end{itemize}

This is depicted in \cref{Fig7} which shows the variation of the dimming fraction with axion mass and photon frequency assuming a magnetic field strength of $B\simeq30G$ and electron number density of $n_e\simeq 10^4\rm cm^{-3}$. We note from \cref{Fig7} that efficient conversion generally takes place for high frequency photons but for the resonant axion mass (detemined by $n_e$ through \cref{mreso}) even low frequency photons exhibit substantial dimming, irrespective of the BH spin.

\iffalse
We further note from \cref{fig:dimming1_c}, \cref{fig:dimming1_f} and \cref{fig:dimming1_i} that for a rapidly rotating Kerr BH, near the cut-off frequency, efficient photon-axion conversion can happen at intermediate and high magnetic fields (the greenish-blue vertical lines in \cref{fig:dimming1_c}, \cref{fig:dimming1_f} and \cref{fig:dimming1_i}) while for higher frequencies the conversion is most efficient at intermediate magnetic fields. This is because, near the cut-off frequency $\mathrm{\Delta_{osc}}$ is dominated by both $\mathrm{\Delta_a}$ and  $\mathrm{\Delta_M}$. Note that when $\mathrm{\Delta_{osc}}$ is dominated by $\mathrm{\Delta_a}$ at low frequencies, it becomes insensitive to changes in $B$ (see \cref{fig:delta_osc} and \cref{fig:delta_osc_mag_field}). The cut-off frequency marks a transition phase such that below this frequency $\mathrm{\Delta_{osc}}$ is dominated by $\mathrm{\Delta_a}$ while above this frequency $\mathrm{\Delta_{osc}}$ is dominated by $\mathrm{\Delta_M}$. 
Thus, in this transition phase, $\mathrm{\Delta_{osc}}$ becomes less sensitive to changes in the magnetic field such that $z\mathrm{\Delta_{osc}}\sim (2n+1)\pi$ is achieved over a larger range of $B$ leading to a more efficient conversion near the cut-off frequency. 
\fi

\begin{figure}[!htbp]
\centering

\begin{subfigure}[b]{0.32\textwidth}
  \centering
  \includegraphics[width=\textwidth]{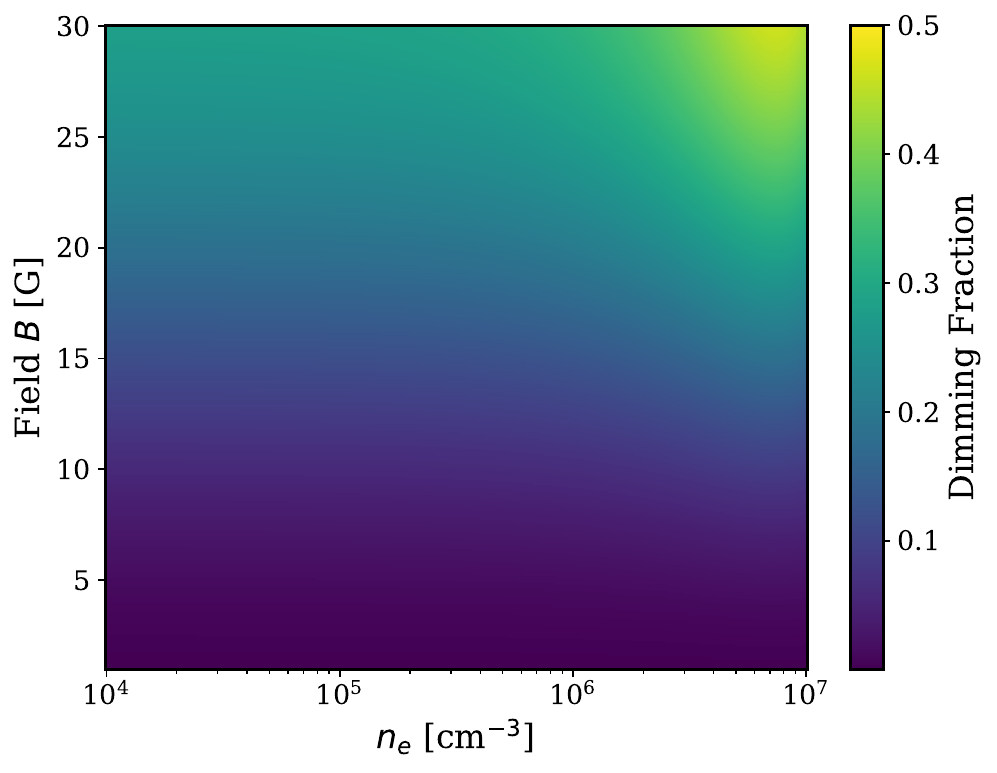}
  \caption{$m_a = 10^{-7}$ eV, $a = 0.3$}
  \label{fig:dimming3_a}
\end{subfigure}
\hfill
\begin{subfigure}[b]{0.32\textwidth}
  \centering
  \includegraphics[width=\textwidth]{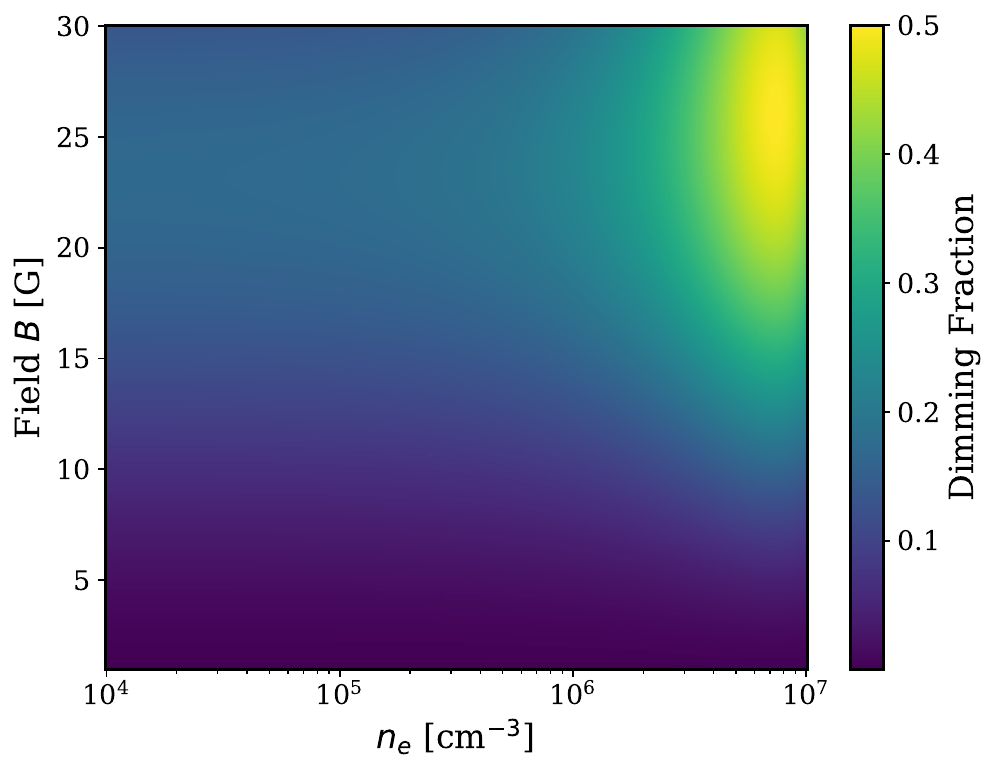}
  \caption{$m_a = 10^{-7}$ eV, $a = 0.6$}
  \label{fig:dimming3_b}
\end{subfigure}
\hfill
\begin{subfigure}[b]{0.32\textwidth}
  \centering
  \includegraphics[width=\textwidth]{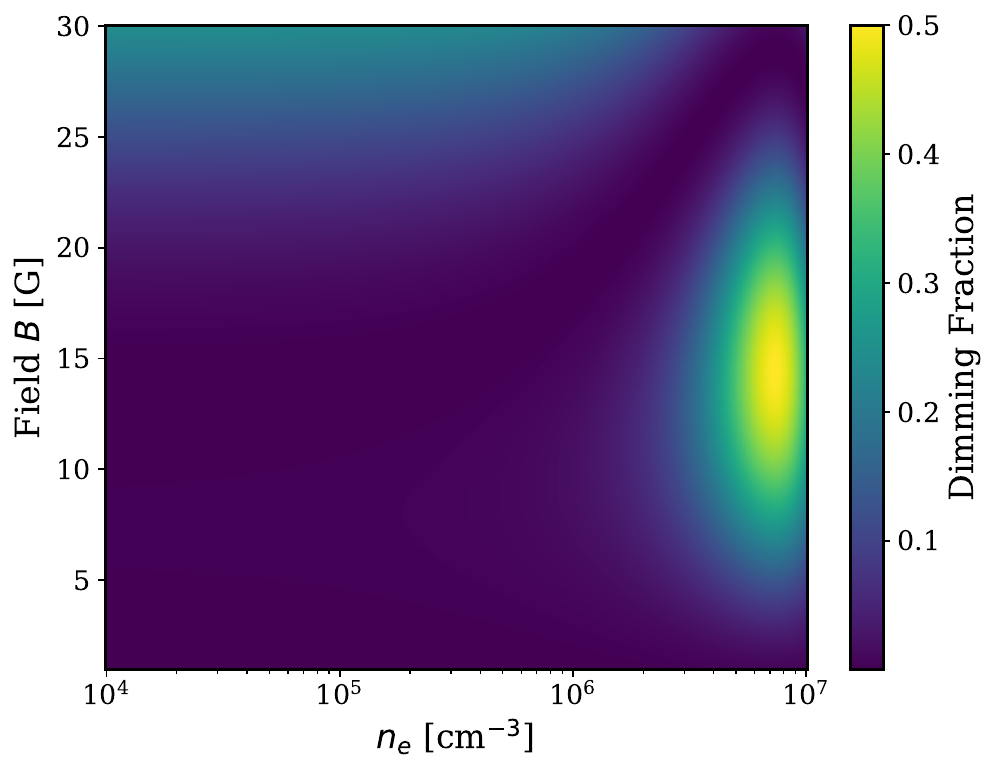}
  \caption{$m_a = 10^{-7}$ eV, $a = 0.99$}
  \label{fig:dimming3_c}
\end{subfigure}

\bigskip

\begin{subfigure}[b]{0.32\textwidth}
  \centering
  \includegraphics[width=\textwidth]{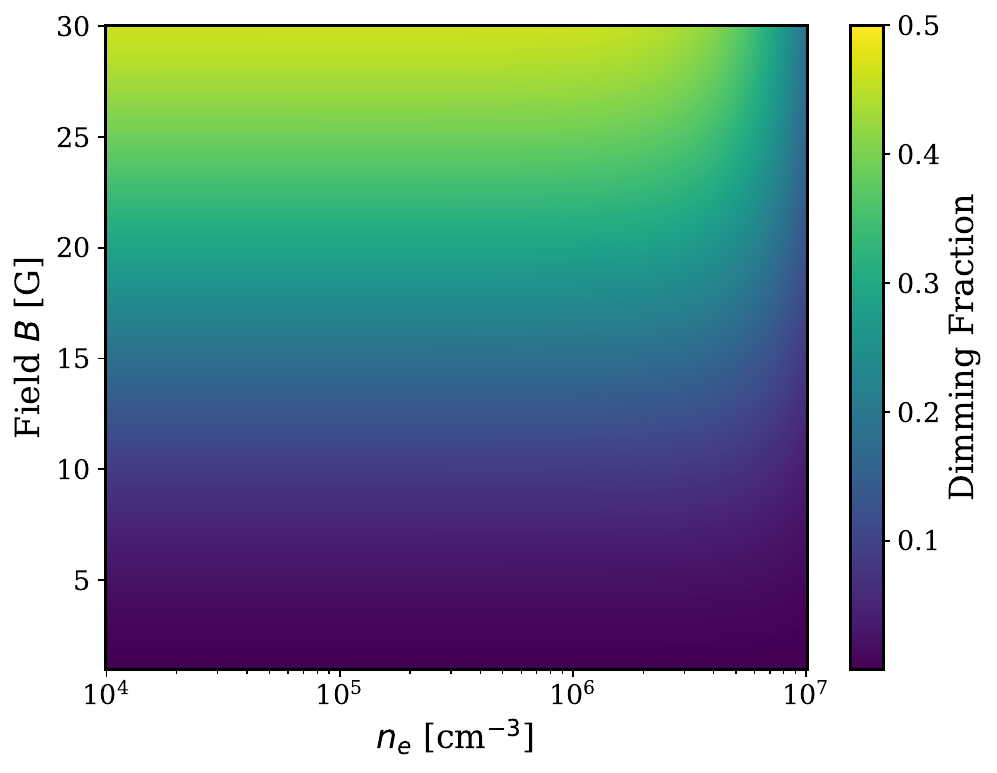}
  \caption{$m_a = 10^{-8}$ eV, $a = 0.3$}
  \label{fig:dimming3_d}
\end{subfigure}
\hfill
\begin{subfigure}[b]{0.32\textwidth}
  \centering
  \includegraphics[width=\textwidth]{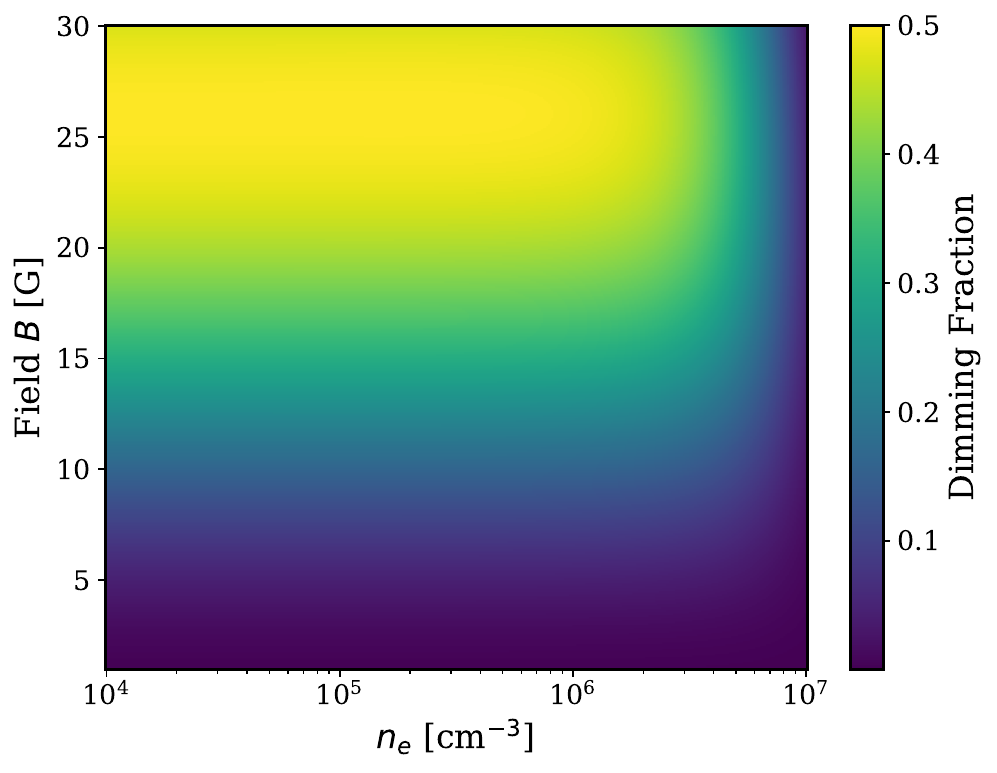}
  \caption{$m_a = 10^{-8}$ eV, $a = 0.6$}
  \label{fig:dimming3_e}
\end{subfigure}
\hfill
\begin{subfigure}[b]{0.32\textwidth}
  \centering
  \includegraphics[width=\textwidth]{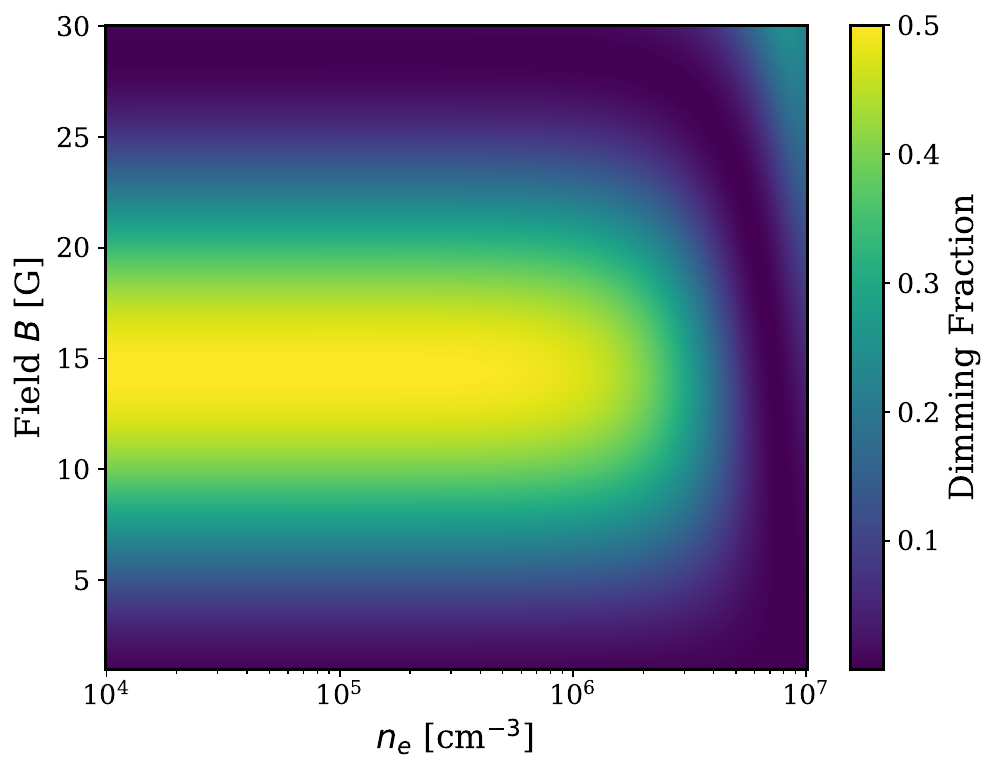}
  \caption{$m_a = 10^{-8}$ eV, $a = 0.99$}
  \label{fig:dimming3_f}
\end{subfigure}

\begin{subfigure}[b]{0.32\textwidth}
  \centering
  \includegraphics[width=\textwidth]{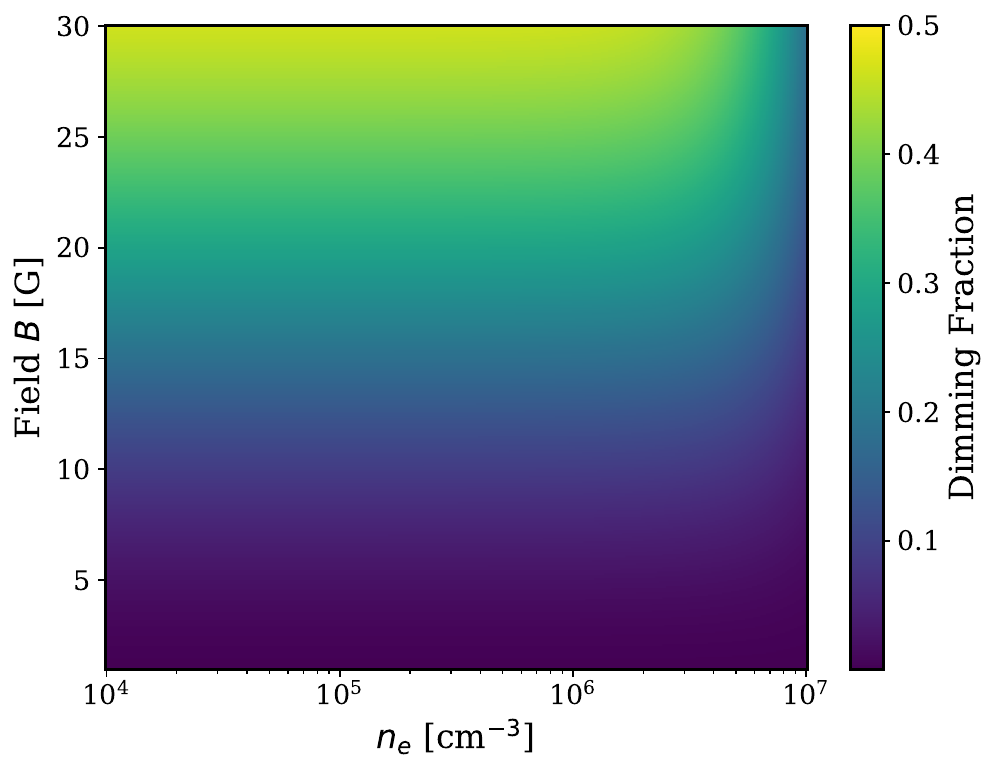}
  \caption{$m_a = 10^{-10}$ eV, $a = 0.3$}
  \label{fig:dimming3_g}
\end{subfigure}
\hfill
\begin{subfigure}[b]{0.32\textwidth}
  \centering
  \includegraphics[width=\textwidth]{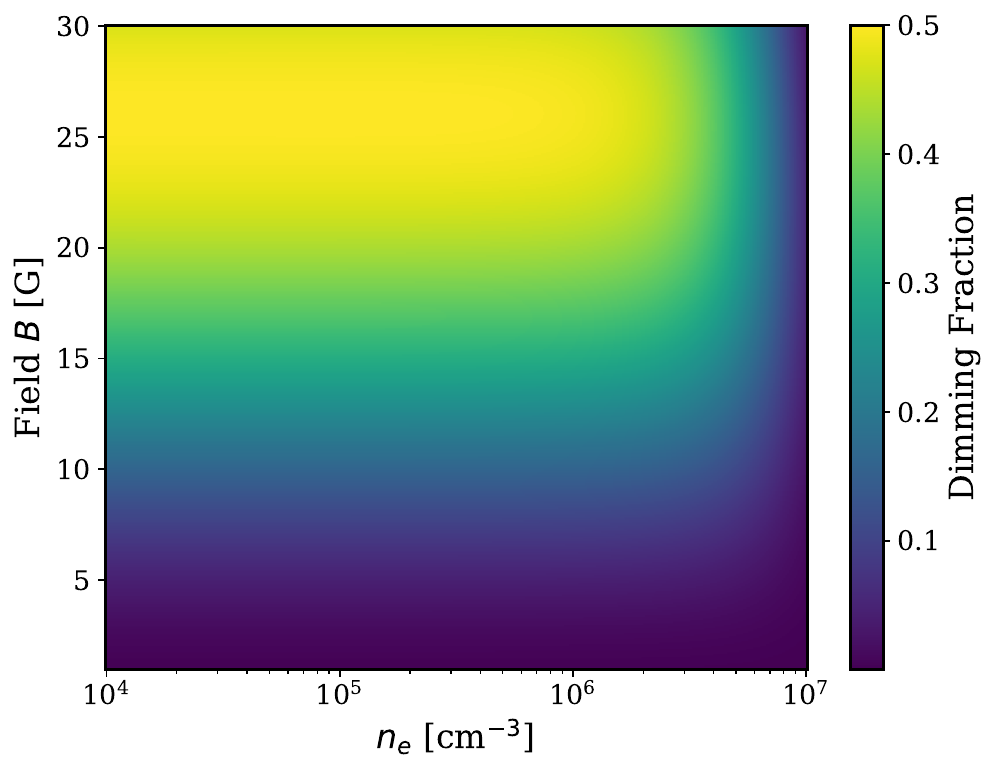}
  \caption{$m_a = 10^{-10}$ eV, $a = 0.6$}
  \label{fig:dimming3_h}
\end{subfigure}
\hfill
\begin{subfigure}[b]{0.32\textwidth}
  \centering
  \includegraphics[width=\textwidth]{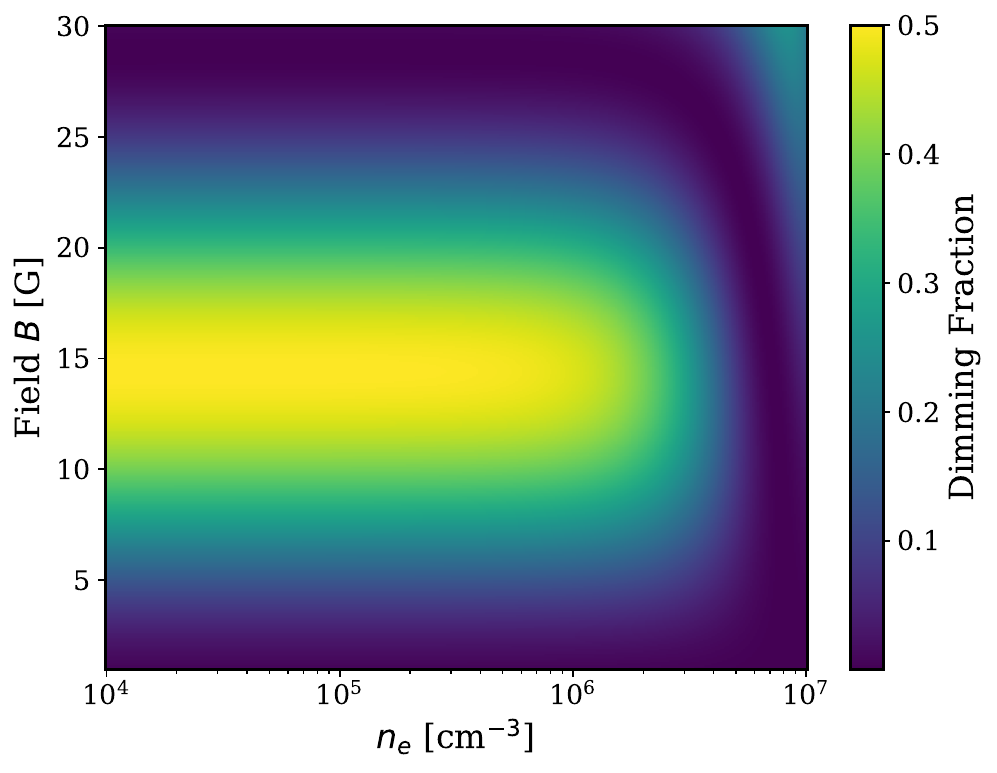}
  \caption{$m_a = 10^{-10}$ eV, $a = 0.99$}
  \label{fig:dimming3_i}
\end{subfigure}

\caption{Variation of dimming fraction with magnetic field strength and electron number density corresponding to $g_{a\gamma} = 10^{-11}$ GeV$^{-1}$, and energy of the photon $\omega = 10^{6}$eV. In each horizontal panel we consider three different BH spins for a fixed axion mass while in each vertical panel we consider variation in dimming percentage with axion mass for a fixed BH spin.
}
\label{fig:dimming_g3}

\end{figure}

\begin{figure}[htbp]
\centering

\begin{subfigure}[b]{0.32\textwidth}
  \centering
  \includegraphics[width=\textwidth]{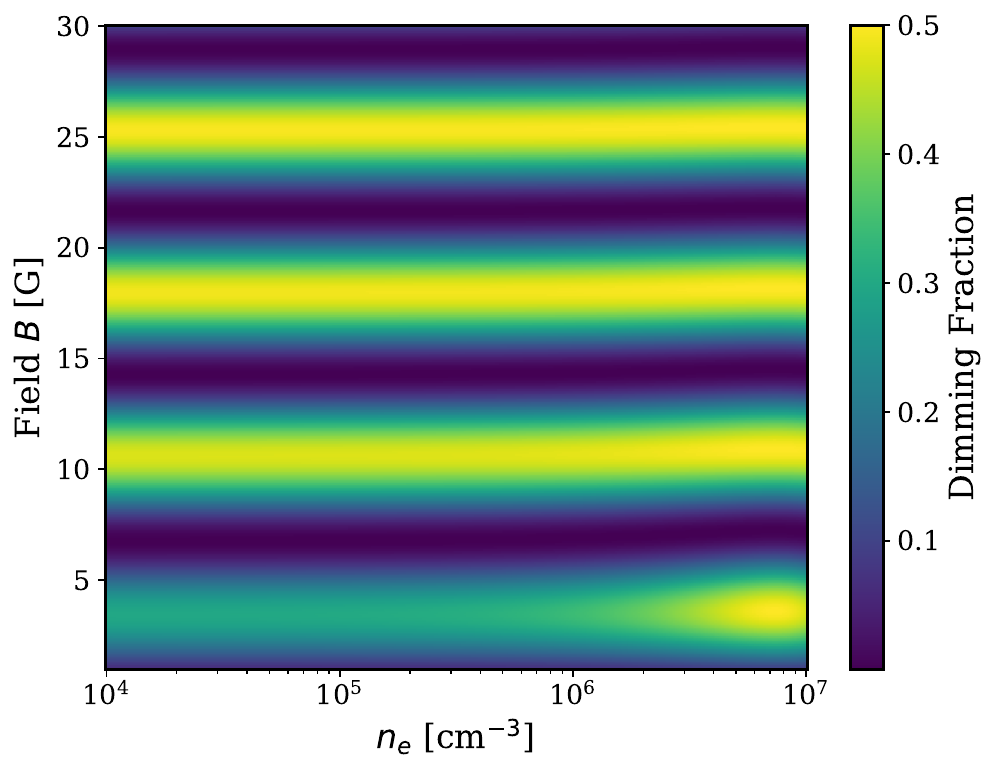}
  \caption{$m_a = 10^{-7}$ eV, $a = 0.3$}
  \label{fig:dimming4_a}
\end{subfigure}
\hfill
\begin{subfigure}[b]{0.32\textwidth}
  \centering
  \includegraphics[width=\textwidth]{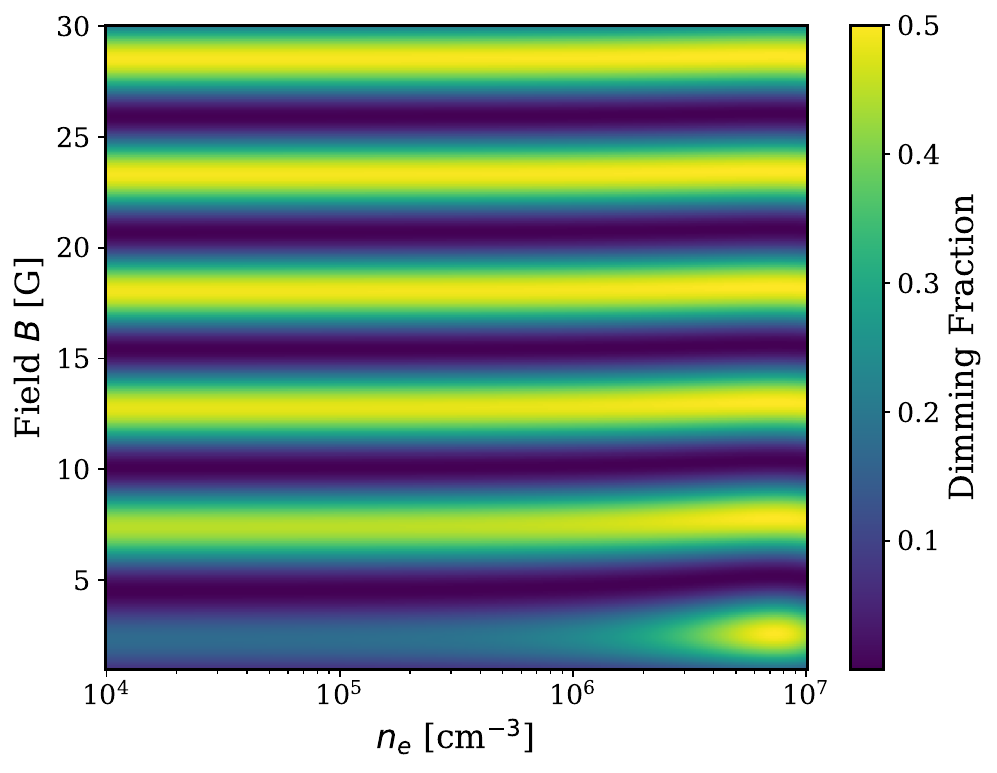}
  \caption{$m_a = 10^{-7}$ eV, $a = 0.6$}
  \label{fig:dimming4_b}
\end{subfigure}
\hfill
\begin{subfigure}[b]{0.32\textwidth}
  \centering
  \includegraphics[width=\textwidth]{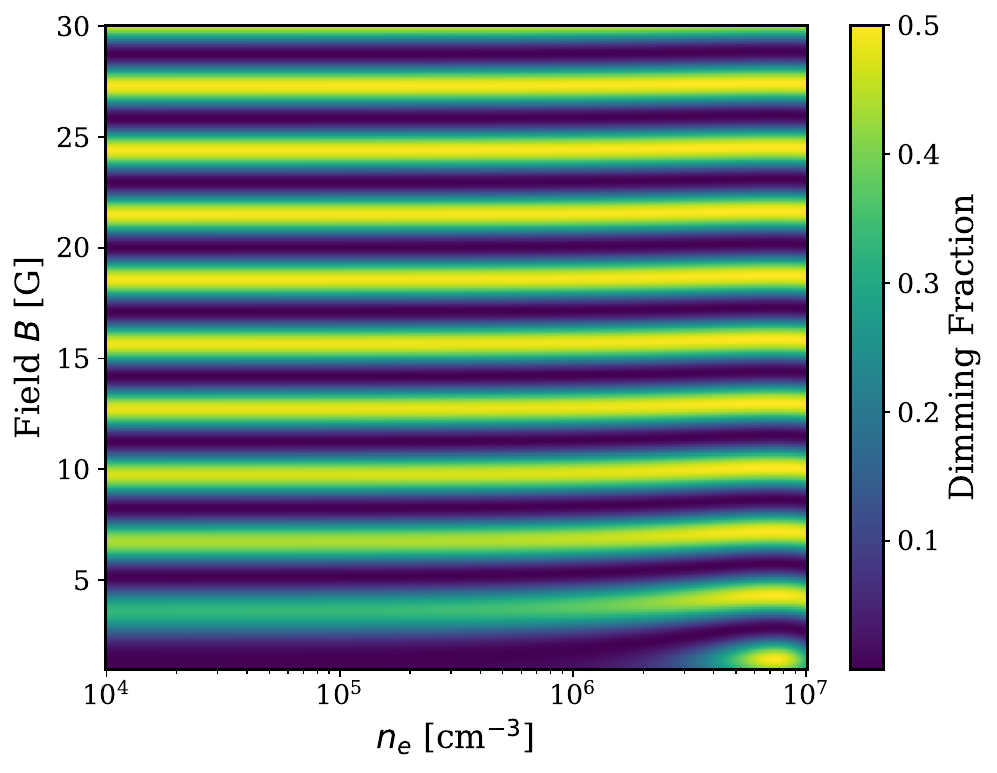}
  \caption{$m_a = 10^{-7}$ eV, $a = 0.99$}
  \label{fig:dimming4_c}
\end{subfigure}

\bigskip

\begin{subfigure}[b]{0.32\textwidth}
  \centering
  \includegraphics[width=\textwidth]{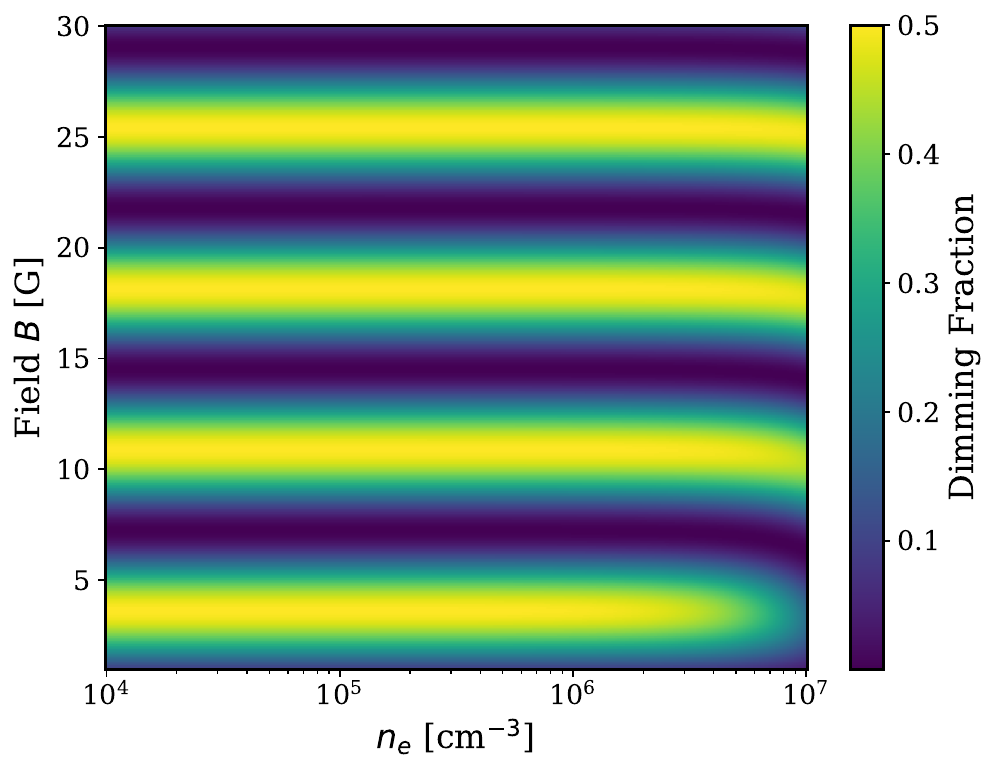}
  \caption{$m_a = 10^{-8}$ eV, $a = 0.3$}
  \label{fig:dimming4_d}
\end{subfigure}
\hfill
\begin{subfigure}[b]{0.32\textwidth}
  \centering
  \includegraphics[width=\textwidth]{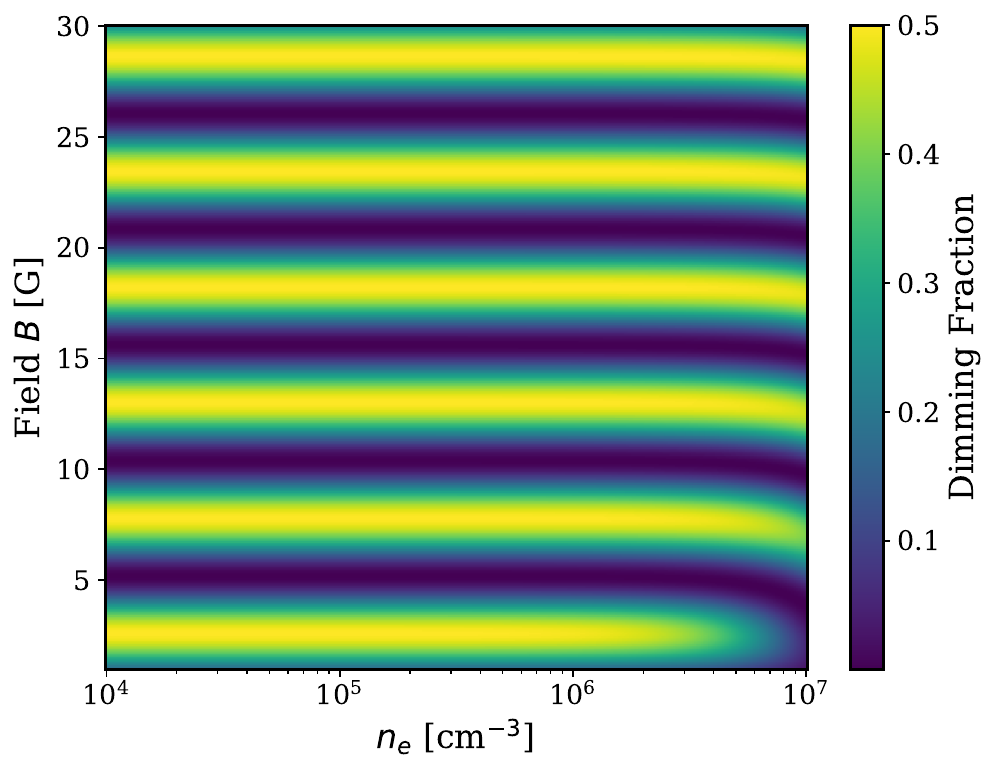}
  \caption{$m_a = 10^{-8}$ eV, $a = 0.6$}
  \label{fig:dimming4_e}
\end{subfigure}
\hfill
\begin{subfigure}[b]{0.32\textwidth}
  \centering
  \includegraphics[width=\textwidth]{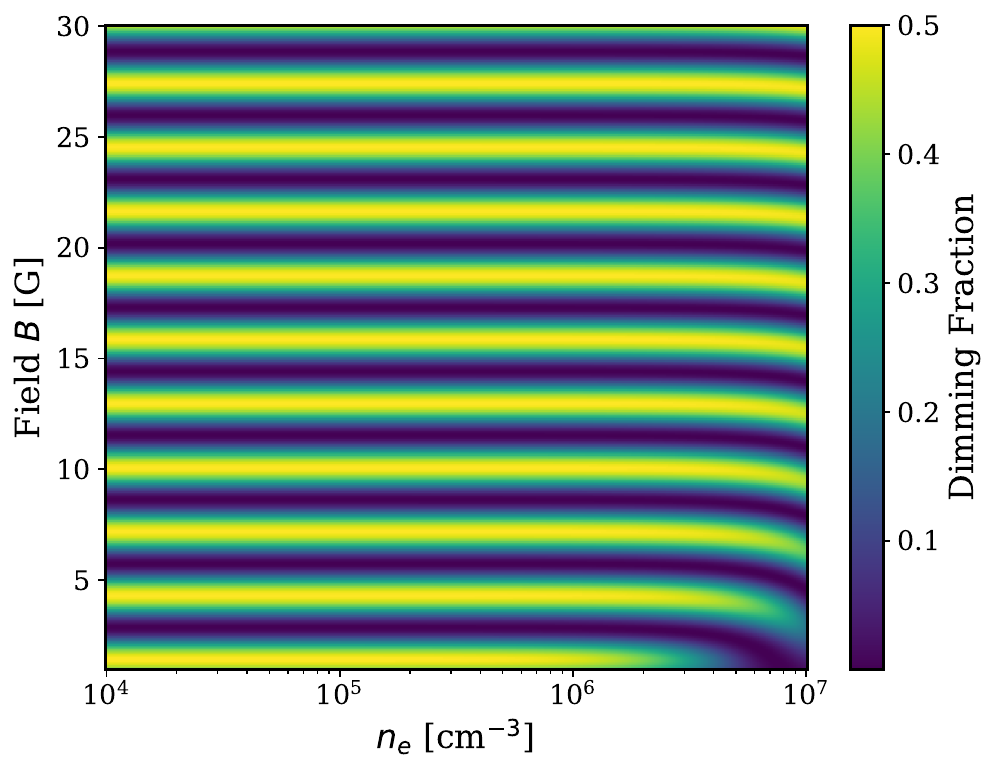}
  \caption{$m_a = 10^{-8}$ eV, $a = 0.99$}
  \label{fig:dimming4_f}
\end{subfigure}

\begin{subfigure}[b]{0.32\textwidth}
  \centering
  \includegraphics[width=\textwidth]{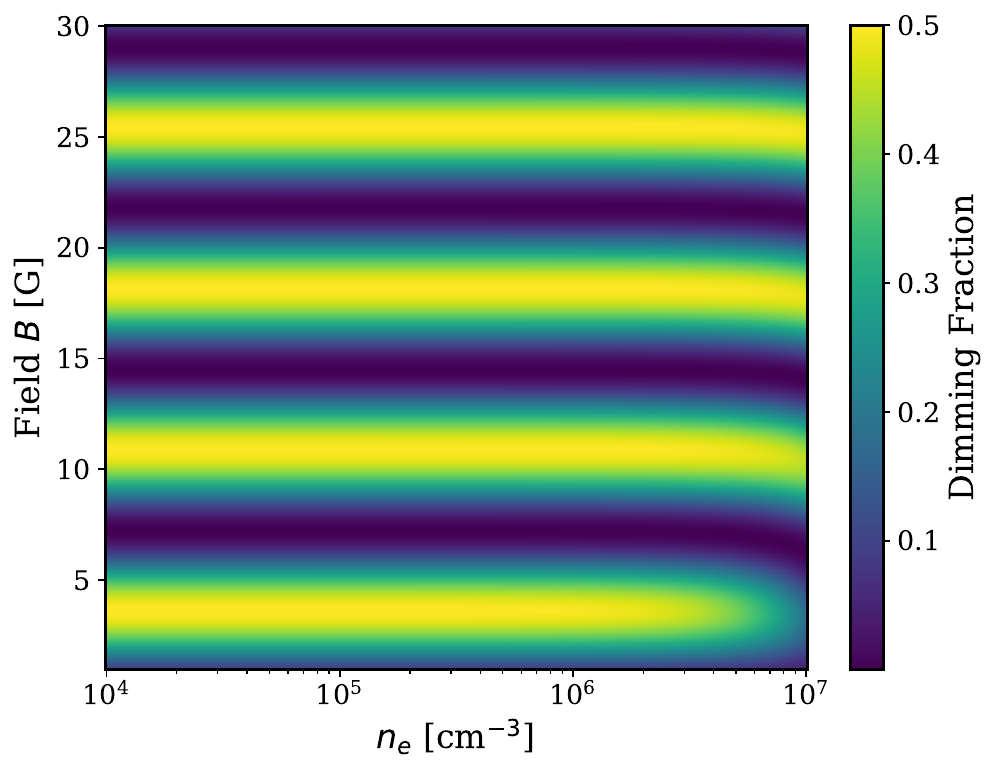}
  \caption{$m_a = 10^{-10}$ eV, $a = 0.3$}
  \label{fig:dimming4_g}
\end{subfigure}
\hfill
\begin{subfigure}[b]{0.32\textwidth}
  \centering
  \includegraphics[width=\textwidth]{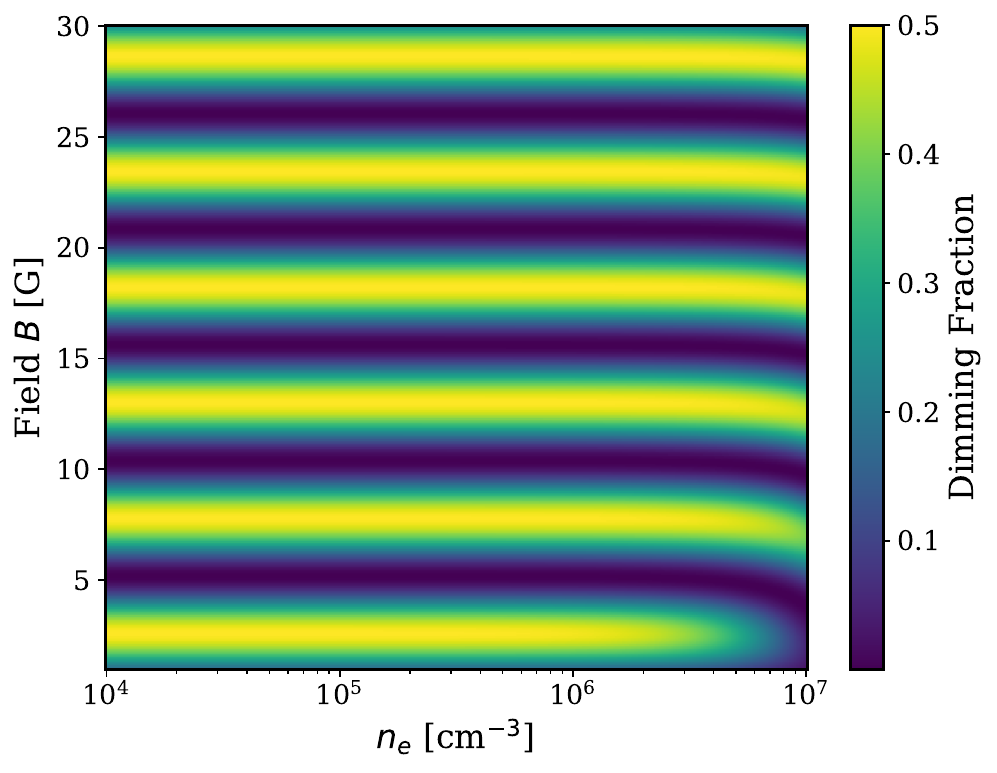}
  \caption{$m_a = 10^{-10}$ eV, $a = 0.6$}
  \label{fig:dimming4_h}
\end{subfigure}
\hfill
\begin{subfigure}[b]{0.32\textwidth}
  \centering
  \includegraphics[width=\textwidth]{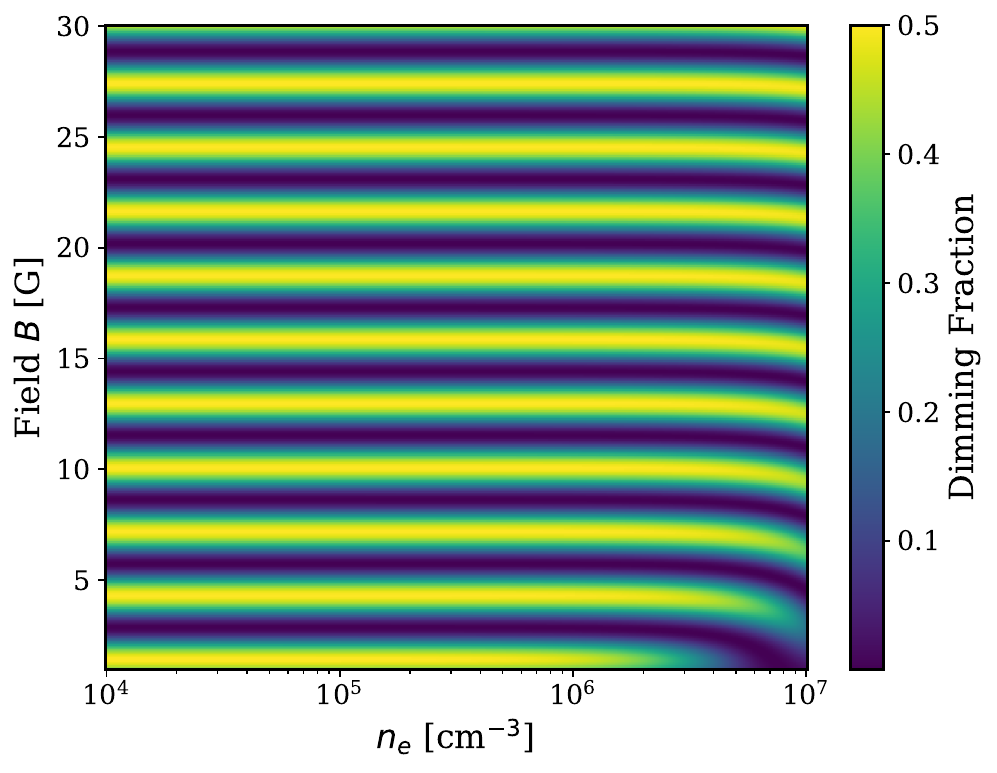}
  \caption{$m_a = 10^{-10}$ eV, $a = 0.99$}
  \label{fig:dimming4_i}
\end{subfigure}

\caption{Variation of dimming fraction with magnetic field strength and electron number density corresponding to $g_{a\gamma} = 10^{-10}$ GeV$^{-1}$, and energy of the photon $\omega = 10^{6}$eV. In each horizontal panel we consider three different BH spins for a fixed axion mass while in each vertical panel we consider variation in dimming percentage with axion mass for a fixed BH spin.}
\label{fig:dimming_g4}

\end{figure}

In \cref{fig:dimming_g3} and \cref{fig:dimming_g4} we show the variation of the percentage dimming for a photon of frequency $\omega\simeq 10^6$eV, due to variation in the magnetic field and the electron number density assuming photon-axion coupling $g_{a\gamma}\sim 10^{-11}\rm GeV^{-1}$ and $g_{a\gamma}\sim 10^{-10}\rm GeV^{-1}$ respectively. We choose the photon frequency to be $\omega\simeq 10^6$eV since we have noted in \cref{fig:dimming_g1}, \cref{fig:dimming_g2} and \cref{Fig7} that efficient conversion can be achieved in this frequency. 

From \cref{fig:dimming_g3} we note that for axions of mass $m_a=0.1\rm neV$ and $m_a=10\rm neV$, the conversion is most efficient at low to intermediate electron densities. This is because, at $\omega=10^6\rm eV$, when the magnetic field $B\sim$ few tens of Gauss, the axion mass is in the aforesaid regime and the electron densities are low, $\mathrm{\Delta_M}$ dominates over $\mathrm{\Delta_a}$, $\mathrm{\Delta_{pl}}$ and $\mathrm{\Delta_{vac}}$ (\cref{eq:delta_M}, \cref{eq:delta_a}, \cref{eq:delta_pl}, \cref{eq:delta_vac}) and consequently $\mathrm{\Delta_{osc}}\simeq 2\mathrm{\Delta_M}$ can be achieved, which increases the amplitude of the conversion probability (\cref{eq:conversion_prob}) to order unity. It is important to note that even though the amplitude of the conversion probability becomes $\sim $ unity, efficient conversion cannot take place if the argument of the sinusoidal term in the conversion probability (\cref{eq:conversion_prob}), $\mathrm{\Delta_{osc}}z \sim (2n\pi)$. Thus, efficient conversion also depends on the path length $z$ traversed by the photon in the photon region of the Kerr BH.
We have seen earlier that path length increases near a rapidly rotating Kerr BH. Thus, near a rapidly rotating Kerr BH, $\mathrm{\Delta_{osc}}z\sim (2n+1)\pi$ can be achieved at lower magnetic fields (\cref{fig:dimming3_c}, \cref{fig:dimming3_f} and \cref{fig:dimming3_i}), since this leads to a smaller $\mathrm{\Delta_{osc}}$ which is compensated by a larger path length. Thus, in \cref{fig:dimming3_f} and \cref{fig:dimming3_i} we see efficient photon-axion conversion in presence of intermediate magnetic fields near maximally rotating Kerr BHs to axions of mass $m_a\lesssim 10^{-8}\rm eV$.  Conversely, for reasons discussed above efficient conversion to axions of mass $m_a\lesssim 10^{-8}\rm eV$ near intermediate to slowly spinning BHs requires higher magnetic fields (\cref{fig:dimming3_d}, \cref{fig:dimming3_e}, \cref{fig:dimming3_g} and \cref{fig:dimming3_h}). Unlike the case for low mass axions ($m_a\lesssim 10^{-8}\rm eV$) when conversion to axions of mass $m_a\simeq 100$neV is considered (\cref{fig:dimming3_a}-\cref{fig:dimming3_c}), the dimming fraction is the highest when the electron density assumes the highest possible value ($n_e\simeq 10^7\rm cm^{-3}$) observed for M87*, irrespective of the BH spins. This is because for photons of frequency $\omega\simeq 10^6$eV and photon-axion coupling $g_{a\gamma}\sim10^{-11}\rm GeV^{-1}$, conversion to axions of mass $m_a\simeq 100$neV can efficiently occur in presence of magnetic field $B\sim $ few tens of Gauss if the electron number density is high. At low electron densities $\mathrm{\Delta_{osc}}\simeq \mathrm{\Delta_a}$ and hence the amplitude of the conversion probability (\cref{eq:conversion_prob}, \cref{eq:oscillation}) is sufficiently lesser than unity.  At high electron densities on the other hand, $\mathrm{\Delta_{pl}}\sim \mathrm{\Delta_a}\gg \mathrm{\Delta_{vac}}$ and as a consequence $\mathrm{\Delta_{osc}}\approx 2\Delta_{M}$ begins to hold. This leads to an enhancement in the amplitude of the conversion probability leading to efficient conversion. Further, as the BH spin is increased, efficient conversion can be achieved at lower magnetic fields (\cref{fig:dimming3_c}), since the effective path length $z$ traversed by the photon increases with an increase in the BH spin (\cref{tab:dimming_results_17_sorted}, \cref{tab:dimming_results_2} and \cref{tab:dimming_results_3}) and as a consequence $\mathrm{\Delta_{osc}}z\sim (2n+1)\pi$ (\cref{eq:conversion_prob}) can be achieved at lower magnetic fields. In this regime $\mathrm{\Delta_{osc}}\sim2\mathrm{\Delta_M}$ holds but $\mathrm{\Delta_M}$ decreases with a decrease in the magnetic field which in turn is compensated by a larger $z$ with an increase in the BH spin.

\cref{fig:dimming_g4} is same as \cref{fig:dimming_g3} but now the photon-axion coupling is $g_{a\gamma}\simeq 10^{-10}\rm GeV^{-1}$, i.e, an order of magnitude larger than that in \cref{fig:dimming_g3}. With an enhancement in $g_{a\gamma}$ by an order, at $\omega\sim 10^6~\rm eV$, $\mathrm{\Delta_{osc}}\simeq 2\mathrm{\Delta_M}$ generally holds irrespective of the axion mass and the electron number density, thereby maximizing the amplitude of the conversion probability (whose highest magnitude can be unity, see \cref{eq:conversion_prob}). Whenever the sinusoidal term in \cref{eq:conversion_prob} is near unity there is maximal conversion and this happens whenever $\mathrm{\Delta_{osc}}z\simeq (2n+1)\pi$. Interestingly, the rate of change of dimming fraction with magnetic field increases with an increase in the BH spin and this holds good irrespective of the axion mass. This may be attributed to an enhanced path length as the BH spin is increased. Moreover, the magnitudes of $g_{a\gamma}$, $B$ and the effective path length $z$ are such that $\mathrm{\Delta_{osc}}z\simeq (2n+1)\pi$ can be achieved for different choices of $n$. As the path length increases near a rapidly rotating Kerr BH, $\mathrm{\Delta_{osc}}z$ becomes more sensitive to changes in the magnetic field thereby leading to a more frequent enhancement and reduction in the dimming fraction. Note that, for a given magnetic field the conversion probability does not vary much with the electron number density (for any BH spin or axion mass). This is again because of the enhanced  photon-axion coupling, such that $\mathrm{\Delta_{osc}}\simeq 2\mathrm{\Delta_M}$ mostly holds, and hence the conversion probability $P_{a\gamma}$ (\cref{eq:conversion_prob} is nearly insensitive to changes in the electron number density in the regime relevant for M87* \cite{EventHorizonTelescope:2021srq}. 
 
 Note that the choice of the axion-photon coupling of $\mathcal{O}(10^{-11})\,\mathrm{GeV}^{-1}$ for the mass range $\mathcal{O}(1-100)$ neV is motivated by existing upper bounds from relevant axion search experiments \cite{Cameron:1993bhr,Minowa:1998sj,CAST:2004gzq,GammeVT-969:2007pci,Povey:2010hs,Ehret:2010zz,OSQAR:2013jqp, OSQAR:2015qdv,CAST:2024eil,ALPSII:2025eri}. While $g_{a\gamma} \simeq 10^{-10}$ is comparatively large for the axion mass range under consideration, it is chosen here to illustrate the correlation of efficient dimming with large $\Delta_{\rm M}$.\footnote{Astrophysical constraints on $g_{a\gamma}$, typically obtained under additional model-dependent assumptions, can be rather stringent. Nevertheless, they are generally subject to significant uncertainties \cite{ParticleDataGroup:2024cfk}.
Recent constraints obtained using white dwarfs may be worth noting in this context \cite{Dessert:2022yqq,ParticleDataGroup:2024cfk,Benabou:2025jcv}. } 
Additionally, the choice of the mass range 
$m_a \simeq \mathcal{O}(1\text{--}100)\,{\rm neV}$ 
ensures that the parameter region satisfying 
$\Delta_a \simeq \Delta_{\rm pl}$ can be realized. 
This, in turn, leads to 
$\Delta_{\rm osc} \simeq 2\Delta_{\rm M}$, 
as discussed earlier, thereby resulting in substantial dimming. Further, given that 
$\omega_{\rm pl} \simeq \mathcal{O}(1\text{--}10)\,\mathrm{neV}$, 
and considering high-energy photons with 
$\omega \gg \omega_{\rm pl}$, efficient photon to axion conversion can occur for relativistic axions in the mass range 
$m_a \simeq \mathcal{O}(1\text{--}100)\,\mathrm{neV}$.

Thus, if the photon region around M87* can be resolved at high frequencies then one may hope to detect signatures of axion through the photon ring dimming.

\section{Estimation of photon flux in the vicinity of the photon region of Kerr black holes}
\label{S5}
Having estimated the percentage dimming of photons, we shall now derive the spectrum of photons and consequently the spectrum of axions obtained due to photon-axion conversion. In order to arrive at the 
the precise spectral shape one requires knowledge of the photon flux reaching the photon region at each frequency. This requires one to evaluate the integrand in \cref{eq:conversion_rate}. This, in turn, depends on the physical origin of these photons. In the case of astrophysical black holes, the radiation is expected to arise from the gas in the surrounding environment. In practice, however, the spatial distribution and dynamics of this gas can be quite complicated, making an exact evaluation of the resulting spectra difficult. To simplify the analysis, we adopt an idealized model in which the photon emitting gas is assumed to be distributed spherically around the black hole. With this assumption, the spectral properties can be estimated analytically without relying on detailed numerical simulations. Observations indicating low radiative efficiency in supermassive black holes such as M87* and Sgr A* further suggest that the emission region is unlikely to be a thin disk; instead, it is more plausibly associated with a geometrically thick, hot accretion flow (see, e.g., \cite{Yuan:2014gma}). Within this context, the spherical gas model employed here should provide a reasonable order-of-magnitude estimate of the expected spectra.

Assuming the spherical gas model, we aim to calculate the number of photons approaching the photon region of the Kerr BH. We consider a light ray denoted by the ray vector $k^\mu$, approaching the photon region. The ray is emitted from the emitting region with coordinates $p_e\equiv (r_e,\theta_e)$. Let $\chi_e$ be the angle made by the ray vector from the radial direction denoted by the unit vector $\hat{e}_{(r)}$ (line joining the emitting point to the centre of the BH) (\cref{emission_point}).  At the emitting point $p_e$ we consider an orthogonal coordinate system defined by the tetrads $(\hat{e}_{(r)},\hat{e}_{(\theta)},\hat{e}_{(\phi)})$. From 
\cref{emission_point} it is clear that $\xi_e$ denotes the azimuthal angle of the ray path such that:
\begin{figure}
    \centering
    \includegraphics[width=0.8\linewidth]{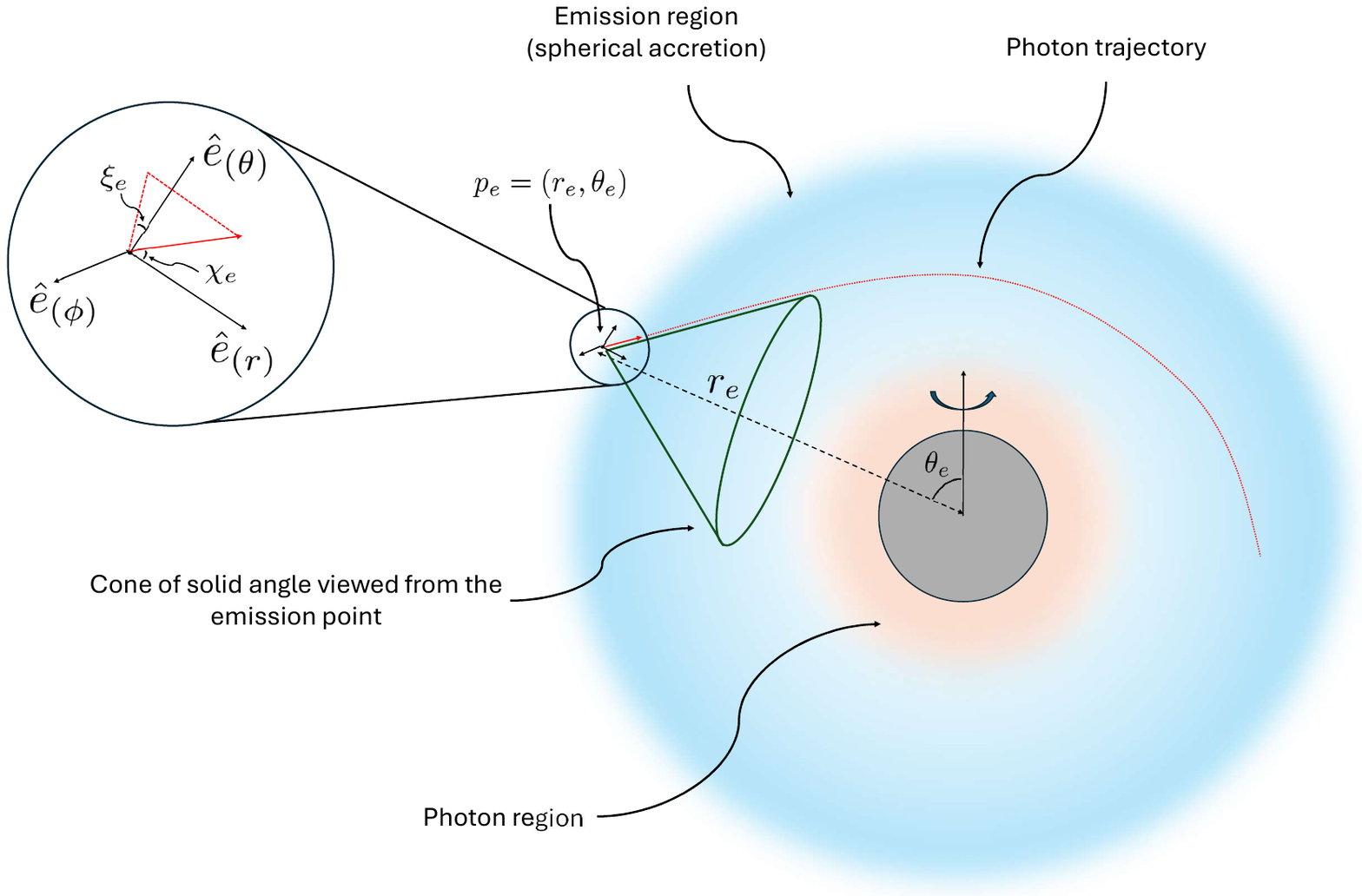}
    \caption{Schematic diagram showing the emission from a point represented by $p_e(r_e,\theta_e)$ in the spherical accretion emission region (denoted by blue). The local non-rotating inertial frame is given by $(\hat{e}_{(r)},\hat{e}_{(\theta)},\hat{e}_{(\phi)})$. The photon travels towards the photon region (denoted by orange) where it gets converted.}
    \label{emission_point}
\end{figure}
\begin{eqnarray}
    \tan \chi_e &=& \frac{\sqrt{ (k^{(\theta)})^2 + (k^{(\phi)})^2   }}{k^{(r)}}, \label{tan_chi} \\
    \tan \xi_e &=& \frac{k^{(\phi)}}{k^{(\theta)}} \label{tan_xi},
\end{eqnarray}
where $k^{(r)}$, $k^{(\theta)}$ and $k^{(\phi)}$ denote the components of the ray vector in the radial, polar and the azimuthal direction respectively. Now,
\begin{align}
    k^{(a)}=\hat{e}^{(a)}_{\mu}k^\mu,
    \end{align}
    where $k^\mu$ correspond to the components of the photon 4-momentum vector obtained from the geodesic equations (\cref{geodesic}, \cref{eq:Radial_geodesic}, \cref{eq:Theta_geodesic}, \cref{eq:Phi_geodesic}) and $\hat{e}^{(a)}_\mu$ correspond to the Bardeen tetrads associated with locally non-rotating observers given as \cite{Bardeen:1972fi},

    \begin{align}
    \hat{e}^{(t)}&= \Bigg\lbrace\frac{4M^2a^2r^2\sin^2\theta}{\Sigma\sqrt{A\Sigma\Delta}} +\Big(1-\frac{2Mr}{\Sigma}\Big)\sqrt{\frac{A}{\Sigma\Delta}}\Bigg\rbrace \partial_t + \nonumber \\ &\Bigg\lbrace\frac{2Mar\sin^2\theta}{\Sigma}\sqrt{\frac{A}{\Sigma\Delta}}-\frac{2Mar}{\sqrt{A\Sigma\Delta}}\left(r^2 + a^2 +\frac{2Mra^2\sin^2\theta}{\Sigma}\right)\sin^2\theta \Bigg \rbrace \partial_\phi\nonumber, \\
        \hat{e}^{(r)}&=\sqrt{\frac{\Sigma}{\Delta}}\partial_r \nonumber, \\
        \hat{e}^{(\theta)}&=\sqrt{\Sigma}~\partial_\theta \nonumber, \\    
        \hat{e}^{(\phi)}&=-\frac{2Mar\sin^2\theta}{\Sigma}\sqrt{\frac{\Sigma}{A}}\frac{1}{\sin\theta} ~\partial_t +   \sqrt{\frac{\Sigma}{A}}\frac{1}{\sin\theta}  \left(r^2 + a^2 +\frac{2Mra^2\sin^2\theta}{\Sigma}\right)\sin^2\theta ~\partial_\phi. \end{align}
where $A=(r^2+a^2)^2-\Delta a^2\sin^2\theta$.

    Photons are emitted isotropically from each point in the emitting region with a certain emission rate. We now provide an estimate of the number of photons approaching the photon region of the Kerr BH. We consider an infinitesimal volume $dV_e$ centered around the point $p_e$ at a radial distance $r_e$ (\cref{emission_point}). We denote the number of photons per unit proper time $\tau_e$, emitted from $dV_e$ in the frequency range $d\omega_e$, towards an infinitesimal solid angle $d\Omega_e$ as,
    \begin{equation}
    d^6\left( \frac{d N}{d \tau_e} \right) = J^{(N)}_e (\omega_e,r_e) d\Omega_e dV_e d\omega_e,
    \label{J_e}
\end{equation}
where $J_e^{(N)}$ denotes the number of photons emitted per unit time, per unit frequency, per unit volume and per unit solid angle. 
Note that in \cref{J_e}, the photon frequency $\omega_e$, the proper time $\tau_e$, the solid angle $\Omega_e$ and the volume $V_e$ are measured with respect to a local inertial frame located at $p_e$. Since the emission is isotropic, $J^{(N)}_e$ does not depend on $\chi_e$ and $\xi_e$. The infinitesimal solid angle $d\Omega_e=\sin\chi_e d\chi_e d\xi_e$. Using \cref{tan_chi} and the geodesic equations we note that,
\begin{equation}
    \sin \chi_e =  \frac{1}{\sqrt{1+\frac{\mathcal{R}(r_e,\lambda,\eta)  A (r_e,\theta_e)\sin^2 \theta_e}{\Theta(\theta_e,\lambda,\eta)  \Delta (r_e) A (r_e,\theta_e)\sin^2 \theta_e + \Sigma^2(r_e,\theta_e) \lambda^2 \Delta(r_e)}}}\Bigg|_{p_e},
\end{equation}
while 
\begin{equation}
    d\chi_e \, d\xi_e
=  |J| \, d\lambda \, d\eta,
\end{equation}
 where $J$ is the Jacobian matrix given by 
 \[J=\begin{bmatrix}
     \frac{\partial \chi_e}{\partial \lambda} & \frac{\partial \chi_e}{ \partial \eta}\\
     \frac{\partial \xi_e }{\partial \lambda} & \frac{\partial \xi_e}{\partial \eta}
 \end{bmatrix}\]
 which can be obtained by writing \cref{tan_chi} and \cref{tan_xi} in terms of the impact parameters $\lambda$ and $\eta$,
 \begin{align}
     \tan\chi_e=\frac{\sqrt{A(r_e,\theta_e)\Theta(\theta_e,\lambda,\eta)+\lambda^2 \rm {cosec^2}\theta_e\Sigma^2(r_e,\theta_e)}}{\sqrt{\frac{A(r_e,\theta_e)\mathcal{R}(r_e,\lambda,\eta)}{\Delta(r_e)}}},~~~~~~~~~~~\tan\xi_e=\frac{\Sigma(r_e,\theta_e)\lambda\rm{cosec}\theta_e}{\sqrt{A(r_e,\theta_e)\Theta(\theta_e,\lambda,\eta)}}.
     \label{tanchi_tanxi}
 \end{align}
 Thus, \cref{J_e} can be written as,
 \begin{align}
   d^6\left( \frac{d N}{d \tau_e} \right) = \frac{1}{2}\times  J^{(N)}_e (\omega_e,r_e) \sin\chi_e |J|d\lambda d\eta dV_e d\omega_e, 
    \label{J_e-2}  
 \end{align}
 where the factor of $\frac{1}{2}$ takes into account the fact that only photons with $0\leq \chi_e\leq \pi/2$ can approach the photon region.

 Further, the proper time interval $d\tau_e$ in the emitting region at $p_e$ is related to the coordinate time interval $dt$ by $d\tau_e=\sqrt{-g_{\rm tt}(r_e,\theta_e)}dt$.
 The 3-d volume element $dV_e$ appearing in \cref{J_e} in a local inertial frame at $p_e$ is given by,
 \begin{equation}
    dV_e = \sqrt{\gamma_{\rm rr} \gamma_{\theta\theta} \gamma_{\phi\phi}} \, dr_e \, d\theta_e \, d\phi_e,
\label{volele}
\end{equation}
where $\gamma_{rr}, \gamma_{\theta\theta}$ and $\gamma_{\phi\phi}$ denote the components of the induced metric as discussed in \cref{induced-metric}. After performing the integration over $\phi_e$, \cref{J_e-2} becomes,
\begin{align}
   d^5\left( \frac{d N}{d t} \right) = \pi J^{(N)}_e (\omega_e,r_e) \sin\chi_e |J|  \sqrt{\gamma_{\rm rr} \gamma_{\theta\theta} \gamma_{\phi\phi}} \sqrt{-g_{\rm tt}(r_e,\theta_e)}d\lambda d\eta dr_e \, d\theta_e  d\omega_e. 
    \label{J-e-3}  
 \end{align}

 \subsection{Photons by thermal bremsstrahlung}
 In order to estimate the total number of photons approaching the photon region we need to know $J^{(N)}_e$.
This requires an understanding of the mechanism of generation of photons. We have seen in the previous section that efficient conversion can happen mainly for high energy photons.
Since we are interested in photons that can efficiently get converted to axions, we will consider mainly X-rays and gamma-rays. For supermassive BHs like M87*, such high energy photons can be generated mainly by thermal bremsstrahlung of plasma \cite{Quataert:2002xn}. The energy radiated by thermal bremsstrahlung per unit volume, per unit time and per unit frequency is given by \cite{Rybicki:2004hfl},
\begin{align}
    \frac{d^3E}{d\omega_e d\tau_e dV_e}=\frac{16 (\alpha \hbar )^3}{3 m_e } \left(\frac{2 \pi}{3 m_e k_B}\right)^{1/2}  T_e^{-1/2} n_e^2 e^{-\hbar\omega_e/k_BT_e} \bar{g}_{\rm ff}.
\label{Rybicki_equation}
   \end{align}
 In \cref{Rybicki_equation} $n_e$ and $T_e$ represent the electron number density and temperature respectively while $\bar{g}_{\rm ff}$ denotes the velocity averaged Gaunt factor. The emission rate is expressed as the classical result multiplied by the free–free Gaunt factor ($\bar{g}_{\rm ff}$), accounting for quantum-mechanical corrections in the Born approximation. The Gaunt factor depends on $T_e$ and $\omega_e$, but for the purpose of an order of magnitude estimation $\bar{g}_{\rm ff}\sim 1$ is considered \cite{Nomura:2022zyy}.
Moreover, in the above equation we assumed that the ion density approximately equals to the electron density. Also, $\omega_e$, $dV_e$ and $\tau_e$ denote the photon frequency, the infinitesimal volume of the emitting region and the time in a local inertial frame at the emission point. Since the radiation is isotropic, we note from \cref{Rybicki_equation} that the number of photons radiated by thermal bremsstrahlung per unit volume, per unit time, per unit frequency and per unit solid angle is given by 
\begin{align}
 \frac{d^3 N}{d\tau_e d\omega_e dV_e} = \frac{1}{4\pi\hbar\omega_e} \frac{d^3E}{d\omega_e d\tau_e dV_e}.
 \label{Rybicki-2}
\end{align}
From \cref{Rybicki_equation} and \cref{Rybicki-2} one can obtain the number of photons per unit proper time $\tau_e$, emitted from $dV_e$ in the frequency range $d\omega_e$, towards an infinitesimal solid angle $d\Omega_e$ as,
\begin{align}
    d^6 \Bigg(\frac{dN}{d\tau_e}\Bigg)=\frac{1}{\hbar\omega_e}\frac{4 (\alpha \hbar )^3}{3 \pi m_e } \left(\frac{2 \pi}{3 m_e k_B}\right)^{1/2}  T_e^{-1/2} n_e^2 e^{-\hbar\omega_e/k_BT_e} \bar{g}_{\rm ff} d\Omega_e dV_e d\omega_e.
\label{Rybicki-3}
\end{align}
When \cref{Rybicki-3} is compared with \cref{J_e} we obtain,
\begin{align}
    J^{(N)}_e(\omega_e, r_e)=\frac{1}{\omega_e}\frac{4 \alpha^3 \hbar ^2}{3\pi m_e } \left(\frac{2 \pi}{3 m_e k_B}\right)^{1/2}  T_e^{-1/2} n_e^2 e^{-\hbar\omega_e/k_BT_e} \bar{g}_{\rm ff}.
    \label{J_e-3}
\end{align}
In terms of photon frequency measured by an observer at infinity, \cref{J_e-3} can be written as,
\iffalse
\begin{align}
    J_e(\omega_c, r_e)=\frac{\sqrt{-g_{\rm tt}(r_e,\theta_e)}}{\omega_c\sqrt{-g_{\rm tt}(\tilde{r},\theta_e)}}\frac{4 \alpha^3 \hbar ^2}{3\pi m_e } \left(\frac{2 \pi}{3 m_e k_B}\right)^{1/2}  T_e^{-1/2} n_e^2 ~{\rm exp}\Bigg[-\frac{\hbar\omega_c}{k_B T_e}\frac{\sqrt{-g_{\rm tt}(\tilde{r},\theta_e)}}{\sqrt{-g_{\rm tt}(r_e,\theta_e)}}\Bigg]\bar{g}_{\rm ff},
    \label{J_e-4}
\end{align}
\fi
\begin{align}
    J^{(N)}_e(\omega_0, r_e,\theta_e)=\sqrt{-g_{\rm tt}(r_e,\theta_e)}\frac{4 \alpha^3 \hbar ^2}{3\pi m_e } \left(\frac{2 \pi}{3 m_e k_B}\right)^{1/2}  \omega_0^{-1}T_e^{-1/2} n_e^2 ~{\rm exp}\Bigg[-\frac{\hbar\omega_0}{k_B T_e}\frac{1}{\sqrt{-g_{\rm tt}(r_e,\theta_e)}}\Bigg]\bar{g}_{\rm ff} ,
    \label{J_e-4}
\end{align}
\iffalse
as the frequency of the photon in the emission region $\omega_e$ and the photon region $\omega_c$ are respectively given by,
\begin{align}
    \omega_e=\frac{\omega_0}{\sqrt{-g_{\rm tt}(r_e,\theta_e)}}~~~~~~~\omega_c=\frac{\omega_0}{\sqrt{-g_{\rm tt}(\tilde{r}, \theta_e)}},
\end{align}
\fi
where $\omega_0$ is the frequency measured by an observer at infinity.
When the plasma is heated to viral temperature, the electron temperature $T_e$ is given by,
\begin{align}
    T_e(r_e)=T_{e,c}\Bigg(\frac{r_e}{\tilde{r}}\Bigg)^{-1},
    \label{T_e}
\end{align}
where $T_{e,c}\approx10^{11}$K denote the electron temperature in the photon region at $\tilde{r}$ \cite{Nomura:2022zyy}. Moreover, since we assumed a spherical gas model for the plasma, the electron number density falls as \cite{Nomura:2022zyy,EventHorizonTelescope:2021srq},
\begin{align}
    n_e(r_e)=n_{e,c}\Bigg(\frac{r_e}{\tilde{r}}\Bigg)^{-3/2},
    \label{n_e}
\end{align}
where $n_{e,c}$ denote the electron number density at $\tilde{r}$ in the photon region . For M87* this has been estimated to be $10^4-10^7\rm ~cm^{-3}$ \cite{EventHorizonTelescope:2021srq}.

\subsection{Spectra of photons and axions}
\label{S5b}
Using \cref{J_e-4}, \cref{T_e} and \cref{n_e} in \cref{J-e-3} we can obtain the number of photons approaching the photon region per unit coordinate time $t$ and per unit frequency $\omega$ (frequency of the photon in the photon region) by evaluating the following integral:

\begin{align}
   \frac{d^2 N}{dt \, d\omega} = \int_{r_{\rm in}}^{r_{out}} \int_{\theta_-}^{\theta_+} \int_{\tilde{\lambda}}^{\tilde{\lambda}(1+\delta \lambda)} \int_{\tilde{\eta}}^{\tilde{\eta}(1+\delta\eta)} dr_e\; d\theta_e\; d\lambda\; d\eta \; \Big[J^{(N)}_e (\omega_0,r_e,\theta_e)\times \nonumber \\ \sin \chi_e \; |J| \; \pi \sqrt{\gamma_{rr}\gamma_{\theta\theta}\gamma_{\phi\phi}}\sqrt{-g_{\rm tt}(\tilde{r},\theta_e)}  \Big],
   \label{d2Ndtdomega}
   \end{align}
where,
   \begin{align}
    \omega_e=\frac{\omega_0}{\sqrt{-g_{\rm tt}(r_e,\theta_e)}}~~~~~~~ {\rm and}~~~~~~~\omega=\frac{\omega_0}{\sqrt{-g_{\rm tt}(\tilde{r}, \theta_e)}}.
\end{align}
Also, $\theta_\pm$ are obtained from the turning points of the polar potential \cite{Gralla:2017ufe},
\begin{align}
   \theta_\pm=\cos^{-1}\Bigg[\mp\sqrt{\frac{1}{2}\Bigg(1-\frac{\tilde{\eta}+\tilde{\lambda}^2}{a^2}\Bigg)+\sqrt{\Bigg(\frac{1}{2}\Bigg(1-\frac{\tilde{\eta}+\tilde{\lambda}^2}{a^2}\Bigg)\Bigg)^2+\frac{\tilde{\eta}}{a^2}}}\Bigg] .
   \label{theta-plus}
\end{align}
Note however, that integrating $\theta_e$ from $0-\pi$ only results in minor changes in the result. Since we consider photons with impact parameters very close to the critical values we evaluate the integrand in \cref{d2Ndtdomega} at the critical impact parameters $\tilde{\lambda}, \tilde{\eta}$ and simply multiply the integrand with $d\lambda=\tilde{\lambda}\delta\lambda$ and $d\eta=\tilde{\eta}\delta\eta$.

   The luminosity of the photons approaching the photon region per unit coordinate time $t$ and per unit frequency $\omega$ is given by,
   \begin{align}
       L_{tot}(\omega_0)=\hbar \omega\frac{d^2 N}{dt \, d\omega}=d\lambda\; d\eta\; \hbar\omega_0\;\int_{r_{\rm in}}^{r_{\rm out}} \int_{\theta_-}^{\theta_+}  dr_e\; d\theta_e\;  \; \Big[J^{(N)}_e (\omega_0,r_e,\theta_e)\times \nonumber \\ \sin \chi_e \; |J| \; \pi \sqrt{\gamma_{\rm rr}\gamma_{\theta\theta}\gamma_{\phi\phi}}  \Big].
       \label{total-spectra}
   \end{align}
   The integral is performed from $r_{\rm in}\simeq 5 R_g$ to $r_{\rm out}\simeq 1000~R_g$ (the extent of the emission region) \cite{Nomura:2022zyy,Roy:2023rjk}. Note that \cref{total-spectra} is proportional to $M^3$ (where $M$ is the mass of the BH), since $J^{(N)}_e$ intrinsically has dimensions of $L^{-3}$ while $\sqrt{\gamma_{\rm rr}\gamma_{\theta\theta}\gamma_{\phi\phi}} dr_e$ has dimensions of $L^3$ and we express length in units of $M$. In this way both the LHS and the RHS of \cref{total-spectra} has the dimensions of energy. 
   Since the photon flux is proportional to $M^3$, photon-axion conversion is more efficient near supermassive BHs compared to stellar mass BHs. Thus,  supermassive BHs
like M87* are excellent candidates to look for photon ring dimming which can in turn be attributed to axion conversion. 
   
   The axion spectral luminosity $L_a(\omega_0)$ is obtained by multiplying $\hbar \omega$ to $\frac{d^2 N_{\gamma \to a}}{dt \, d\omega}$ in \cref{DF} and using \cref{total-spectra} we obtain
    \begin{align}
  L_a(\omega_0)=L_{\rm tot}(\omega_0)\frac{1}{2} \int_{\theta_-}^{\theta_+}P_{\gamma \to a}(\omega_0,\theta_e,z(\tilde{\lambda}, \tilde{\eta})) \,d\theta_e,
  \label{axion-spectra}
  \end{align}
  where,
   \begin{align}
  P_{\gamma \to a}(\omega_0,\theta_e,z(\tilde{\lambda},\tilde{\eta})) = \left( \frac{\mathrm{\Delta_M}}{\Delta_{\text{osc}}(\omega_0/\sqrt{-g_{\rm tt}(\tilde{r},\theta_e)})/2} \right)^2 \sin^2 \left( \frac{\Delta_{\text{osc}}(\omega_0/\sqrt{-g_{\rm tt}(\tilde{r},\theta_e)})}{2} z (\tilde{\lambda},\tilde{\eta})\right),
\label{eq:conversion_probability}
  \end{align}
The photon spectral luminosity is obtained by subtracting the axion spectral luminosity from the total luminosity.
We normalize the luminosity evaluated above with the infrared photon spectral luminosity $\omega_0\approx 10^{12}~$Hz which yields the relative luminosities,
   \begin{align}
       RL_{\rm tot}=\frac{L_{\rm tot}(\omega_0)}{L_{\rm tot}(10^{12}\rm ~Hz)},~~~~~RL_{a}=\frac{L_{a}(\omega_0)}{L_{\rm tot}(10^{12}\rm ~Hz)},~~~~~RL_{\rm ph}=\frac{L_{\rm tot}(\omega_0)-L_{a}(\omega_0)}{L_{\rm tot}(10^{12}\rm ~Hz)}.~~~~~
       \label{RL}
   \end{align}
In the following section, we will discuss the 
implications of the axion-photon conversion on the photon flux 
coming from the vicinity of the Kerr BH, leading to  
depletion of the flux and dimming of the image.
 
\section{Results \& Discussion}
\begin{figure}[!htbp]
\centering

% -------- Row 1 --------
\begin{subfigure}[b]{0.32\textwidth}
    \centering
    \includegraphics[width=\linewidth]{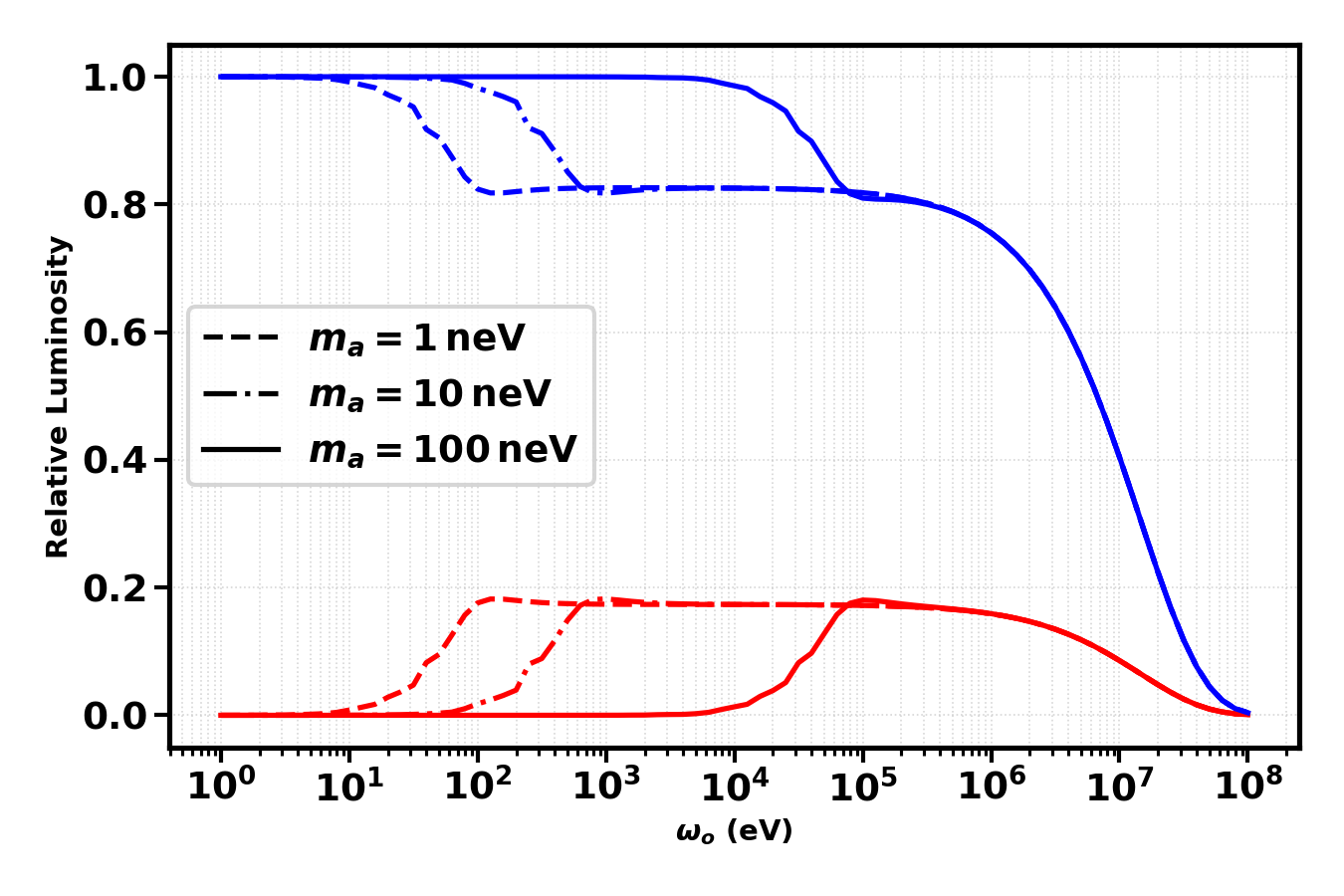}
    \caption{$a=0.3$ and $g_{a\gamma}=10^{-10}$}
    \label{11a}
\end{subfigure}\hfill
\begin{subfigure}[b]{0.32\textwidth}
    \centering
    \includegraphics[width=\linewidth]{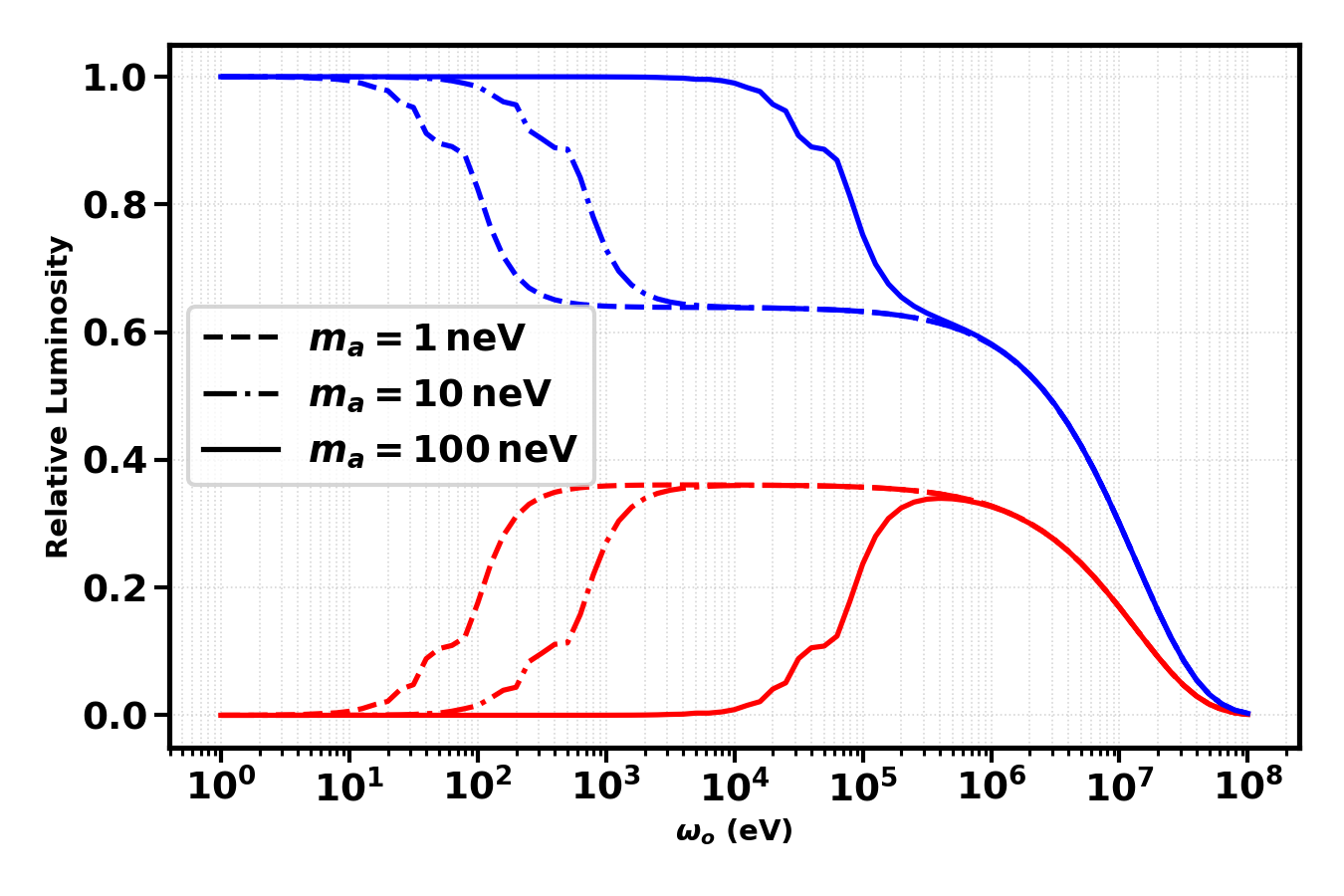}
    \caption{$a=0.6$ and $g_{a\gamma}=10^{-10}$}
    \label{11b}
\end{subfigure}\hfill
\begin{subfigure}[b]{0.32\textwidth}
    \centering
    \includegraphics[width=\linewidth]{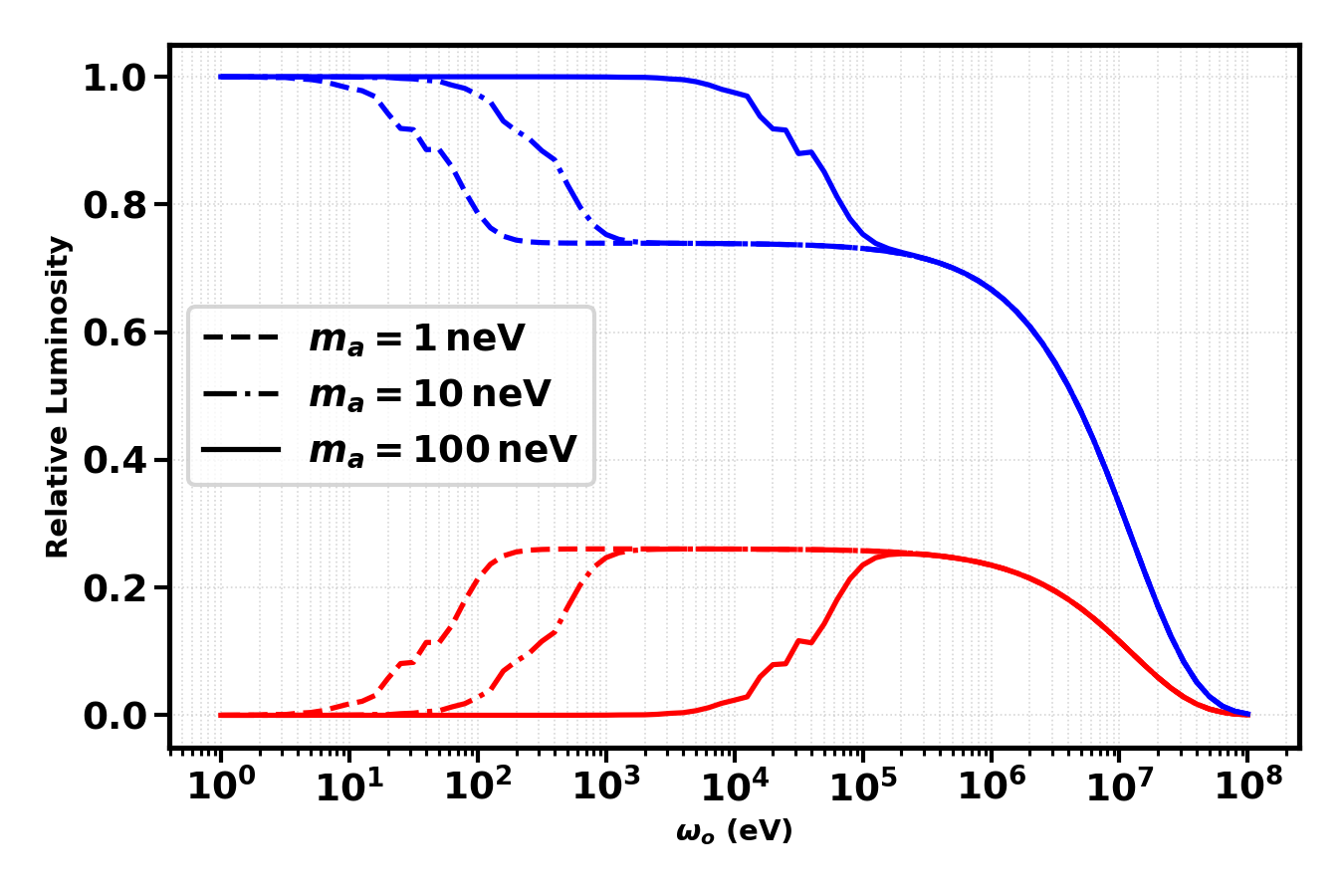}
    \caption{$a=0.99$ and $g_{a\gamma}=10^{-10}$}
    \label{11c}
\end{subfigure}

\vspace{0.3cm}

% -------- Row 2 --------
\begin{subfigure}[b]{0.32\textwidth}
    \centering
    \includegraphics[width=\linewidth]{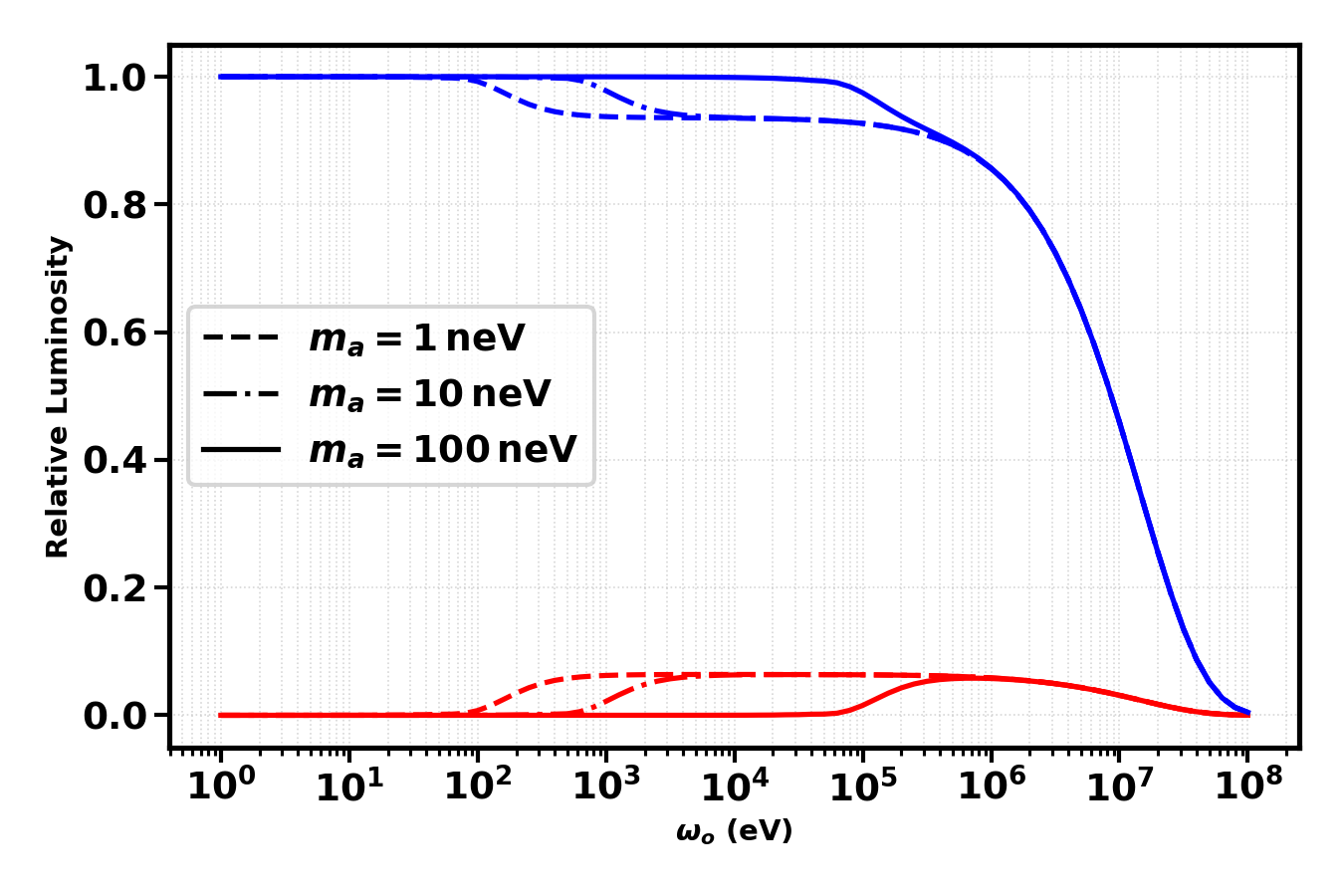}
    \caption{$a=0.3$ and $g_{a\gamma}=10^{-11}$}
    \label{11d}
\end{subfigure}\hfill
\begin{subfigure}[b]{0.32\textwidth}
    \centering
    \includegraphics[width=\linewidth]{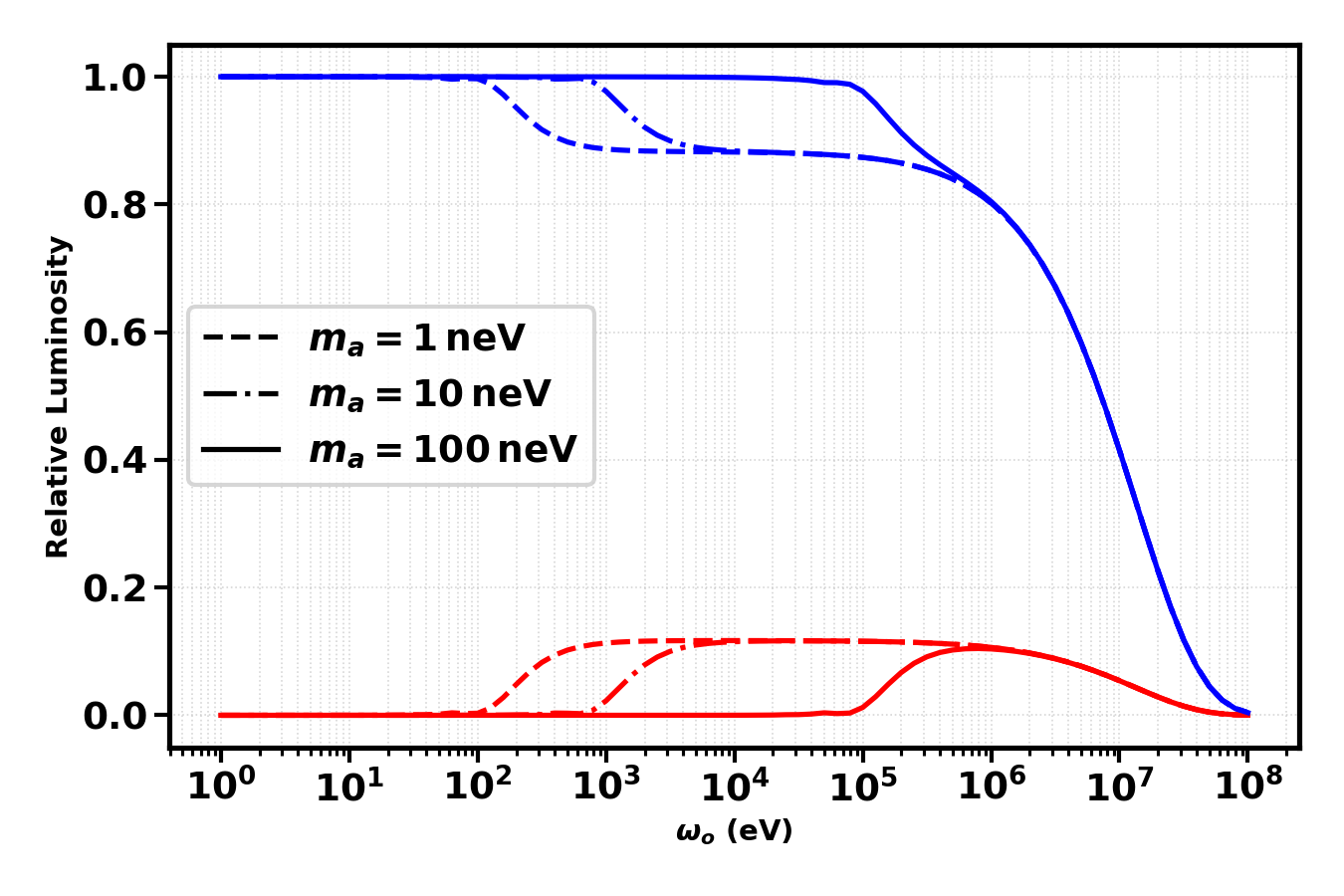}
    \caption{$a=0.6$ and $g_{a\gamma}=10^{-11}$}
    \label{11e}
\end{subfigure}\hfill
\begin{subfigure}[b]{0.32\textwidth}
    \centering
    \includegraphics[width=\linewidth]{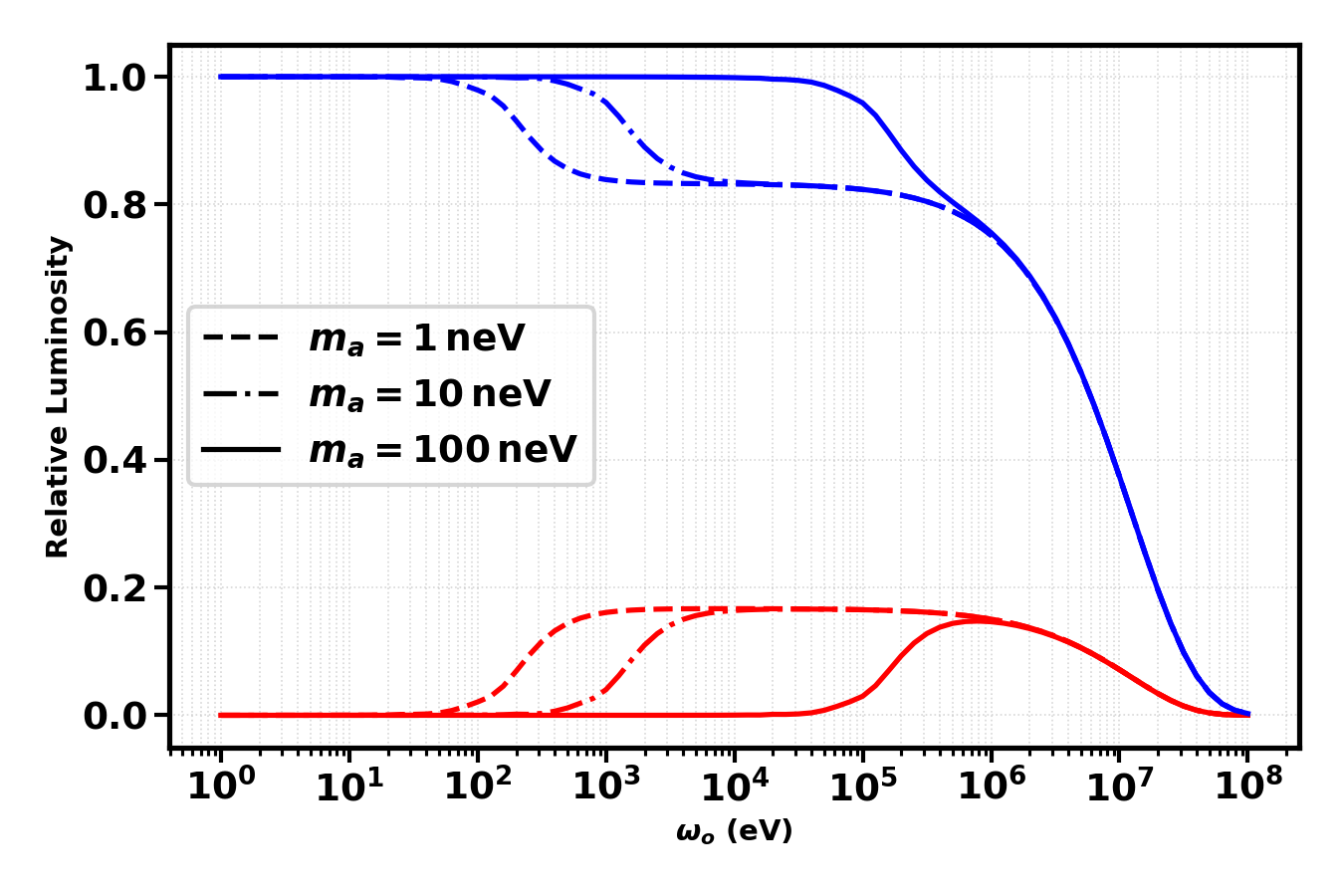}
    \caption{$a=0.99$ and $g_{a\gamma}=10^{-11}$}
    \label{11f}
\end{subfigure}

\vspace{0.3cm}

% -------- Row 3 --------
\begin{subfigure}[b]{0.32\textwidth}
    \centering
    \includegraphics[width=\linewidth]{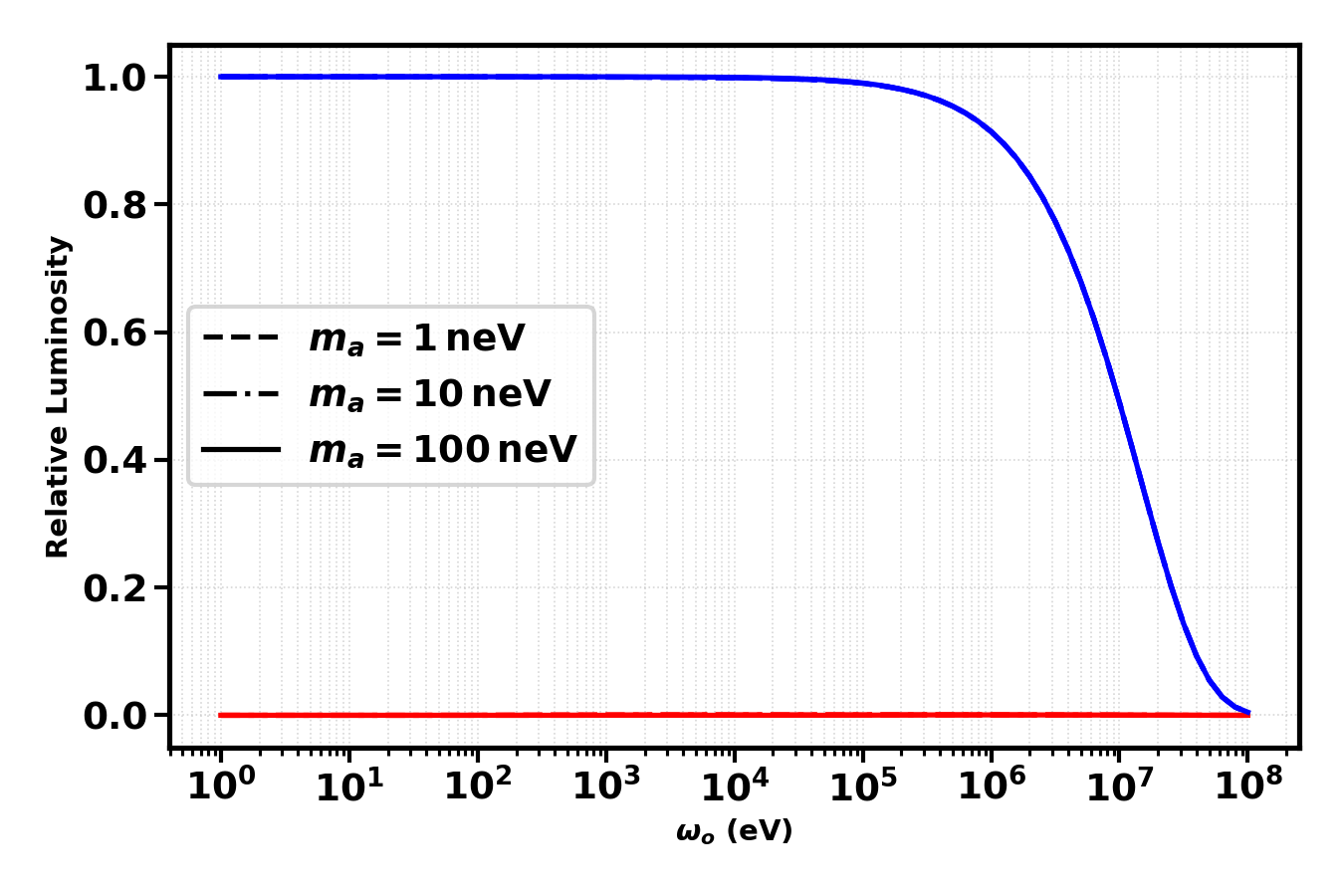}
    \caption{$a=0.3$ and $g_{a\gamma}=10^{-12}$}
    \label{11g}
\end{subfigure}\hfill
\begin{subfigure}[b]{0.32\textwidth}
    \centering
    \includegraphics[width=\linewidth]{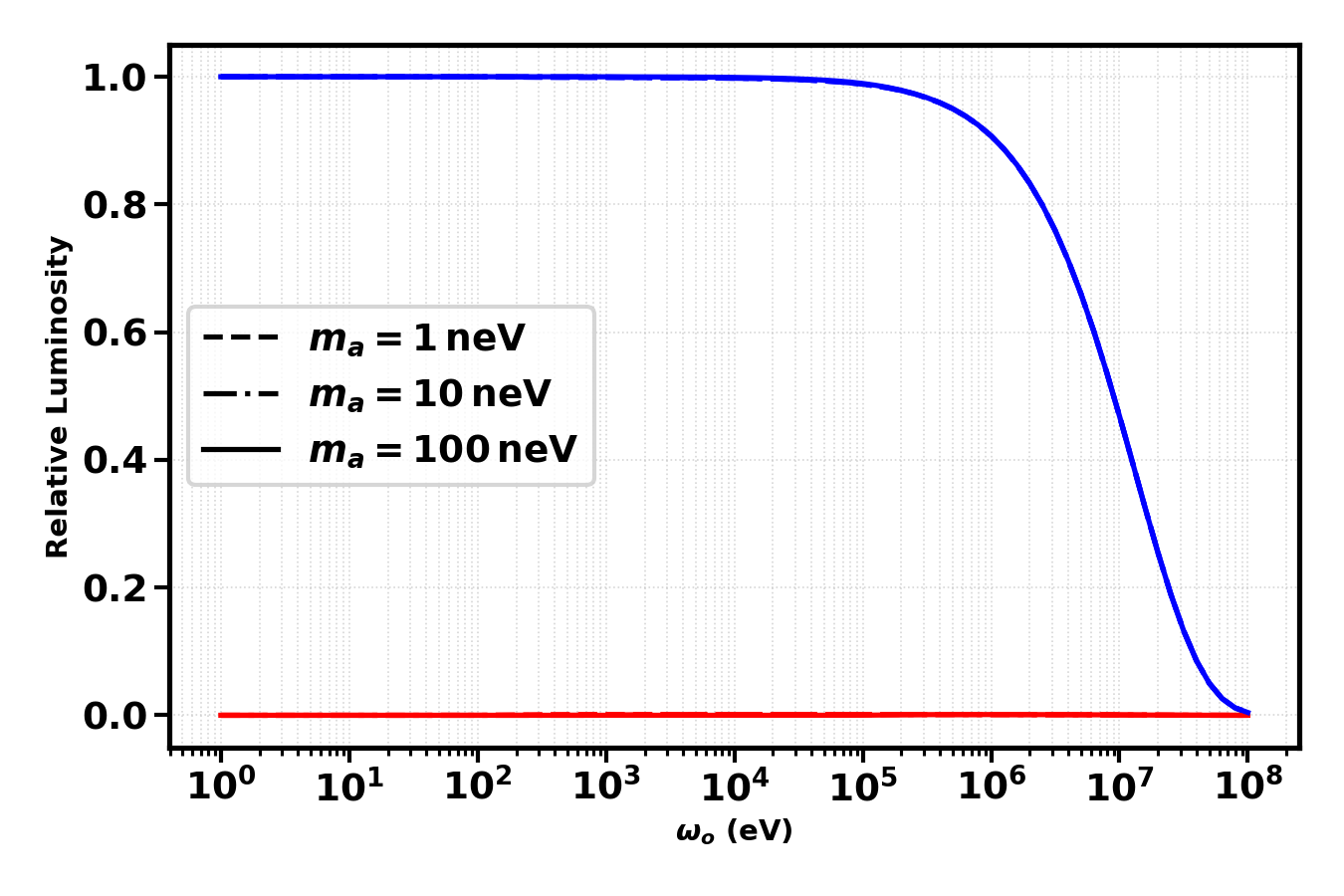}
    \caption{$a=0.6$ and $g_{a\gamma}=10^{-12}$}
    \label{11h}
\end{subfigure}\hfill
\begin{subfigure}[b]{0.32\textwidth}
    \centering
    \includegraphics[width=\linewidth]{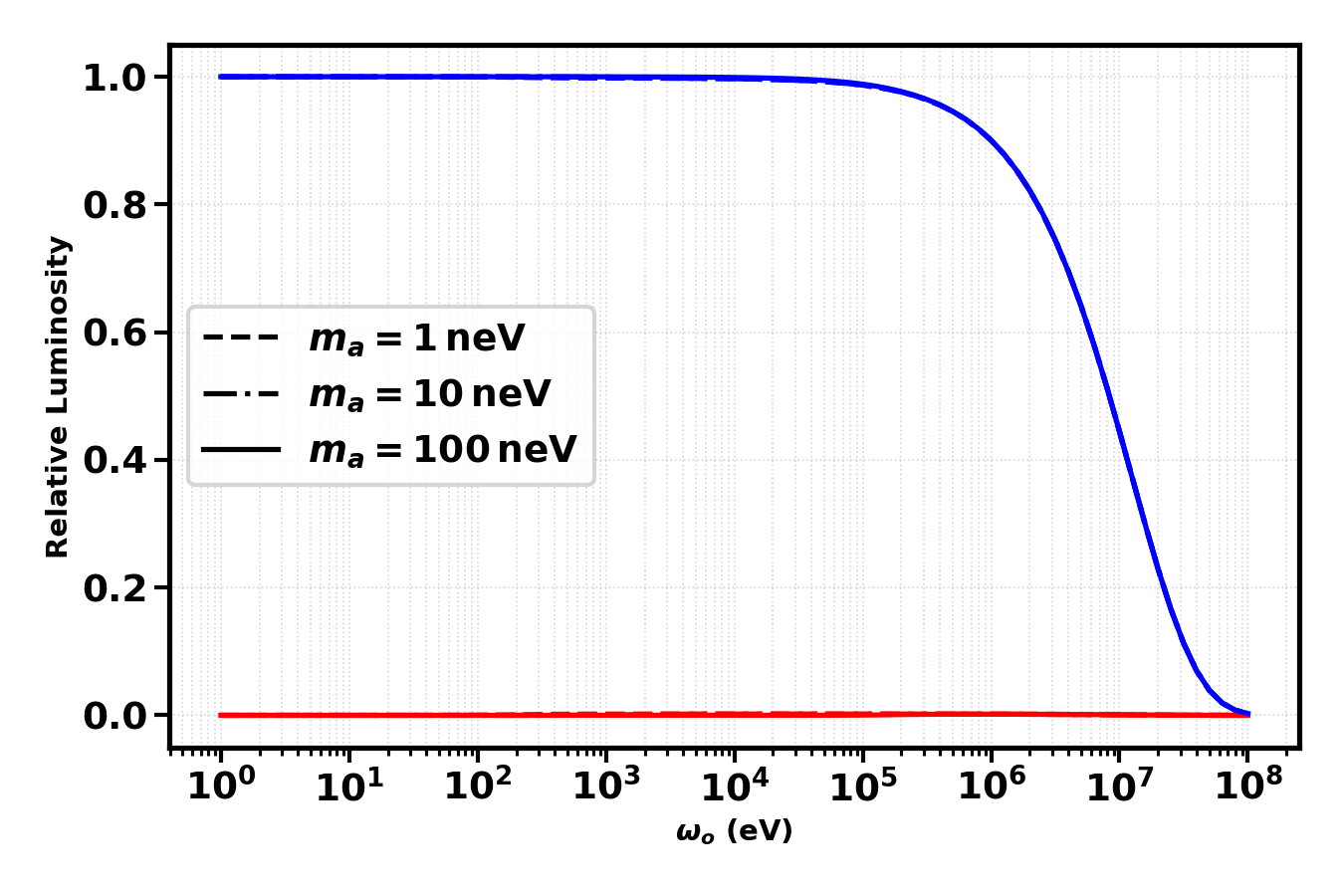}
    \caption{$a=0.99$ and $g_{a\gamma}=10^{-12}$}
    \label{11i}
\end{subfigure}

\vspace{0.3cm}

\caption{The above figure shows the variation of the relative luminosity of photons and axions with the observed frequency for three different masses of axions (1 neV, 10 neV, 100 neV) and varying three coupling parameter values $g_{a\gamma}$ = ($10^{-10}$, $10^{-11}$, $10^{-12}$) GeV$^{-1}$. For a given $g_{a\gamma}$, the photon and axion spectral luminosities are shown near three different BH spins a=0.3, a=0.6 and a=0.99.
In the above plots the magnetic field is taken to be $B \simeq 30 G$ and the electron number density $n_e \simeq 10^{4}$ $\text{cm}^{-3}$. Blue color denotes the photon spectra and red color denotes the axion spectra. We use $M=6.2\times 10^9 M_\odot$ (corresponding to the mass of M87*) to obtain the above spectra.  }
\label{fig11}
\end{figure}

\begin{figure}[!htbp]
\centering

% -------- Row 1 --------
\begin{subfigure}[b]{0.32\textwidth}
    \centering
    \includegraphics[width=\linewidth]{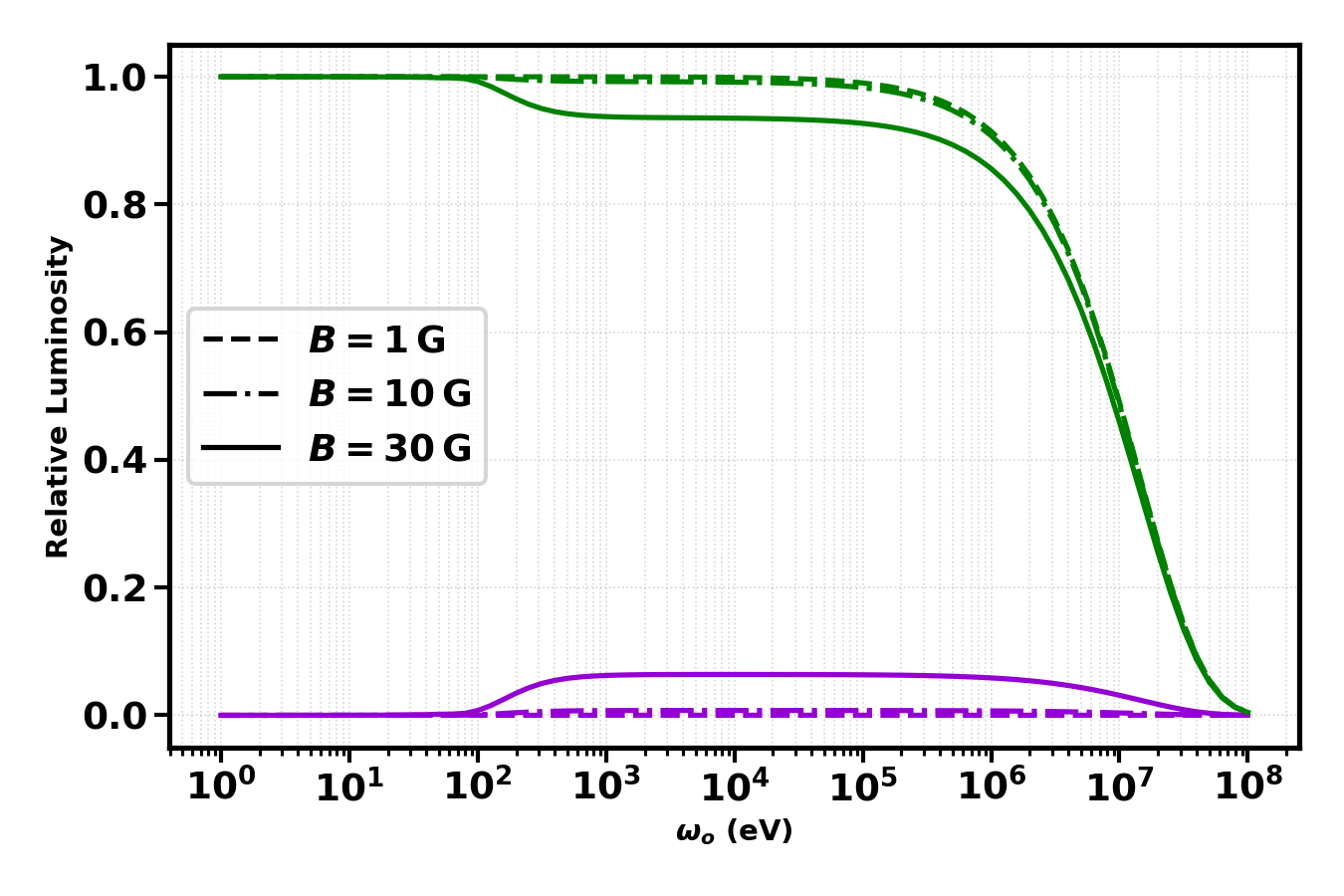}
    \caption{$a=0.3$ and $n_e=10^{4}\text{ cm}^{-3}$}
\end{subfigure}\hfill
\begin{subfigure}[b]{0.32\textwidth}
    \centering
    \includegraphics[width=\linewidth]{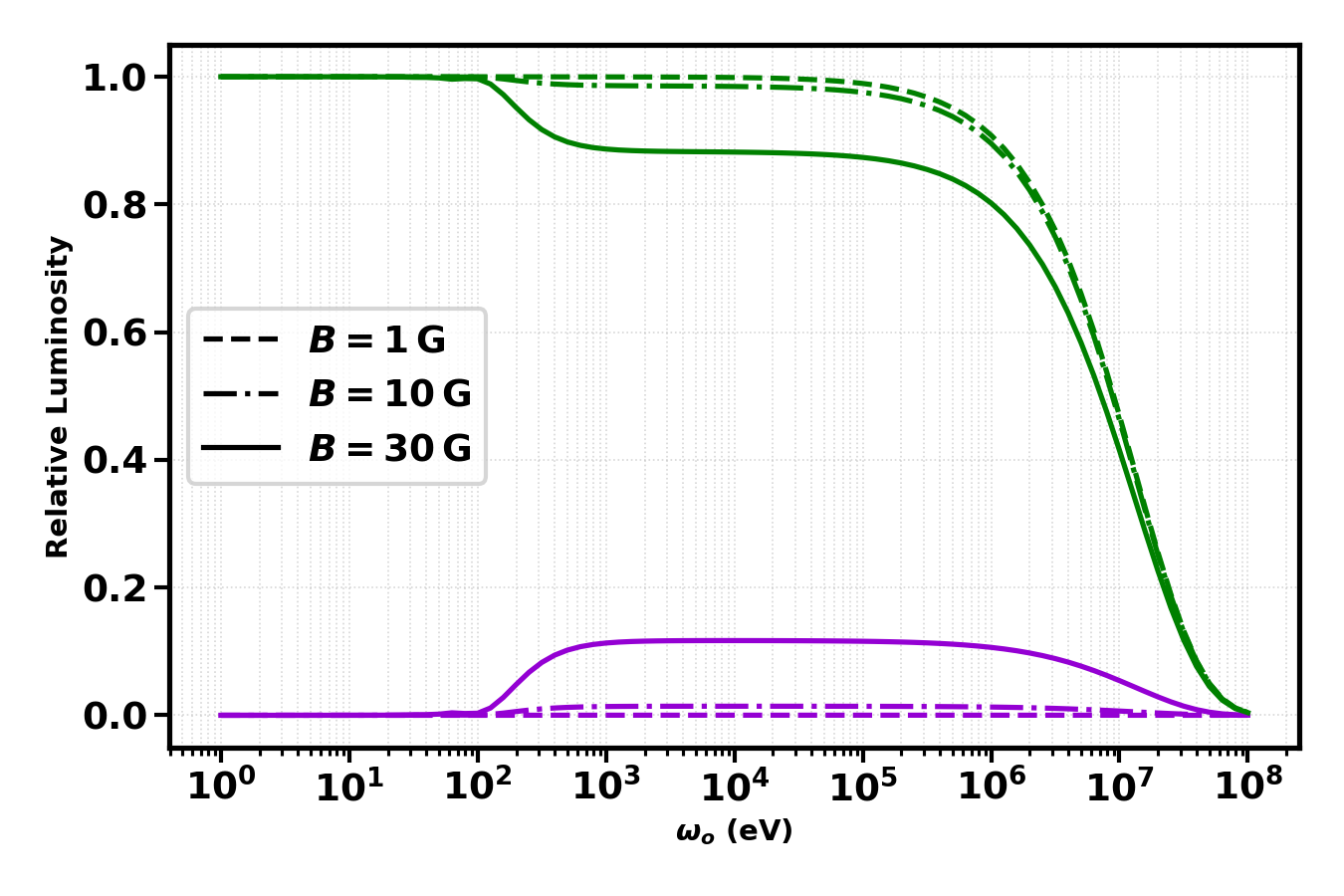}
    \caption{$a=0.6$ and $n_e=10^{4}\text{ cm}^{-3}$}
\end{subfigure}\hfill
\begin{subfigure}[b]{0.32\textwidth}
    \centering
    \includegraphics[width=\linewidth]{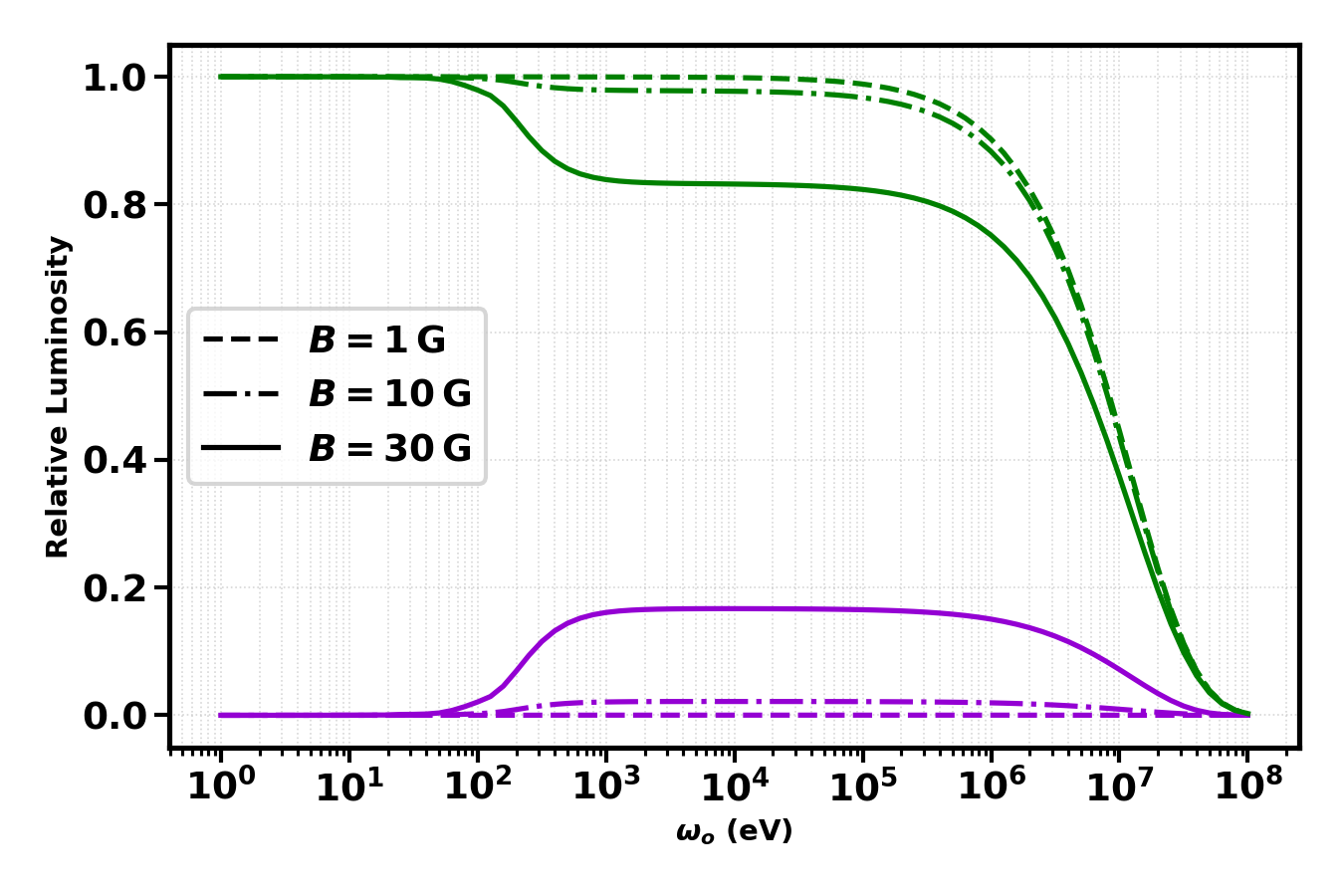}
    \caption{$a=0.99$ and $n_e=10^{4}\text{cm}^{-3}$}
\end{subfigure}

\vspace{0.3cm}

% -------- Row 2 --------
\begin{subfigure}[b]{0.32\textwidth}
    \centering
    \includegraphics[width=\linewidth]{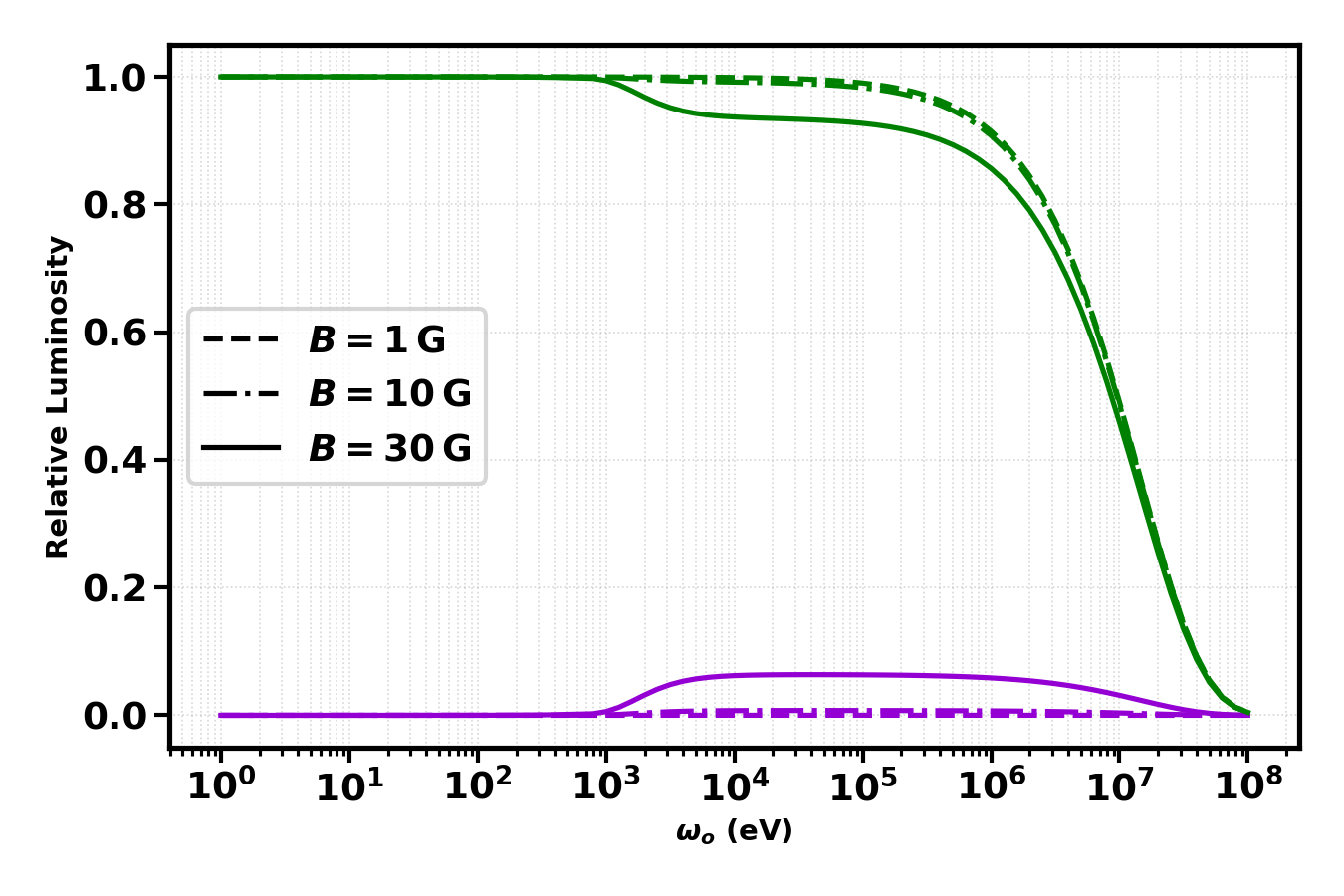}
    \caption{$a=0.3$ and $n_e=10^{5}\text{ cm}^{-3}$}
\end{subfigure}\hfill
\begin{subfigure}[b]{0.32\textwidth}
    \centering
    \includegraphics[width=\linewidth]{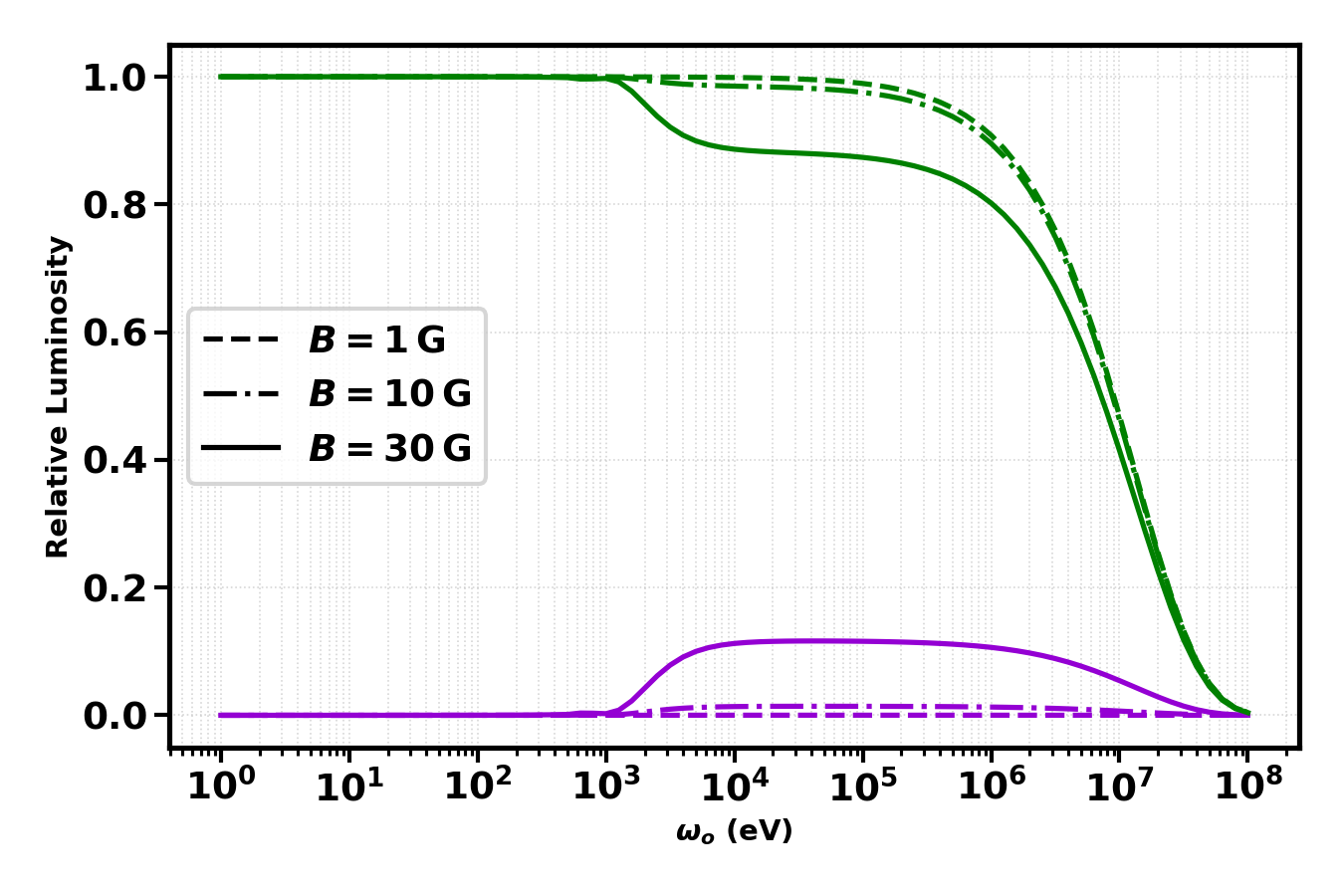}
    \caption{$a=0.6$ and $n_e=10^{5}\text{ cm}^{-3}$}
\end{subfigure}\hfill
\begin{subfigure}[b]{0.32\textwidth}
    \centering
    \includegraphics[width=\linewidth]{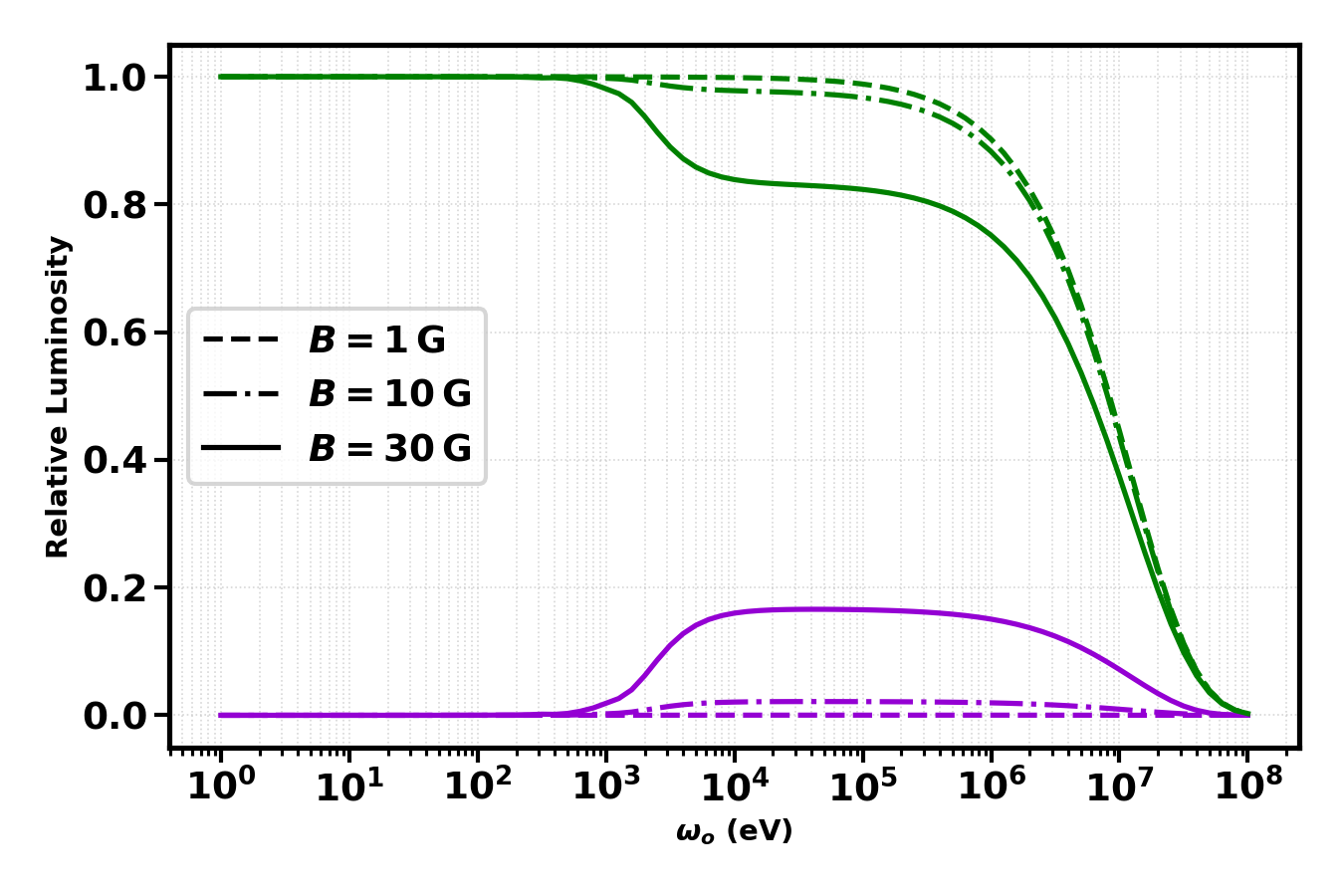}
    \caption{$a=0.99$ and $n_e=10^{5}\text{ cm}^{-3}$}
\end{subfigure}

\vspace{0.3cm}

% -------- Row 3 --------
\begin{subfigure}[b]{0.32\textwidth}
    \centering
    \includegraphics[width=\linewidth]{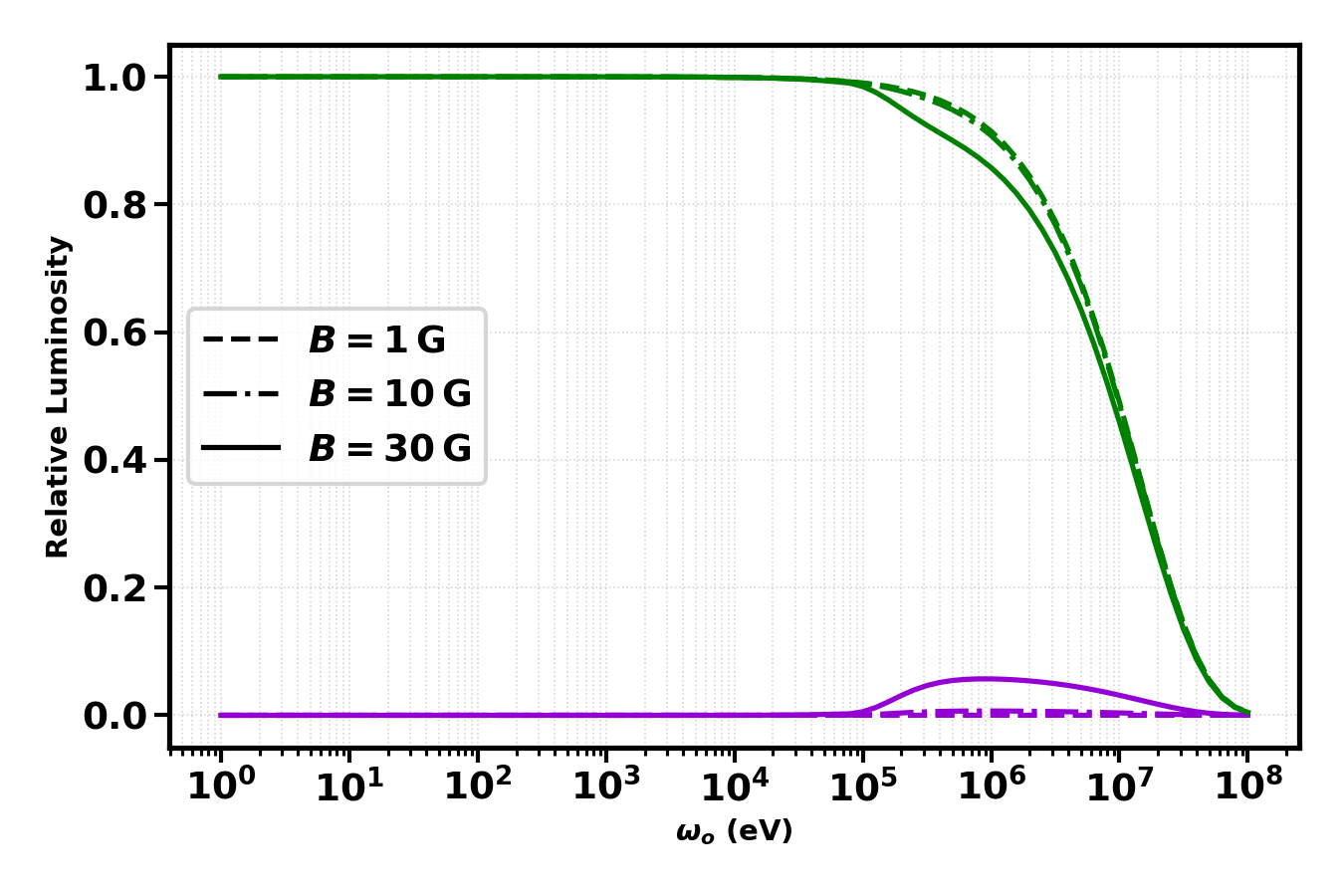}
    \caption{$a=0.3$ and $n_e=10^{7}\text{ cm}^{-3}$}
\end{subfigure}\hfill
\begin{subfigure}[b]{0.32\textwidth}
    \centering
    \includegraphics[width=\linewidth]{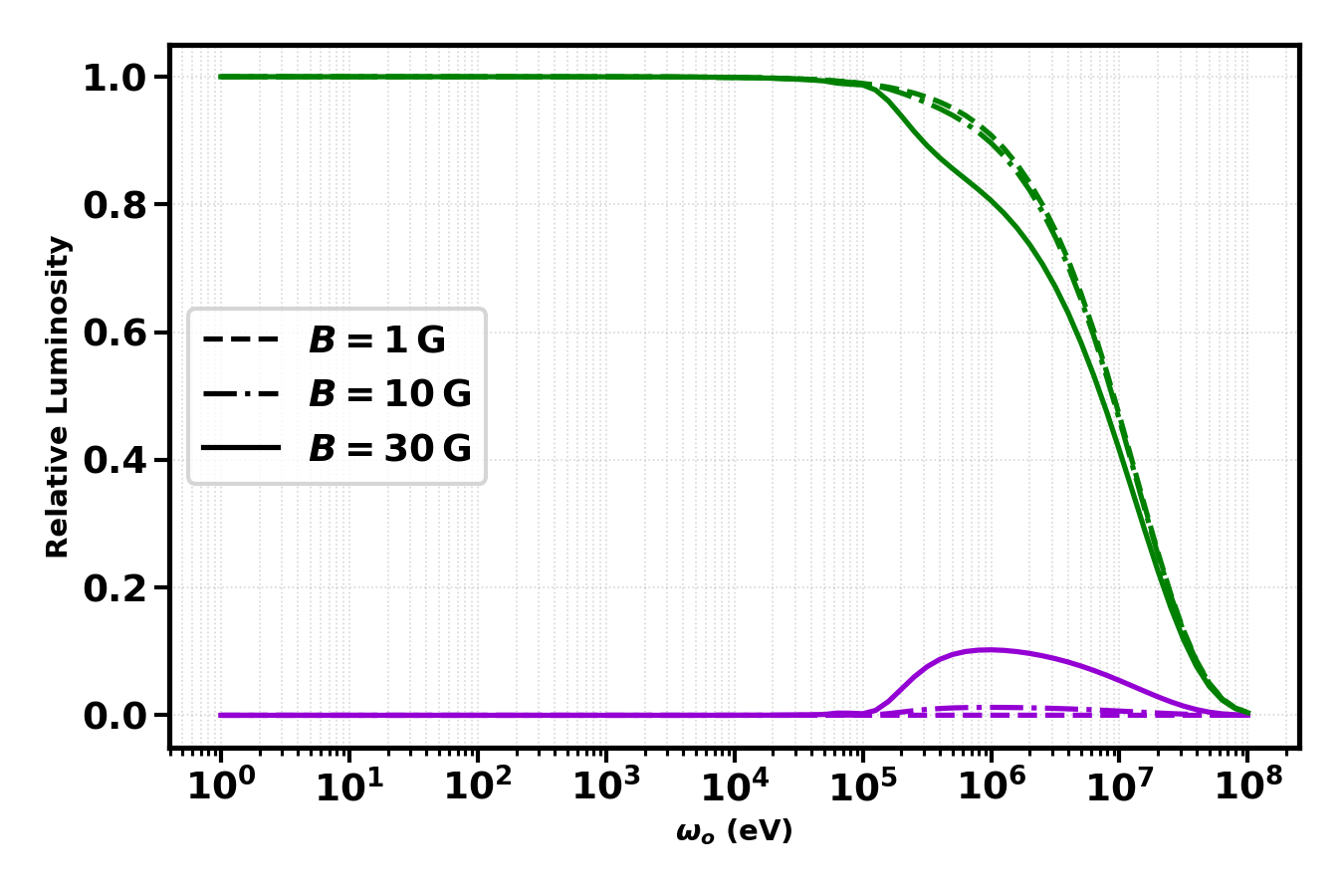}
    \caption{$a=0.6$ and $n_e=10^{7}\text{ cm}^{-3}$}
\end{subfigure}\hfill
\begin{subfigure}[b]{0.32\textwidth}
    \centering
    \includegraphics[width=\linewidth]{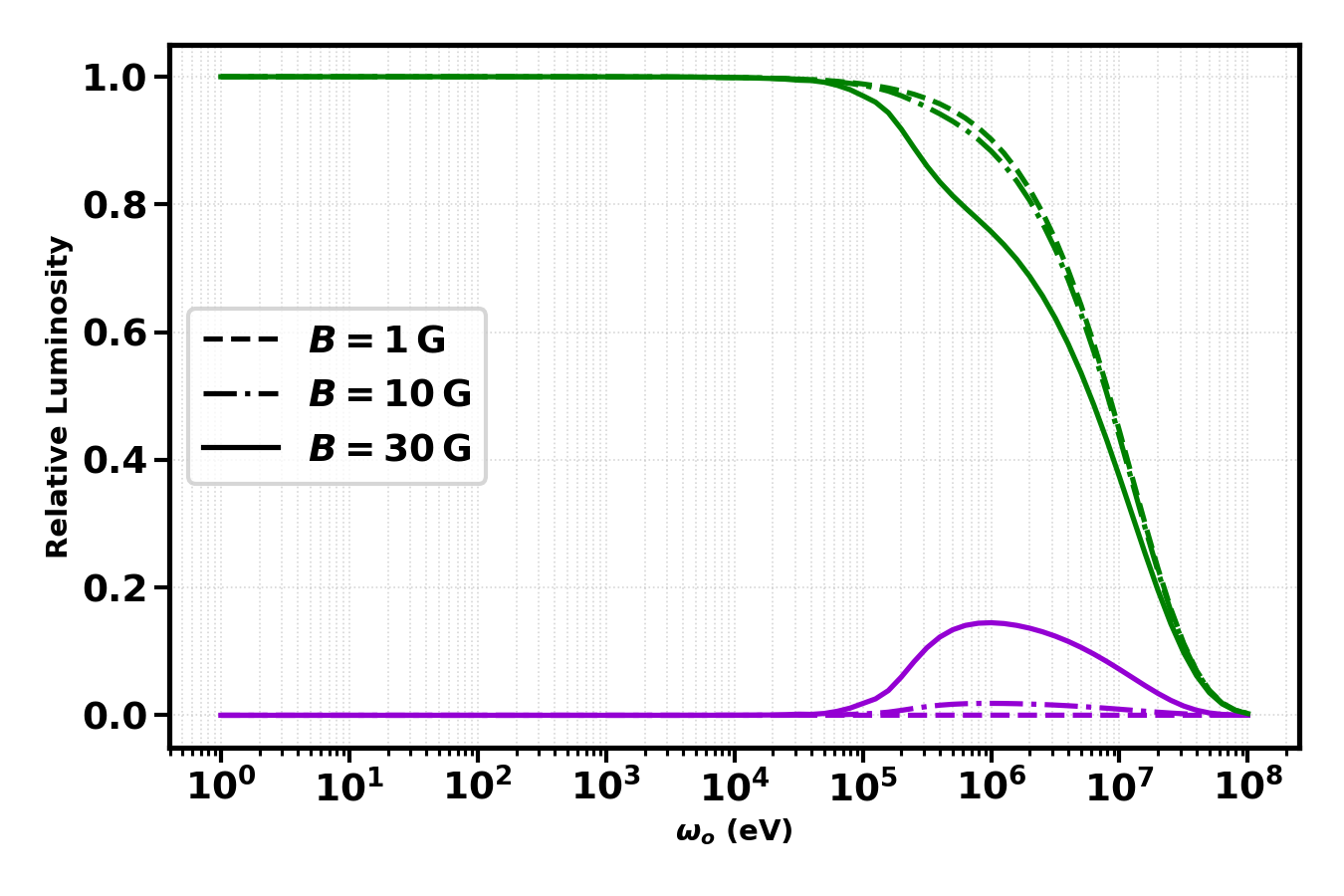}
    \caption{$a=0.99$ and $n_e=10^{7}\text{ cm}^{-3}$}
\end{subfigure}

\vspace{0.3cm}

\caption{The above figure shows the variation of the relative luminosity of photons and axions with the observed frequency for three different magnetic fields (1 G, 10 G, 30 G) and varying over three values of number density $n_e\simeq$  ($10^{4}\text{ cm}^{-3}$, $10^{5}\text{ cm}^{-3}$, $10^{7}\text{ cm}^{-3}$). For a given $n_e$, the conversion is studied around BHs of spin a=0.3, a=0.6 and a=0.99. Here the coupling parameter $g_{a\gamma} \simeq 10^{-11}\rm ~GeV^{-1}$ and axion mass $m_a \simeq 1$ $\text{neV}$. Green colour denotes the photon spectra and purple colour denotes the axion spectra. We use $M= 6.2\times 10^9 M_\odot$ (corresponding to the mass of M87*) to obtain the above spectra.} 
\label{fig12}
\end{figure}

\begin{figure}[!htbp]
\centering

% -------- Row 1 --------
\begin{subfigure}[b]{0.32\textwidth}
    \centering
    \includegraphics[width=\linewidth]{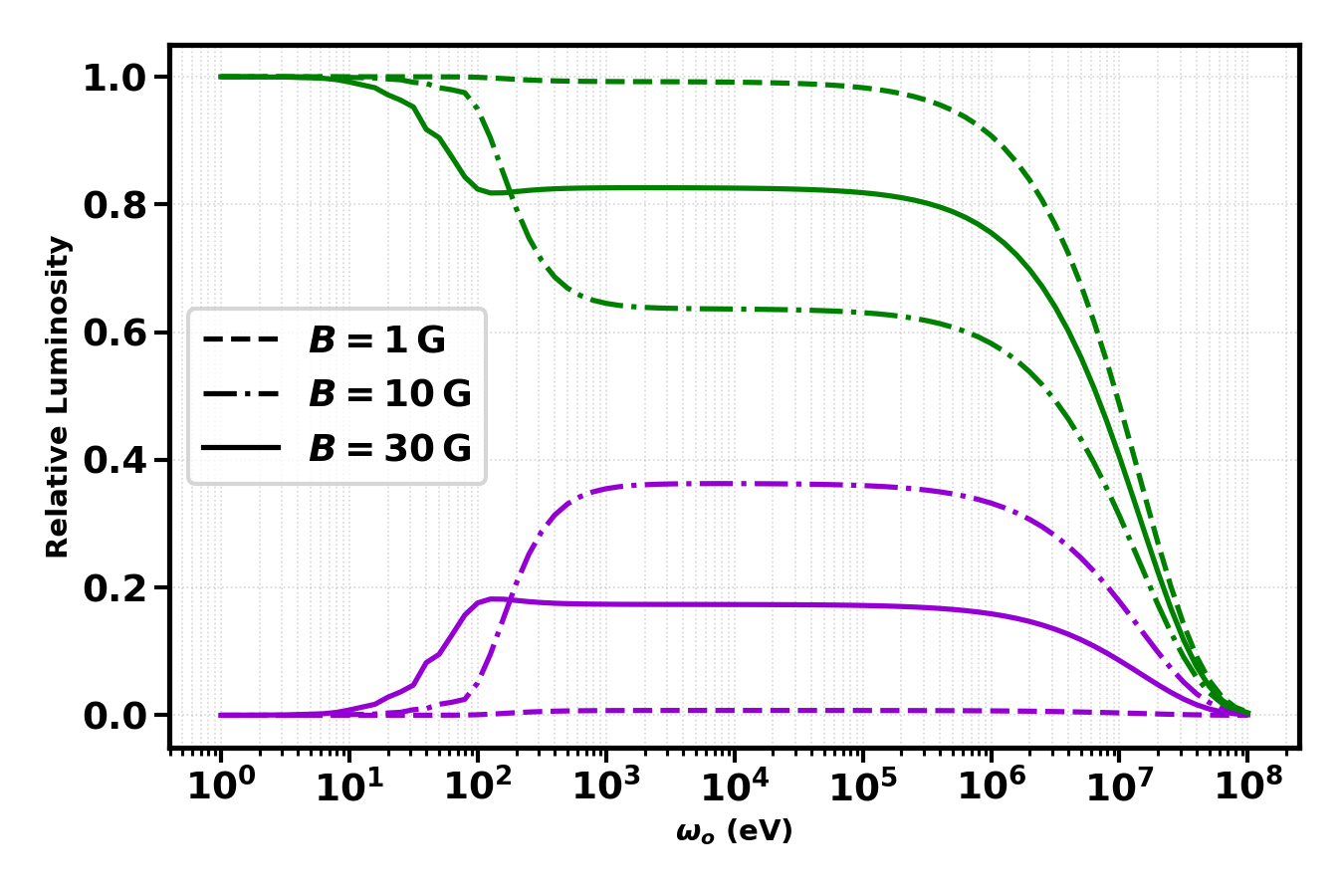}
    \caption{$a=0.3$ and $n_e=10^{4}\text{ cm}^{-3}$}
\end{subfigure}\hfill
\begin{subfigure}[b]{0.32\textwidth}
    \centering
    \includegraphics[width=\linewidth]{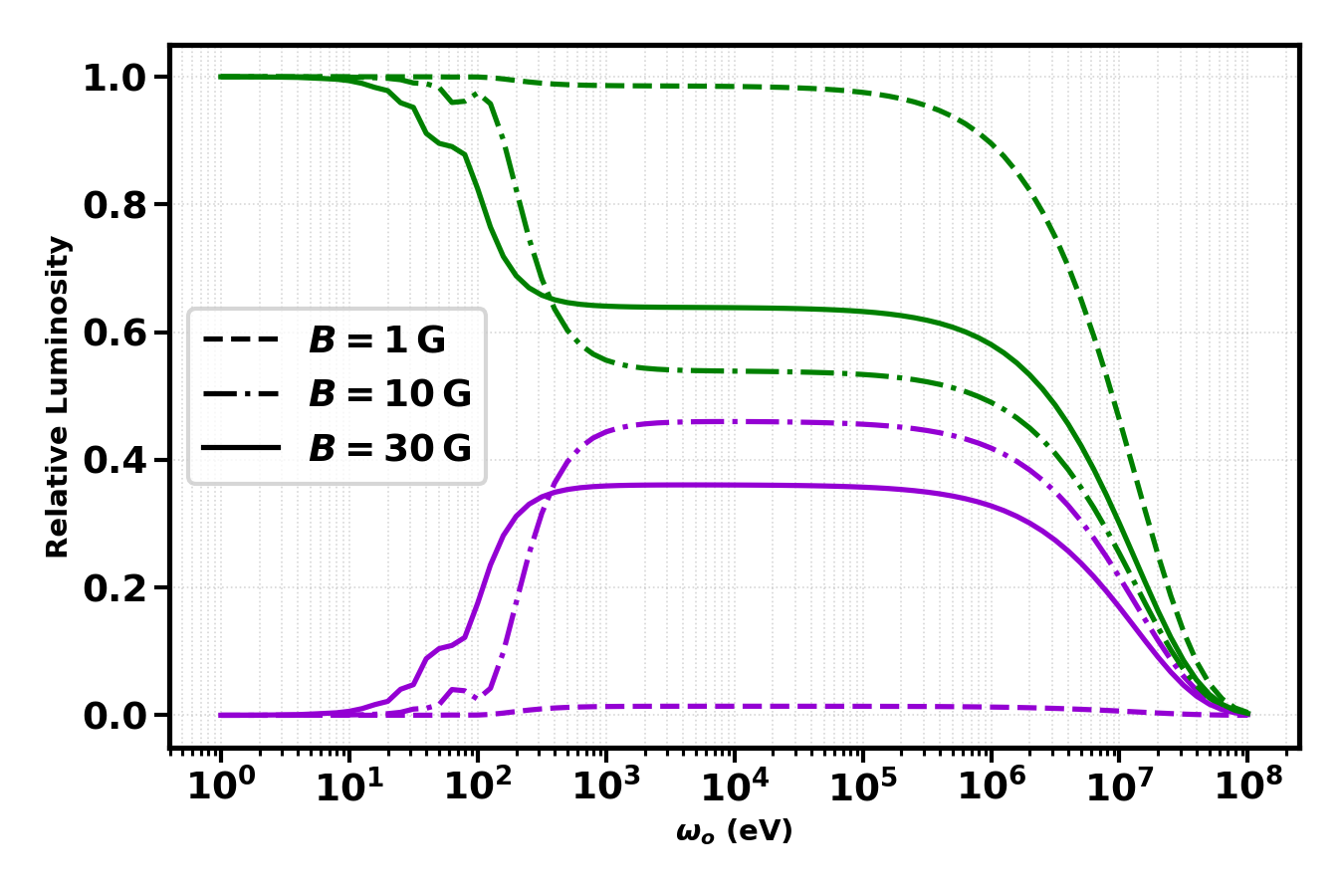}
    \caption{$a=0.6$ and $n_e=10^{4}\text{ cm}^{-3}$}
\end{subfigure}\hfill
\begin{subfigure}[b]{0.32\textwidth}
    \centering
    \includegraphics[width=\linewidth]{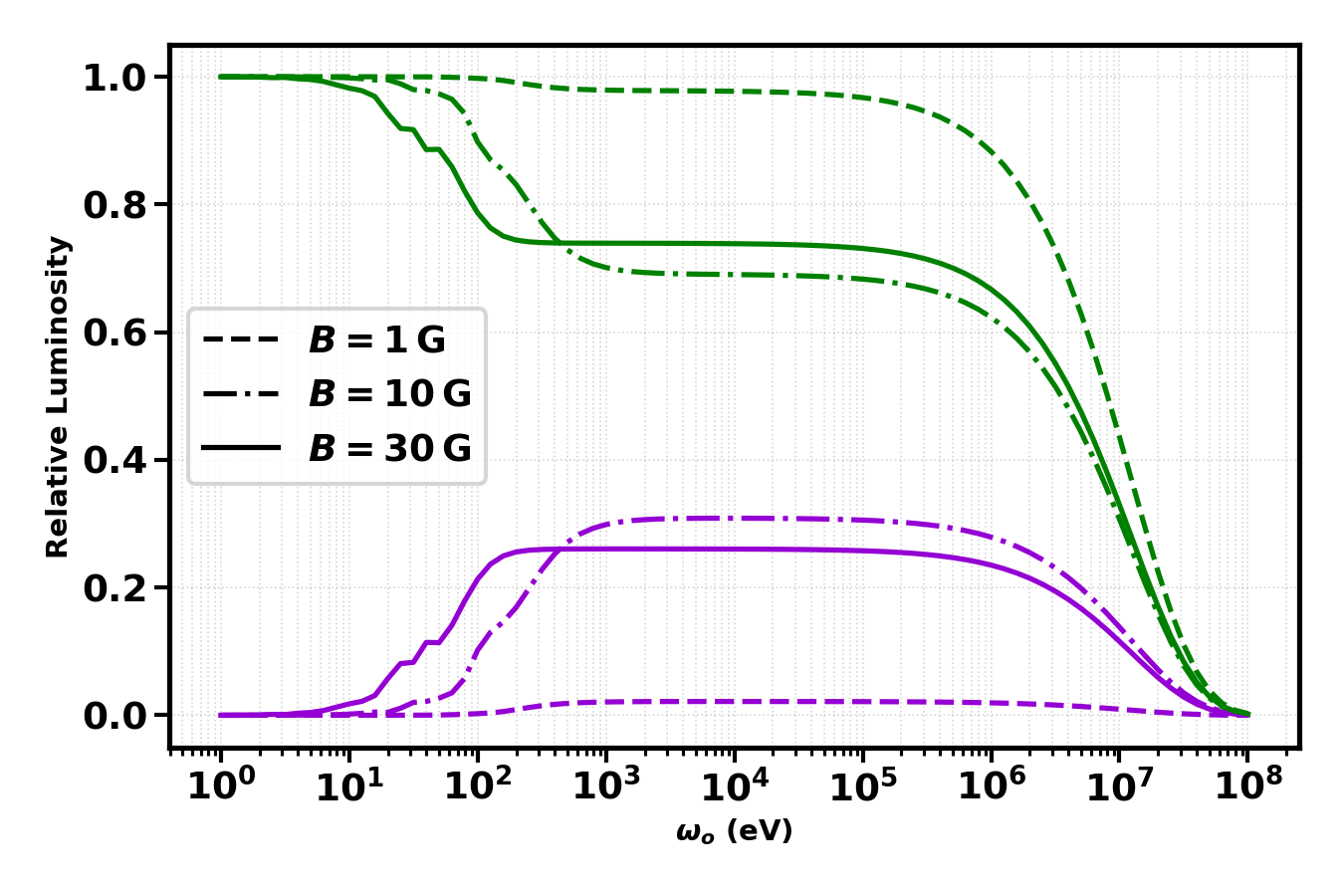}
    \caption{$a=0.99$ and $n_e=10^{4}\text{cm}^{-3}$}
\end{subfigure}

\vspace{0.3cm}

% -------- Row 2 --------
\begin{subfigure}[b]{0.32\textwidth}
    \centering
    \includegraphics[width=\linewidth]{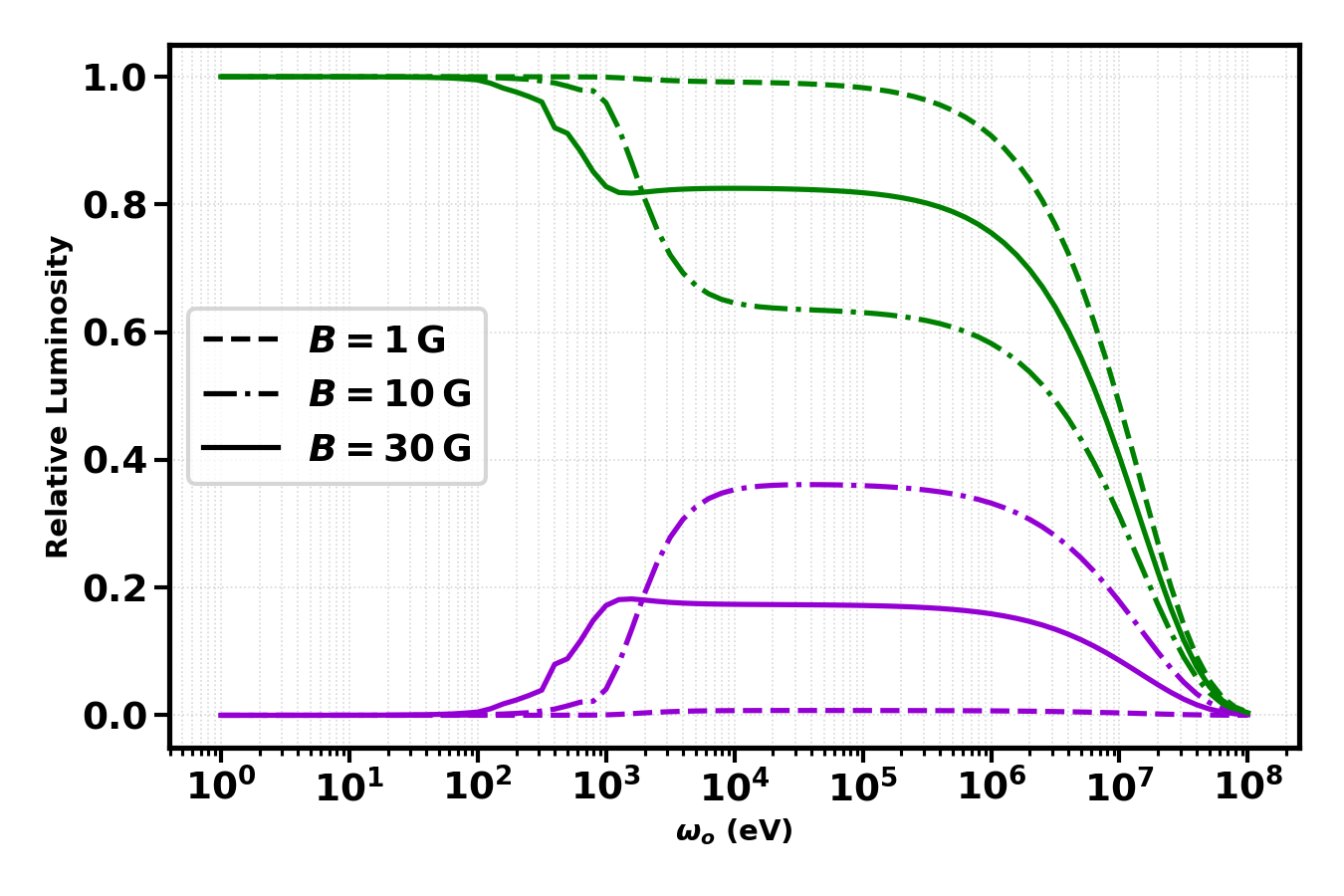}
    \caption{$a=0.3$ and $n_e=10^{5}\text{ cm}^{-3}$}
\end{subfigure}\hfill
\begin{subfigure}[b]{0.32\textwidth}
    \centering
    \includegraphics[width=\linewidth]{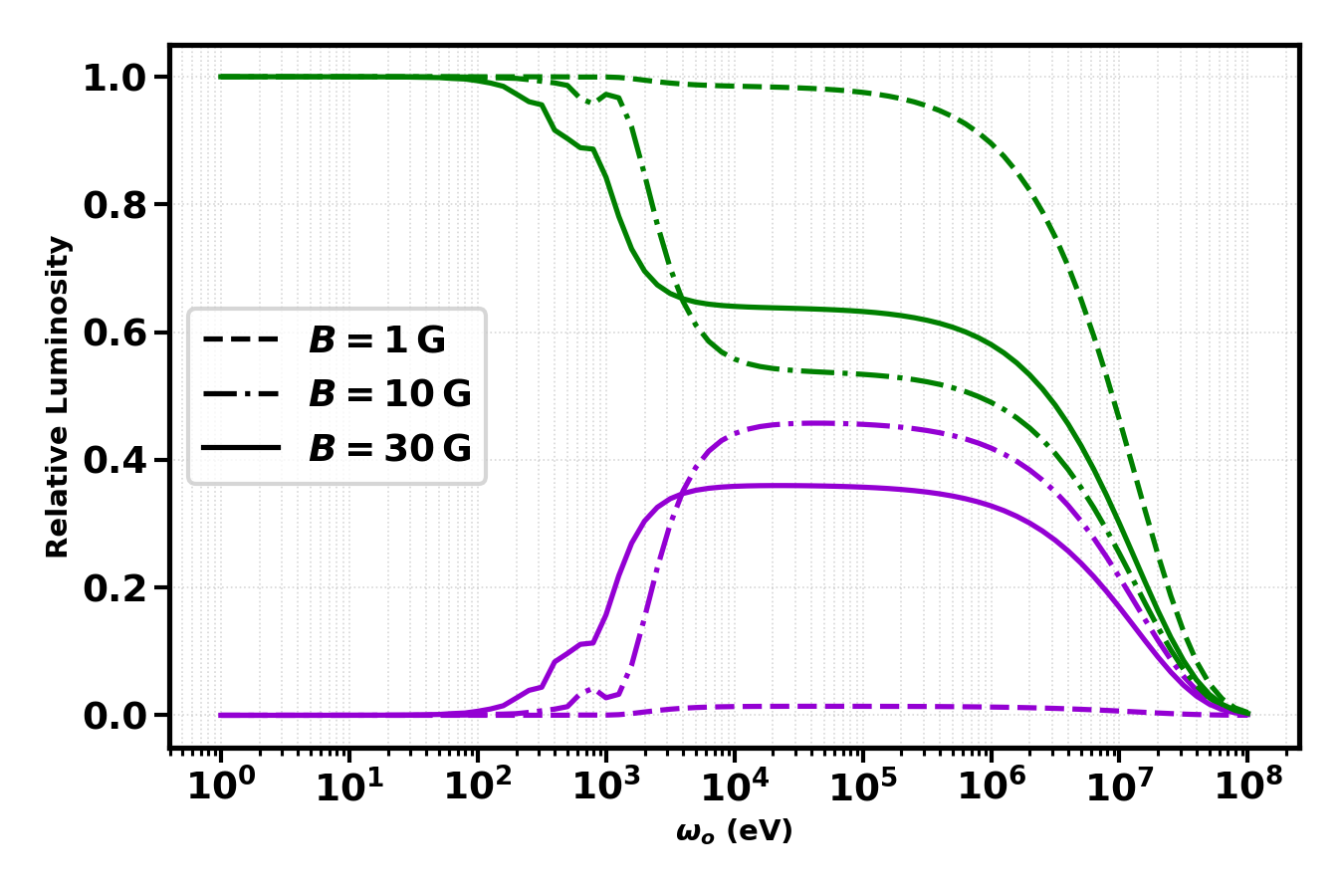}
    \caption{$a=0.6$ and $n_e=10^{5}\text{ cm}^{-3}$}
\end{subfigure}\hfill
\begin{subfigure}[b]{0.32\textwidth}
    \centering
    \includegraphics[width=\linewidth]{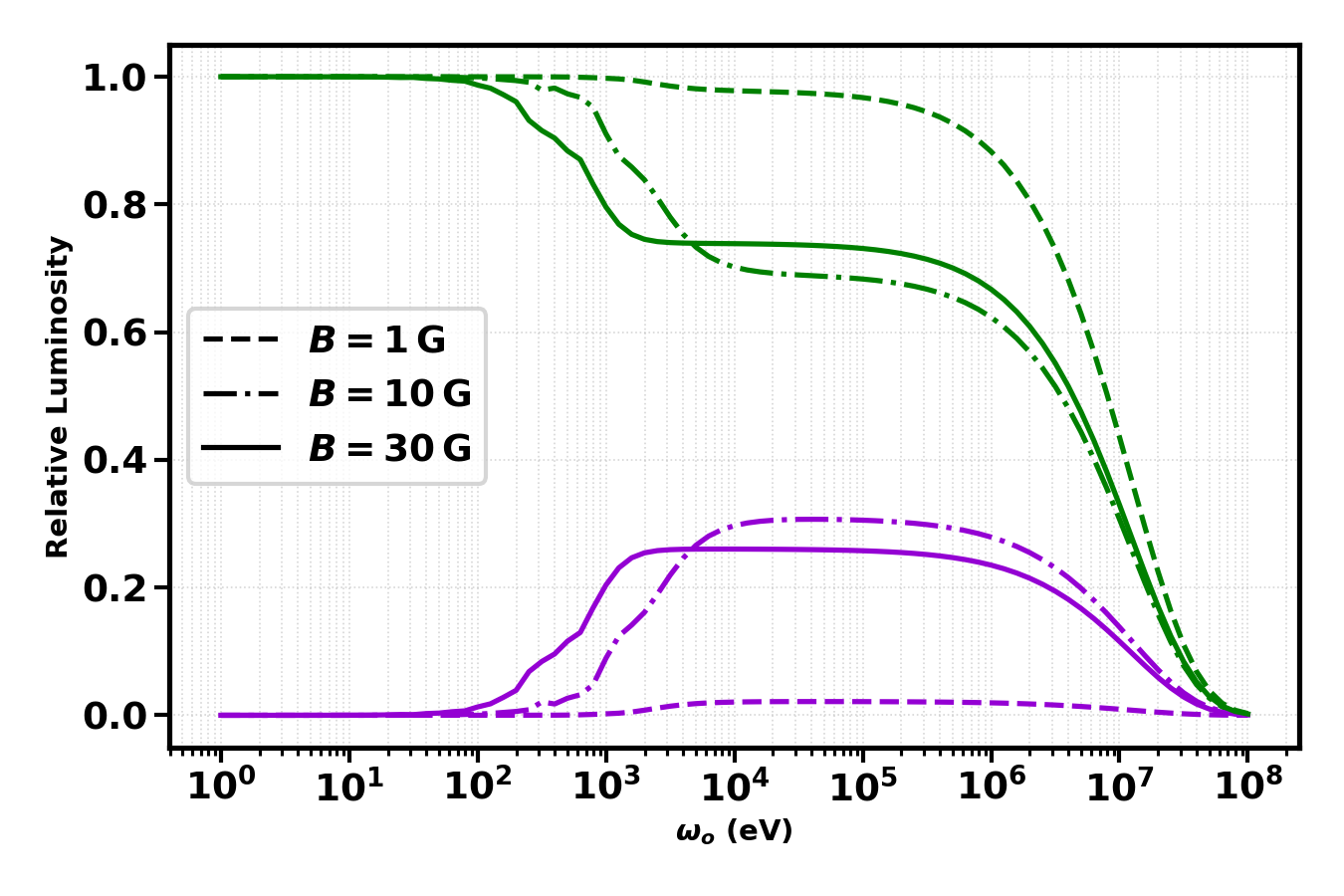}
    \caption{$a=0.99$ and $n_e=10^{5}\text{ cm}^{-3}$}
\end{subfigure}

\vspace{0.3cm}

% -------- Row 3 --------
\begin{subfigure}[b]{0.32\textwidth}
    \centering
    \includegraphics[width=\linewidth]{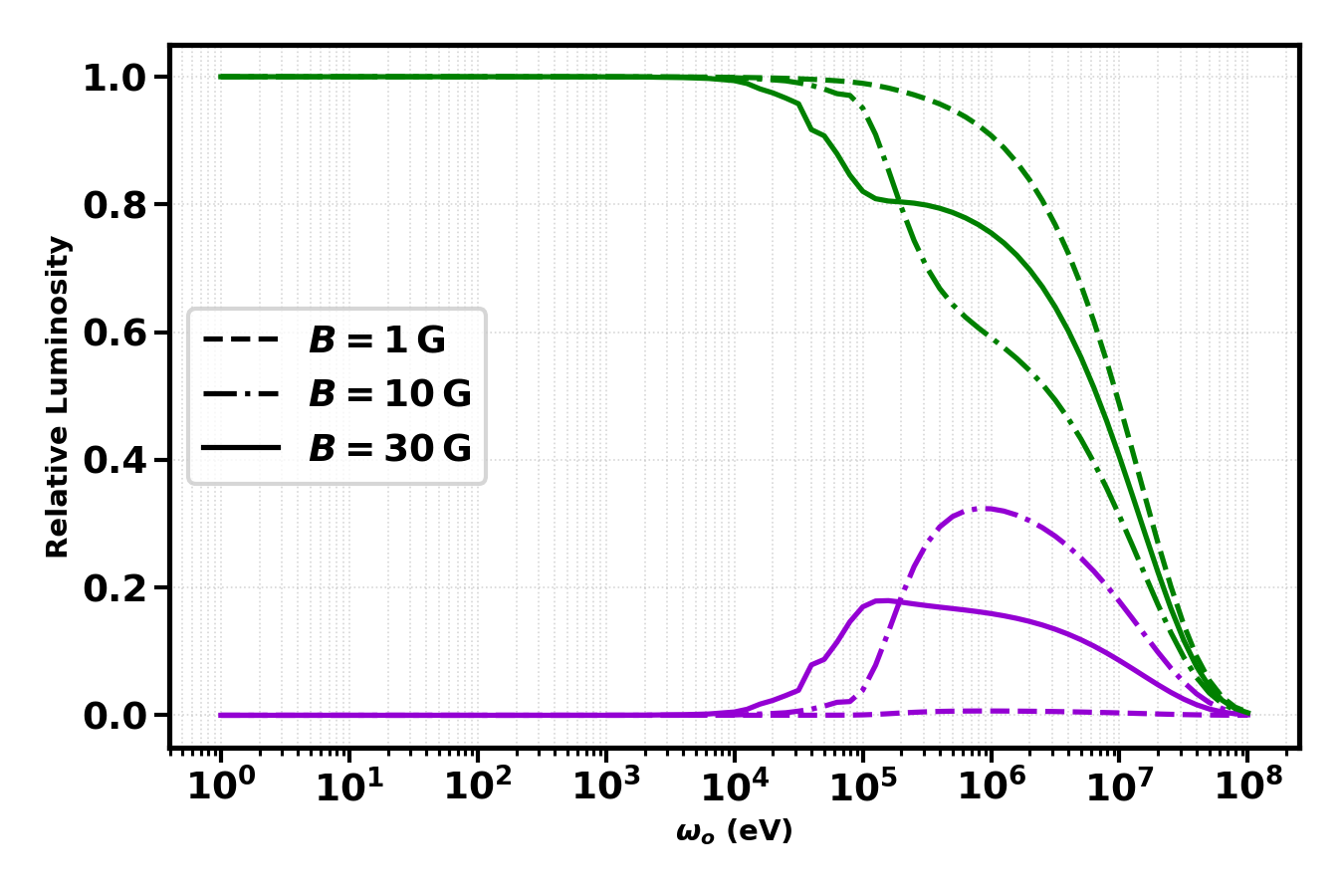}
    \caption{$a=0.3$ and $n_e=10^{7}\text{ cm}^{-3}$}
\end{subfigure}\hfill
\begin{subfigure}[b]{0.32\textwidth}
    \centering
    \includegraphics[width=\linewidth]{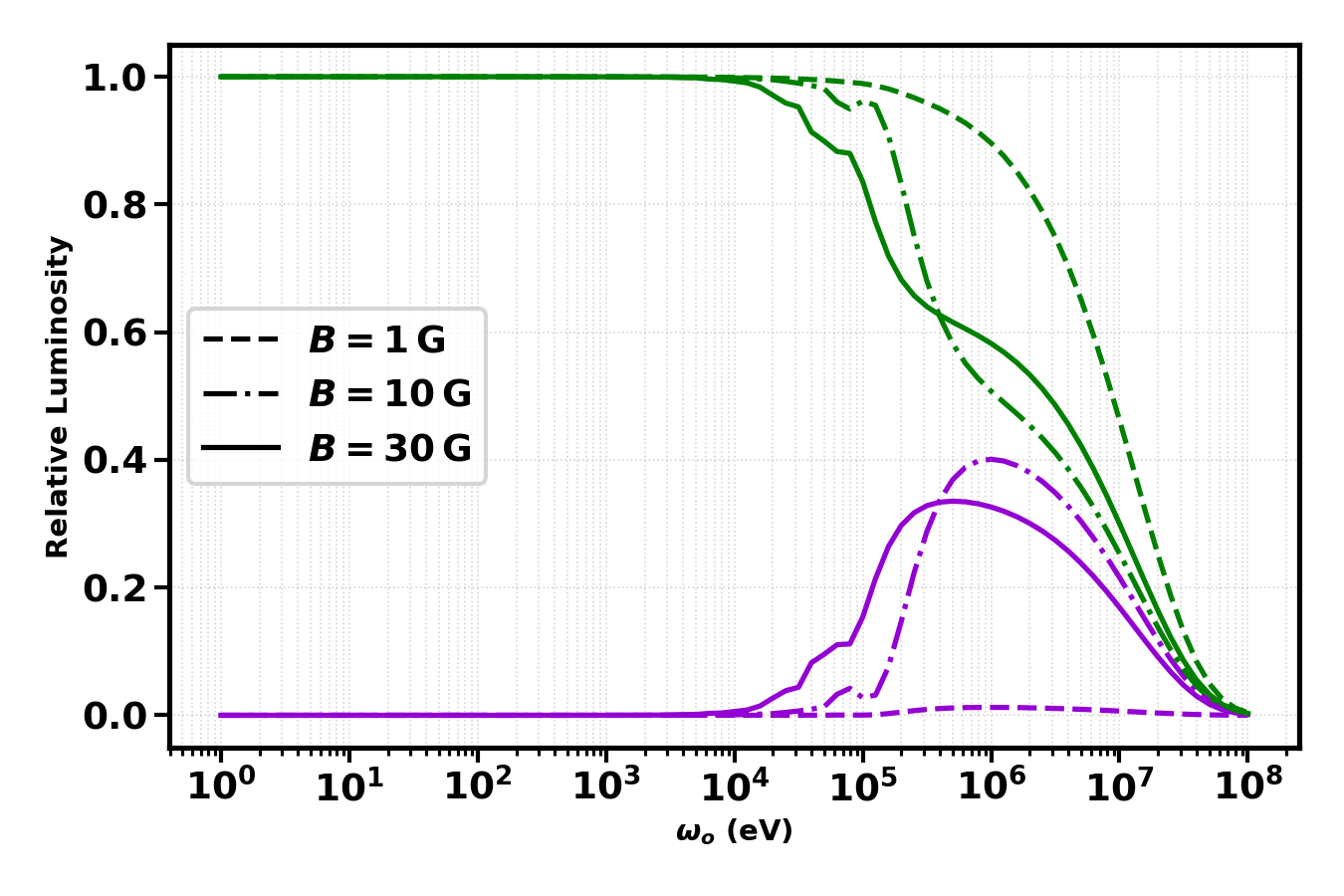}
    \caption{$a=0.6$ and $n_e=10^{7}\text{ cm}^{-3}$}
\end{subfigure}\hfill
\begin{subfigure}[b]{0.32\textwidth}
    \centering
    \includegraphics[width=\linewidth]{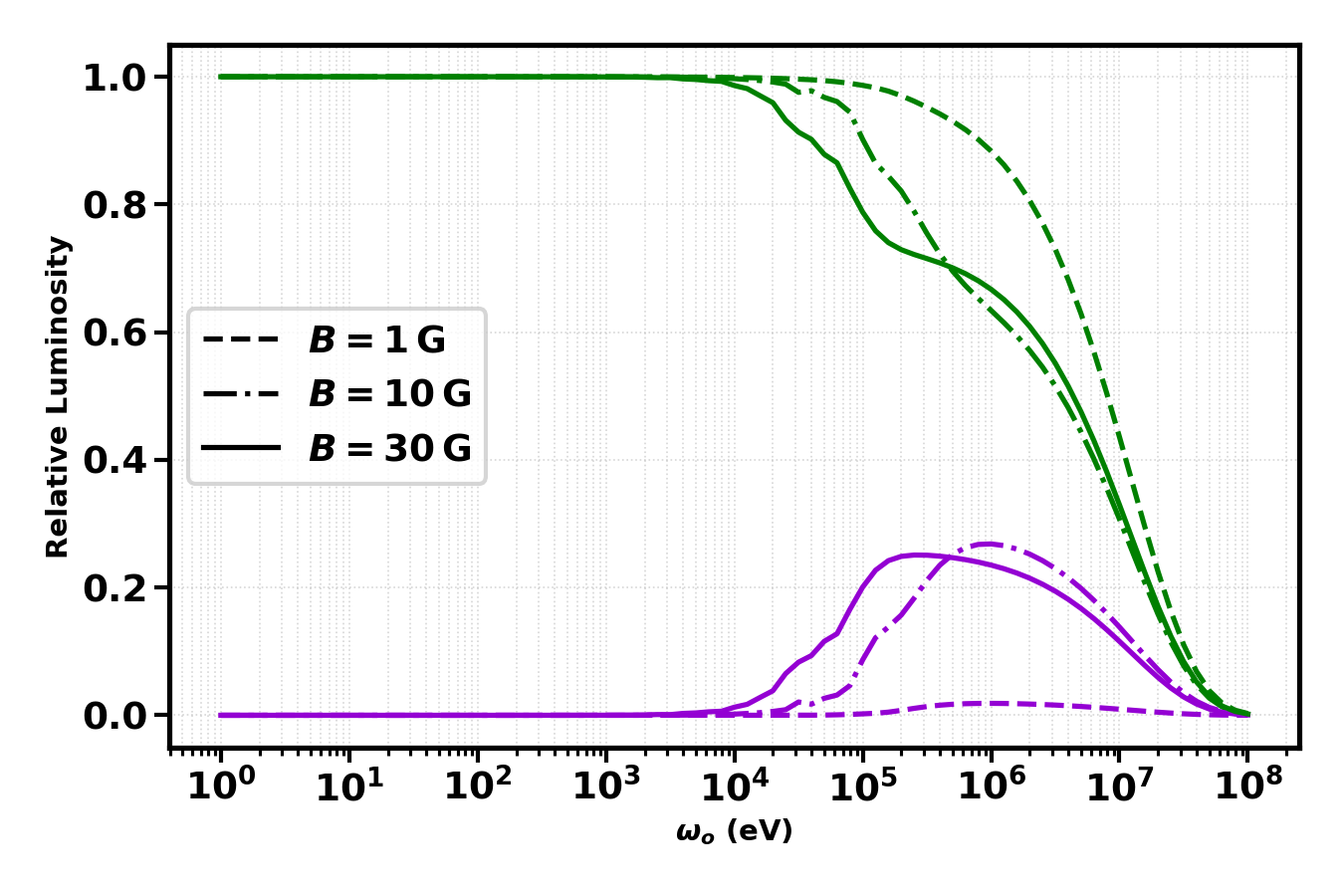}
    \caption{$a=0.99$ and $n_e=10^{7}\text{ cm}^{-3}$}
\end{subfigure}

\vspace{0.3cm}

\caption{The above figure shows the variation of the relative luminosity of photons and axions with the observed frequency for three different magnetic fields (1 G, 10 G, 30 G) and varying over three values of number density $n_e\simeq$  ($10^{4}\text{ cm}^{-3}$, $10^{5}\text{ cm}^{-3}$, $10^{7}\text{ cm}^{-3}$). For a given $n_e$, the conversion is studied around BHs of spin a=0.3, a=0.6 and a=0.99. Here the coupling parameter $g_{a\gamma} \simeq 10^{-10}\rm ~GeV^{-1}$ and axion mass $m_a \simeq 1$ $\text{neV}$. Green colour denotes the photon spectra and purple colour denotes the axion spectra. We use $M= 6.2\times 10^9 M_\odot$ (corresponding to the mass of M87*) to obtain the above spectra.}
\label{fig13}
\end{figure}

In \cref{fig11}-\cref{fig13} we plot the variation of the relative luminosities of photons and axions with the observed frequency $\omega_0$ for BHs of spin $a=0.3$ (low), $a=0.6$ (intermediate) and $a=0.99$ (high). For a given spin we have enlisted eight critical impact parameters $\tilde{\lambda},\tilde{\eta}$ corresponding to the Boyer-Lindquist radii $\tilde{r}$ in the photon region in \cref{tab:dimming_results_17_sorted}, \cref{tab:dimming_results_2} and \cref{tab:dimming_results_3}. For each pair of such impact parameters, we have taken a slight deviation in them ($\delta\lambda$ and $\delta\eta$) and accordingly estimated the path length for four sets of ($\delta\lambda$, $\delta\eta$).
The choice of $\delta\lambda$ and $\delta\eta$ are discussed in \cref{Sec4}. For a given $\omega_0$ and a given BH spin we calculate the luminosities of photons and axions by averaging over these thirty-two ($\tilde{\lambda}, \tilde{\eta},\delta{\lambda},\delta{\eta}$) sets. This is then repeated over the entire frequency range for a fixed BH spin to arrive at the photon and the axion spectra illustrated in \cref{fig11}-\cref{fig13}.

In \cref{fig11} we show the variation of the relative luminosities of photons and axions with the observed frequency $\omega_0$ in the vicinity of BHs of spin $a=0.3$, $a=0.6$ and $a=0.99$ for different axion masses and photon-axion couplings observed at magnetic field of 30G and electron number density $n_e\sim 10^4 ~\rm cm^{-3}$ (as estimated for M87*). We note from \cref{fig11} that the conversion is highly sensitive to $g_{a\gamma}$ and when $g_{a\gamma}\lesssim 10^{-12}~\rm GeV^{-1}$, no conversion takes place (\cref{11g}, \cref{11h} and \cref{11i}). This is because as $g_{a\gamma}$ is lowered, $\mathrm{\Delta_M}<\mathrm{\Delta_{osc}}/2$ and the amplitude of the conversion probability (\cref{eq:conversion_prob}) becomes substantially lower than unity. When $g_{a\gamma}$ is increased by an order we note that the conversion increases 
with an increase in the BH spin (\cref{11d}-\cref{11f}). This, however, is not a generic trend since when the photon-axion coupling is further increased to $g_{a\gamma}\approx 10^{-10}~\rm GeV^{-1}$, the conversion is maximum for the intermediate spin BH (\cref{11b} and \cref{11c}).
This is because the conversion probability (\cref{eq:conversion_prob}) is not a monotonic function of the path length $z$ (which in turn is sensitive to the BH spin), rather it varies sinusoidally. Therefore, once the photon-axion coupling exceeds a certain threshold ($g_{a\gamma}\gtrsim 10^{-11}\rm ~GeV^{-1}$) such that $\mathrm{\Delta_{osc}}\simeq 2\Delta_{M}$ is achieved above a cut-off frequency, the conversion is most efficient when $\mathrm{\Delta_{osc}}z\sim (2n+1)\pi $.   An enhanced coupling also leads to more efficient conversion and greater dimming which is evident from the increased luminosity of the axions when $g_{a\gamma}$ is increased from $10^{-12}\rm ~GeV^{-1}-10^{-10}\rm ~GeV^{-1}$.  
We further note from \cref{fig11} that the conversion does not occur at low frequencies (unless the axion mass equals the resonant mass (\cref{mreso})). We have also noted this in \cref{fig:dimming_g1}.

If an attenuation in the photon spectrum is observed in accordance with \cref{fig11}-\cref{fig13} then one can constrain the axion mass and the coupling from the observed spectra.
\cref{fig11} suggests that the cut-off frequency and the frequency window across which the conversion is efficient
is sensitive mainly to the axion mass and its coupling with the photons. The BH spin does not significantly affect the cut-off frequency and the conversion window, but it does increase the axion spectral luminosity, thereby leading to an enhanced dimming. This may not be monotonic with the BH spin but a spinning BH does cause a greater dimming compared to the Schwarzschild case \cite{Nomura:2022zyy}.

For a given $g_{a\gamma}$,
the conversion starts from a lower frequency and continues over a larger range of frequencies if the axion mass is low. On the other hand, for a given axion mass, the cut-off frequency lowers and hence the frequency window of conversion widens if the coupling is increased. If there are estimates of the electron number density and the magnetic field surrounding the BH, then the cut-off frequency can be determined from \cref{ineq1} and \cref{ineq2}. If the axion-mass happens to be lower than the resonant mass (given in \cref{mreso}) then the photon-axion coupling $g_{a\gamma}$ can be determined from the observed cut-off frequency (\cref{ineq2}). On the other hand if the axion-mass is greater than the resonant mass, then the cut-off frequency depends both on $m_a$ and $g_{a\gamma}$ (from \cref{ineq1}). In this case, if $g_{a\gamma}$ is known then the axion mass can be estimated from the cut-off frequency. Now, $g_{a\gamma}$ not only changes the cut-off frequency but also the axion spectral luminosity (\cref{fig11}). Thus, from the magnitude of dimming of the photon spectral luminosity
one can estimate $g_{a\gamma}$ provided the mass and spin of the BH have been estimated independently (since both the BH mass and spin affect the photon spectral luminosity, see \cref{S5b}). Once $g_{a\gamma}$ is determined, the axion mass can also be determined from the observed lower cut-off frequency. 
For M87*, there exist estimates of the BH mass \cite{EventHorizonTelescope:2019dse,Gebhardt:2009cr,Gebhardt:2000fk} and the BH spin $a\sim 0.5-0.9$ \cite{Tamburini:2019vrf,EventHorizonTelescope:2019pgp}. Hence, if future telescopes sensitive to the X-ray/gamma-ray band can achieve angular resolutions $\sim 10^{-5}$ arcsec (like the EHT), then the $m_a-g_{a\gamma}$ parameter space can be strongly constrained based on the observed photon spectra.

We further note from \cref{fig11} -\cref{fig13} that at very high frequencies the luminosity of the photons and also the axions drop as the exponential suppression in the luminosity becomes dominant (\cref{total-spectra} and \cref{J_e-4}). 
Thus, from the observed photon spectra one can indirectly infer about the axion mass and coupling and plausibly the BH spin. If the mass and spin of the source are known from other independent observations, then the degeneracy in the estimation of $m_a$ and $g_{a\gamma}$ can be further reduced. Note that, as mentioned in Sec.\ref{Sec2}, $g_{a\gamma}$ cannot be arbitrarily large, as different experiments and observations severely constrain $g_{a\gamma}\lesssim \mathcal{O}(10^{-11}) {\rm GeV^{-1}}$ \cite{Cameron:1993bhr,Minowa:1998sj,CAST:2004gzq,CAST:2024eil,GammeVT-969:2007pci, Ehret:2010zz,Povey:2010hs,OSQAR:2013jqp,OSQAR:2015qdv,ParticleDataGroup:2024cfk,ALPSII:2025eri}. Also, for efficient production of relativistic 
axions/ALPs $m_{a}\ll \omega$ has been considered.

\cref{fig12} and \cref{fig13} illustrate the sensitivity of the photon and the axion spectra on the magnetic field, the electron number density and their interplay with the BH spin. We assume axion mass $m_a\simeq $1 neV while the photon-axion coupling is taken to be $g_{a\gamma}\simeq 10^{-11}\rm ~GeV^{-1}$ (in \cref{fig12}) and $g_{a\gamma}\simeq 10^{-10}\rm ~GeV^{-1}$ (in \cref{fig13}). From the figures it is clear that the cutoff frequency from where the conversion starts becoming efficient decreases with a decrease in the electron density, i.e., the frequency window of conversion widens when the electron density is low. This is because for low electron densities $\mathrm{\Delta_{osc}}\simeq 2\mathrm{\Delta_M}$ (which ensures maximum conversion) starts holding good from a lower frequency (see \cref{eq:delta_M}, \cref{eq:delta_a}, \cref{eq:delta_pl} and \cref{eq:delta_vac}). For $g_{a\gamma}\simeq 10^{-11}\rm ~GeV^{-1}$, the conversion is most efficient in presence of high magnetic fields near maximally spinning BHs (\cref{fig12}) while for $g_{a\gamma}\simeq 10^{-10}\rm ~GeV^{-1}$ (\cref{fig13}), maximum conversion is achieved in intermediate magnetic fields around moderately spinning BHs. This is again because $\mathrm{\Delta_{osc}}z$ (the argument of the sinusoidal function in   \cref{eq:conversion_prob})  $\sim (2n+1)\pi$ for such values of magnetic field and path length (dependent on spin). When the coupling is increased by an order (to $g_{a\gamma}\simeq 10^{-10}\rm ~GeV^{-1}$), $\mathrm{\Delta_M}\simeq \mathrm{\Delta_{osc}}/2$ increases by an order and hence moderate magnetic field with a smaller path length can satisfy the criterion $\mathrm{\Delta_{osc}}z\simeq (2n+1)\pi$ required for maximum conversion. As photon path length increases with an increase in the BH spin, a smaller path length near an intermediate spin BH compensates for the enhancement in $\mathrm{\Delta_{osc}}$ due to an increase in the photon-axion coupling by an order. Note that the frequency window of conversion is still larger (and the cut-off frequency lower) (\cref{fig13}) for a higher magnetic field as efficient conversion necessarily requires fulfillment of the condition $\mathrm{\Delta_{osc}}\simeq 2\mathrm{\Delta_M}$ which is achieved from a lower frequency when the magnetic field in enhanced (\cref{eq:conversion_prob}, \cref{eq:oscillation}, \cref{eq:delta_a}, \cref{eq:delta_M}, \cref{eq:delta_pl} and \cref{eq:delta_vac}).  
Thus, efficient conversion requires low electron density and moderate to high magnetic fields (depending on the photon-axion coupling and the BH spin). Thus, if the mass and spin of the BH are known from independent observations, then the observed spectra can be used to constrain the $B-n_e-m_a-g_{a\gamma}$ parameter space.

\section{Conclusion}
\label{S6}
In this work we investigate the signatures of photon-axion conversion near Kerr BHs. Such conversion can only happen in the presence of magnetic fields and is much more efficient near supermassive BHs (SMBHs) compared to stellar mass BHs. Supermassive BHs like M87* have been imaged by the Event Horizon Telescope (EHT) and is reported to host magnetic fields $\sim 1-30 \rm G$ in its vicinity. 
The characteristic distance over which photon-axion conversion occurs is generally comparable to, or exceeds, the event-horizon scale of a supermassive black hole, which might imply that the magnetic field must persist over similarly large radial distances. However, the influence of strong gravity allows photons to remain confined near the black hole for prolonged durations. In particular, photons can execute multiple orbits within the photon region along unstable spherical photon orbits before eventually escaping to distant observers. During this stage, they travel at nearly fixed radii, effectively experiencing a sustained magnetic field along their trajectories. If such photons undergo conversion into axions, this process could lead to a reduction in the luminosity of the observed photon spectra.
Thus, sources like M87* are ideal astrophysical laboratories to explore the footprints of photon-axion conversion. The present work investigates such conversion around rotating Kerr BHs which is an extension over earlier works \cite{Nomura:2022zyy, Roy:2023rjk,Hazarika:2024nrj} which studied the implications of the photon-axion conversion around 
static BHs. 

In order to obtain the conversion probability we need to evaluate the path length traversed by the photon in the photon region/close to the photon sphere.
It is important to note that rotating BHs have a photon region compared to a photon sphere for static BHs. We consider photons entering the photon region at near critical impact parameters and numerically evaluate the path length traversed by the photon using the first order geodesic equations. 
We systematically study the effect of the photon frequency, the photon-axion coupling, the mass of the converted axion, the magnetic field, the electron density of the plasma and the BH spin on the conversion probability which in turn leads to a reduction in the photon spectral luminosity.  

Our study reveals that photon-axion conversion is generally most efficient for high frequency photons like X-rays and gamma rays, although conversion in the radio frequency can happen if the axion assumes the resonant mass (\cref{mreso}), determined by the electron density. Since distortion in the photon spectra from M87* or Sgr A* has not been observed in the radio band, then we may infer that the mass of the axion is different from the resonant mass. We further note that no conversion takes place if the photon-axion coupling $g_{a\gamma}\lesssim 10^{-12}~\rm GeV^{-1}$ (\cref{fig11}). Also, $g_{a\gamma}$ cannot be arbitrarily large as there exist several astrophysical constraints restricting the photon-axion coupling to $g_{a\gamma}\lesssim 10^{-10}~\rm GeV^{-1}$ \cite{Noordhuis:2022ljw, Dessert:2022yqq,Dolan:2022kul}. Thus, in this work we mainly consider $g_{a\gamma}\sim 10^{-11}~{\rm and} ~10^{-10}~\rm GeV^{-1}$. We further report that the conversion becomes effective only above a cut-off frequency which depends primarily on the axion mass and the electron density and moderately on the photon-axion coupling. A lower axion mass and electron density along with an enhanced coupling leads to a reduction in the lower cut-off frequency and  widens the frequency window of conversion. In this work we consider conversion to axion of mass $m_a\lesssim 10^{-7}\rm eV$ based on previously reported constraints \cite{Noordhuis:2022ljw, Dessert:2022yqq,Dolan:2022kul}.

The magnitude of dimming of the photon spectral luminosity depends primarily on the magnetic field, the photon-axion coupling and the BH spin. When the photon-axion coupling is $g_{a\gamma}\sim 10^{-11} \rm GeV^{-1}$, the conversion is most efficient near rapidly rotating BHs in presence of high magnetic field ($B\sim 30\rm G$, \cref{fig12}). Interestingly, with an enhanced coupling i.e., $g_{a\gamma}\sim 10^{-10} \rm GeV^{-1}$, maximum conversion is achieved in intermediate magnetic fields around moderately spinning BHs (\cref{fig13}). This is because, when the coupling is increased by an order, $\mathrm{\Delta_{osc}}\simeq 2\mathrm{\Delta_M}$ subsequently increases, such that even a lower magnetic field and a smaller path length $z$ (which decreases with a decrease in the BH spin) can fulfill the condition $\mathrm{\Delta_{osc}}z\simeq (2n+1)\pi$ (\cref{eq:conversion_prob}), necessary for maximum conversion. In general however, the dimming percentage is more around a rotating BH than a static one which can be understood by comparing our \cref{fig11} with the results of Nomura et al. \cite{Nomura:2022zyy}.

Thus, if telescopes like Chandra (with present resolution of 1 arcsec) can resolve the photon region of supermassive BHs (e.g. M87*) (resolution of $10^{-5}$ arcsec required) or if the EHT becomes sensitive in the X-rays or gamma rays, then an observed dimming in the photon spectral luminosity can be used probe the presence of axions near these objects. In this regard one is referred to Uttley et al. \cite{Uttley:2019ngm}
which proposed high-resolution X-ray interferometry.
From the dimming percentage, the cut-off frequency and the frequency window of conversion, we can constrain the parameter space associated with the BH mass and spin, magnetic field strength, electron density, the axion mass and the photon-axion coupling. The degeneracy in the estimates of the aforesaid parameters (in particular, the axion mass and the photon-axion coupling) can be further reduced if the BH mass and spin are known from independent observations and there exist independent estimates of the electron density and the magnetic field strength around the BHs (as is known for the case of M87* and Sgr A*). This can then provide valuable insights into the mass and coupling of axions.

The present analysis is mainly performed with parameters (mass, magnetic field and electron density) associated with M87* but can be easily extended to other SMBHs, e.g. Sgr A*. M87*, however, turns out to be a better candidate to study photon-axion conversion near SMBHs since the spectral flux of photons (which in turn gets converted to axions) is substantially larger for M87* compared to Sgr A*. This is because the photon spectral luminosity $\propto M^3$ (\cref{S5b}) while the spectral flux of photons $\propto M^3/D^2$ \cite{Roy:2023rjk}. Thus, Sgr A* has a lesser photon spectral flux compared to M87* since both its distance and mass are lesser than M87* by three orders of magnitude.

In the present work we consider photon-axion conversion in the presence of a background magnetic field. However, 
it would also be worthwhile to analyze this conversion not only in the presence of a magnetic-field background, but also within an axion background \cite{Masaki:2019ggg}. Indeed, axions may constitute dark matter \cite{preskill:1983,Hu:2000abc,Hui:2016ltb,Chadha-Day:2021szb,Marsh:2015xka} or be generated through superradiance around black holes \cite{Arvanitaki:2009fg, Brito:2015oca}.
Another important direction would be to investigate how photon–axion conversion influences the polarization of radiation emitted from the photon sphere \cite{Dessert:2022yqq}. By investigating these polarized emissions one can gain valuable insights into axion properties.  
In the present work we estimated the axion flux from the photon region of a single BH. This flux can be substantially enhanced if we consider an ensemble of BHs. This is realistic as the Universe hosts an enormous population of black holes and comprises quasars which are much more luminous than low-luminosity active galactic nuclei such as M87*, and their corresponding axion output could be significantly higher. The cumulative contribution from such sources may form a component of the cosmic axion background \cite{Dror:2021nyr}. A detailed evaluation of these effects remains an interesting avenue which can be explored in the future.

\section*{Acknowledgements}
Research of I.B. is funded by the Start-Up
Research Grant from SERB, DST, Government of India
(Reg. No. SRG/2021/000418). S.S. acknowledges financial support from the Shiv Nadar Institution of Eminence and the Council of Scientific and Industrial Research (CSIR), Government of India, for a Direct-SRF fellowship under grant  09/1128(18274)/2024-EMR-I. RD acknowledges CSIR, India for financial support through Senior Research Fellowship (File no. 09/ 1128 (13346)/ 2022 EMR-I) and Shiv Nadar IoE (Deemed to be University).

\section*{Appendix: Photon-Axion Mixing Formalism}
\appendix
\label{AppendixA}

\setcounter{equation}{0}
\renewcommand{\theequation}{A.\arabic{equation}}

\subsection*{A.1. The Minimal Lagrangian and Equations of Motion}

We begin by considering the minimal action describing the interaction between a photon and a pseudo-scalar axion field. The Lagrangian density is given by:
\begin{equation}
    \mathcal{L}_0 = -\frac{1}{4}F_{\mu\nu}F^{\mu\nu} - \frac{1}{2}\partial_\mu \phi \partial^\mu \phi - \frac{1}{2} m_a^2 \phi^2 - \frac{1}{4} g_{a\gamma}\phi F_{\mu\nu}\tilde{F}^{\mu\nu},
\end{equation}
where $\phi$ is the axion field, $m_a$ is the axion mass, $g_{a\gamma}$ is the axion-photon coupling constant, $F_{\mu\nu}$ is the electromagnetic field strength tensor, and $\tilde{F}_{\mu\nu} = \frac{1}{2}\epsilon_{\mu\nu\rho\sigma}F^{\rho\sigma}$ is the dual tensor. The resulting Euler-Lagrange equations of motion are:
\begin{align}
    \Box\phi - m_a^2 \phi &= \frac{1}{4} g_{a\gamma}F_{\mu\nu}F^{\mu\nu}, \\
    \partial_\mu F^{\mu\nu} &= -g_{a\gamma}\tilde{F}_{\mu\nu} \partial_\mu \phi.
\end{align}

To accurately model the propagation in a strong magnetic field, we must account for vacuum polarization effects (the Heisenberg-Euler term). The effective Lagrangian for the photon-axion system becomes \cite{Raffelt:1988}:
\begin{align}
    \mathcal{L} =& -\frac{1}{4}F_{\mu\nu}F^{\mu\nu} - \frac{1}{2}\partial_\mu\phi\partial^\mu\phi -\frac{1}{2}m_a^2\phi^2 -\frac{1}{4}g_{a\gamma}\phi F_{\mu\nu}\tilde{F}^{\mu\nu} \nonumber \\
    &+ \frac{\alpha^2}{90m_e^4}\left[(F_{\mu\nu}F^{\mu\nu})^2 + \frac{7}{4}(F_{\mu\nu}\tilde{F}^{\mu\nu})^2\right],
\end{align}
where $\alpha \approx 1/137$ is the fine-structure constant and $m_e$ is the electron mass. 

From this effective Lagrangian, the modified equations of motion are obtained as:
\begin{align}
    \partial_\mu\partial^\mu\phi - m_a^2\phi &= \frac{1}{4}g_{a\gamma}F_{\mu\nu}\tilde{F}^{\mu\nu}, \\
    \partial_\mu F^{\mu\nu} &= -g_{a\gamma}\tilde{F}^{\mu\nu}\partial_\mu\phi + \frac{4\alpha^2}{45m_e^4}\partial_\mu\left(F^2 F^{\mu\nu} + \frac{7}{4}(F\tilde{F})\tilde{F}^{\mu\nu}\right),
\end{align}
where we denote $F^2 \equiv F_{\rho\sigma}F^{\rho\sigma}$ and $F\tilde{F} \equiv F_{\rho\sigma}\tilde{F}^{\rho\sigma}$.

We decompose the field into a strong constant background and a propagating wave: $F_{\mu\nu} = \bar{F}_{\mu\nu} + (\partial_\mu A_\nu - \partial_\nu A_\mu)$. The background magnetic field is defined as $\bar{F}_{0i}=0$ and $\tilde{\bar{F}}_{0i}=B_i$. We adopt the Coulomb gauge, $\nabla \cdot \mathbf{A} = 0$. Keeping terms to linear order in the propagating fields $A_\mu$ and $\phi$, the equations of motion take the form:
\begin{align}
    \Box \mathbf{A} - \nabla\dot{A}^0 &= g_{a\gamma}\mathbf{B}\dot{\phi}, \\
    \nabla^2 A^0 &= -g_{a\gamma}\mathbf{B}\cdot\nabla\phi, \\
    (\Box - m_a^2)\phi &= -g_{a\gamma}\mathbf{B}\cdot(\dot{\mathbf{A}}+\nabla A^0).
\end{align}

We define the geometry such that the wave propagates along the $z$-axis. The constant magnetic field is oriented at an angle $\theta$ relative to the $z$-axis:
\begin{equation}
    \mathbf{B} = (B \sin\theta, 0, B \cos\theta).
\end{equation}
Due to the Coulomb gauge condition, the photon polarization states are transverse. We can define the polarization components as:
\begin{equation}
    \mathbf{A} = (A_{||}(z,t), A_{\perp}(z,t), 0),
\end{equation}
where $A_{||}$ is parallel to the transverse component of $\mathbf{B}$. Substituting this geometry into the linearized equations (Eqs. A.7-A.9), we find that $A_{\perp}$ decouples, while $A_{||}$ mixes with the axion. 

Adding a phenomenological term $-\omega_{pl}^2 A_{||}$ to account for the plasma effective mass, the coupled equations for $A_{||}$ and $\phi$ are:
\begin{align}
    (\Box - m_a^2)\phi + \omega g_{a\gamma} B \sin\theta A_{||} &= 0, \\
    (\Box - \omega_{pl}^2) A_{||} + \frac{7\omega^2\alpha}{45\pi}\left(\frac{eB}{m_e^2}\right)^2\sin^2\theta A_{||} + \omega g_{a\gamma} B \sin\theta \phi &= 0.
\end{align}

To simplify notation, we define the mixing parameters:
\begin{equation}
    Q_{||} = \frac{7\omega^2\alpha}{45\pi}\left(\frac{eB}{m_e^2}\right)^2\sin^2\theta - \omega_{\rm pl}^2, \quad Q_a = -m_a^2, \quad Q_{a\gamma} = \omega g_{a\gamma}B\sin\theta.
\end{equation}
Assuming the fields depend only on $z$ and $t$, the system can be written in matrix form:
\begin{equation}
    (\mathbf{D} + \mathbf{M})\mathbf{\Psi} = 0,
\end{equation}
where $\mathbf{D} = (-\partial_t^2 + \partial_z^2)\mathbb{I}$, and the mixing matrix and state vector are:
\begin{equation}
    \mathbf{M} = \begin{pmatrix} Q_{||} & Q_{a\gamma} \\ Q_{a\gamma} & Q_a \end{pmatrix}, \qquad \mathbf{\Psi}(z,t) = \begin{pmatrix} A_{||} \\ \phi \end{pmatrix}.
\end{equation}

We assume plane-wave solutions of the form $A_{||}(z,t) = \tilde{A}(z)e^{-i(\omega t - kz)}$ and $\phi(z,t) = \tilde{\phi}(z)e^{-i(\omega t - kz)}$. In the relativistic limit, the envelope functions $\tilde{A}$ and $\tilde{\phi}$ vary slowly compared to the oscillation frequency ($|\partial_z^2 \tilde{A}| \ll k|\partial_z \tilde{A}|$). The second-order system reduces to a first-order Schrödinger-like equation:
\begin{equation}
    i\frac{d}{dz}\tilde{\mathbf{\Psi}}(z) = \mathbf{M}_{\rm eff}\,\tilde{\mathbf{\Psi}}(z), \qquad \text{where} \quad \mathbf{M}_{\rm eff} = \begin{pmatrix} \Delta_{||} & -\mathrm{\Delta_M} \\ -\mathrm{\Delta_M} & \mathrm{\Delta_a} \end{pmatrix}.
\end{equation}
Here, the specific wave-number shifts are defined as:
\begin{align}
    \mathrm{\Delta_M} &= \frac{1}{2} g_{a\gamma} B \sin\theta, \\
    \mathrm{\Delta_a} &= \frac{m_a^2}{2\omega}, \\
    \Delta_{||} &= \Delta_{\text{pl}} - \Delta_{\text{vac}} = \frac{\omega_{\rm pl}^2}{2\omega} - \frac{7\alpha}{90\pi}\frac{\omega B^2 \sin^2\theta}{m_e^4}.
\end{align}

The mixing matrix is diagonalized by the angle $\theta_{\rm mix}$, satisfying $\tan(2\theta_{\rm mix}) = 2\mathrm{\Delta_M} / (\mathrm{\Delta_a} - \Delta_{||})$. The eigenvalues are $\lambda_{\pm} = \frac{1}{2}(\mathrm{\Delta_a} + \Delta_{||} \pm \mathrm{\Delta_{osc}})$, where the oscillation wave number is:
\begin{equation}
    \mathrm{\Delta_{osc}} = \sqrt{(\mathrm{\Delta_a} - \Delta_{||})^2 + 4\mathrm{\Delta_M}^2}.
\end{equation}
For an initial state consisting of a pure photon ($\tilde{A}(0)=1, \tilde{\phi}(0)=0$), the probability of conversion into an axion after a distance $z$ is:
\begin{equation}
    P_{\gamma \rightarrow a}(z) = |\tilde{\phi}(z)|^2 = \left(\frac{2\mathrm{\Delta_M}}{\mathrm{\Delta_{osc}}}\right)^2 \sin^2\left(\frac{\mathrm{\Delta_{osc}} z}{2}\right).
\end{equation}
Substituting the explicit form of the oscillation parameter:
\begin{equation}
    \mathrm{\Delta_{osc}}^2 = \left(\mathrm{\Delta_{pl}} - \mathrm{\Delta_{vac}} - \mathrm{\Delta_a}\right)^2 + 4\mathrm{\Delta_M}^2.
\end{equation}

\bibliographystyle{apsrev4-2}
\bibliography{biblio.bib}

\end{document}